\documentclass[namedreferences,hyperref,optionalrh,solaromanenum]{spr-sola}

\usepackage{graphicx}                    
\usepackage{color}                       
\usepackage{amsmath}
\usepackage{graphicx}
\usepackage{txfonts}
\usepackage{hyperref}
\usepackage{textcomp}
\usepackage{gensymb}

\newcommand{\aap}{{\it Astron. Astrophys.}}

\newcommand{\apjl}{{\it Astrophys. J. Lett.}}

\chardef\us=`\_

\begin{document}

	\begin{frontmatter}
		
		\title{Solar Photospheric Flows and the Origin of Active Regions}
		
		%
		\author[addressref={aff1},corref,email={A.Getling@mail.ru}]{\inits{A.V.}\fnm{Alexander V. }\snm{Getling}\orcid{0000-0002-1766-2620}}
		
		\address[id=aff1]{Skobeltsyn Institute of Nuclear Physics, Lomonosov Moscow State
			University, Moscow, 119991 Russia}

		%
		\runningauthor{Getling}
		\runningtitle{\textit{Solar Physics} Flow Dynamics and the Origin of Active Regions}
		
		
		\begin{abstract}
			The early evolutionary stages of some active regions (ARs) are analysed with the aim of resolving one of the paramount-importance problems of heliophysics: what plays a primary, active role in the formation of active regions and sunspots---magnetic fields or motion of solar plasma? In particular, the first view underlies the popular model of rising $\Omega$-shaped, twisted flux-tube loop of strong magnetic field. In this study, optical images obtained by the Helioseismic and Magnetic Imager (HMI) of the Solar Dynamics Observatory are used along with full-vector maps of magnetic field based on HMI data. Maps of fluid-particle tracks are constructed using local correlation tracking of photospheric brightness elements. Comparisons of optical and magnetic maps for three developing active regions confirm the conclusion made by Bumba and Howard in the mid-1960s that the supergranulation pattern is not destroyed by the emerging magnetic field of the developing AR, which would be inevitable if a loop of the strong-magnetic-field flux tube rose through the subphotospheric layers. There are no signs of spreading flows on scales comparable to the size of the whole active region. Various forms of dissymmetry between the leading and trailing parts of ARs are revealed. Next, an example of a well-developed sunspot group is considered, which demonstrates that the twist of the magnetic configuration may develop much later than the group starts growing.  It is concluded that plasma flows appear to be determinant with respect to magnetic fields. The considered scenarios of the origin of ARs are in sharp contradiction with the rising-tube model. 
		\end{abstract}

		%
		\keywords{Active Regions --- Supergranulation}

	\end{frontmatter}
	
	%
	\section{Introduction} \label{sec:intro}

    It can hardly be disputed that the origin of the entire complex of solar-activity phenomena is rooted in the convection zone. The formation of active regions (ARs) and sunspot groups results from the inherently deep interaction between the solar-plasma flows and magnetic fields. Therefore, understanding this interaction is an issue of primary importance in heliophysics.
    
    The basic question of heliophysical ``philosophy'' is whether (A) the magnetic field plays an active, primary role in the AR development or (B) this role belongs to the plasma flow. The first view underlies the widely known rising-tube model (RTM), which assumes that an $\Omega$-shaped loop of a flux tube of strong magnetic field (typically assumed to be twisted), with the plasma this field is frozen in, emerges on the photospheric surface due to magnetic buoyancy. The second approach implies consideration of various possible mechanisms of \textit{in situ} magnetic-field amplification and structuring by plasma motion. These are in particular local dynamo mechanisms. 

   	The RTM has a number of attractive properties. It is especially interesting because it fits the sunspot-formation process into the overall spatiotemporal pattern of solar activity as described by the Hale and Sp\"orer laws. Largely due to this fact, it became a paradigm in the studies of AR-formation processes for several decades. The review article by \citet{Fan_2009}, written from the position of RTM-based views, presents a number of references to relevant papers.
    
    From today's point of view, however, it seems to be a legacy of the era of low spatial and temporal resolutions. As observations became progressively more detailed, new facts contradicting the RTM were revealed that make the common adequacy of this model doubtful. Moreover, in an updated version of her review article, \citet{Fan_2021} writes: ``new results from the global simulations of the convective dynamos, and from the near-surface layer simulations of active region formation, together with helioseismic investigations of the pre-emergence active regions, are calling into question the picture of active regions as buoyantly rising flux tubes''. 
    
    Let us mention some points of criticism regarding the RTM. \citet{Jabbari_etal_2016} note that ``so far neither numerical \citep{Guerrero_Kapyla_2011} nor observational \citep{Fan_2009,Zhao_etal_2013,Getling_etal_2016} studies have confirmed this scenario''. Also, no signs of rising tubes have been revealed by helioseismology---see discussions by \citet{Leka_etal_2013,Birch_etal_2013,Warnecke_etal_2013, Warnecke_etal_2016}. As \citet{Weber_etal_2023} recognise, new investigations suggest that ``flux emergence may be a more passive process (at least in the upper convection zone); dominated to a greater extent by the influence of convection and to a lesser extent by buoyancy and the Coriolis force acting on rising magnetic flux tubes than previously thought''; thus, they directly address view (B).
    
    Alan Title in his lecture at the 26th IAU General Assembly (Prague, 2006; unpublished) called attention to the fact that, as an AR originates, magnetic elements of both polarities may emerge within the same supergranule. Obviously, if these nuclei of a growing bipolar magnetic region of an AR belonged to a rising $\Omega$-shaped loop of a magnetic tube, they would not preserve their supergranule intact but the apex of the emerging loop would distinctly change it.
    
    Next, in a discussion at a ``Living with a Star'' workshop in September 2007, Marc DeRosa commented on the development of one particular AR as follows (we quote a NASA report, which has later been removed from the Internet): ``the emergence of the sunspot magnetism progressed in a very complex manner, with small pieces appearing to self-assemble into larger, more coherent structures''. A view of the AR- and sunspot-forming process as the self-assemblage of small magnetic elements into large ones 
    has repeatedly been expressed \citep[][etc.]{Parker_1992,Zelenyi_Milovanov_1992}. In particular,
    \citet{Chumak_2013}, based on a theoretical, plasma-physical standpoint, simulated such a process solving a set of stochastic equations of the motion of ``elementary'' (i.e., filamentary) magnetic flux tubes. 
    
    \citet{Stein_Nordlund_2012} emphasise \citep[with reference to][]{Cheung_etal_2007,Martinez-Sykora_etal_2008, Weber_etal_2011,Fang_etal_2012} that convective motions are important to the magnetic-flux emergence. \citet{Sadykov_Krasotkin_2024} report the predominance of material downflows in developing ARs. Among the remarkable features of the AR development scenarios is the asymmetry of magnetic fluxes of the two polarities \citep{Bumba1965, Getling_Buchnev_2019, Grigoryev_etal_2020}. We will return to this particularity in Subsection~\ref{asym}.
    
    In the context of our discussion, the paper by \citet{Stein_Nordlund_2012} is of particular importance. It demonstrates that convection itself can produce bipolar magnetic configurations without a previously formed coherent strong-magnetic-flux tube as a necessary prerequisite of such a process. This is definitely a weighty argument in favour of view (B) on AR formation.
    
    A critical discussion of the RTM is given, in particular, in a comprehensive review by \citet{Zagainova_2022}. They note that the RTM-based possibility of sunspot formation is questioned and even ``closed'' by numerous observations. Nevertheless, the treatment of the AR formation as the result of the buoyant emergence of $\Omega$-shaped loops of flux tubes can still be encountered in the current literature \citep[][etc.]{Levens_etal_2023, Poisson_etal_2024,Grigoryev_Ermakova_2025}. 

    To approach the understanding of the AR- and sunspot-formation mechanisms, discriminating between possibilities (A) and (B) for the interaction between plasma motion and magnetic field, fairly detailed information on the dynamics of plasma flows and magnetic fields is needed. The structure of solar-plasma flows on supergranular and larger scales can successfully be traced using the local correlation tracking of photospheric brightness elements (granules). The construction of their trajectories visualises the velocity field to an acceptable accuracy, being highly informative in this respect.
    
    From the standpoint of our approach, it is important to give much attention to an observational finding made six decades ago by \citet{Bumba1965} \citep[see also][]{Bumba1967}, which has, in essence, been completely ignored by heliophysicists in the subsequent years. They found that ``the supergranular pattern is of fundamental importance in the formation and structure of individual active regions'', this pattern controls the development of the magnetic field in the AR, and the ordered motion within supergranules can even ``greatly influence the positions of spots''. Accordingly, the magnetic field in a developing AR \emph{does not break the pattern of supergranulation} but penetrates, or seeps, through it. This fact is in sharp contradiction with the RTM, is an important argument against the RTM-based ideas about the formation mechanisms for ARs and attributes these mechanisms to class (B).

    Another feature that seems directly related to making an adequate choice between the alternatives is the frequently observed asymmetry of polarities in the growing AR. It has been repeatedly noted in the literature \citep[again,][and others---see Subsection~\ref{asym}]{Bumba1965,Bumba1967}, and its apparent inconsistency with the RTM should be taken into account.
     
	In this paper, the interaction of the material flow and magnetic field at the early development stages of some ARs will be considered. After a brief characterisation of the observational source data in Section~\ref{data}, a comparative data analysis for three ARs will be presented in Section~\ref{results}. Specifically, by means of comparing the optical images, the maps of the velocity field and full-vector maps of the  magnetic field, some remarkable features of the sunspot-forming process will be revealed. In addition, an example of a well-developed sunspot group will be considered, which demonstrates that the twist of the magnetic configuration can arise much later than the group starts developing (Subsection~\ref{11875}). In Subsection~\ref{asym}, attention will be given to the asymmetry between the fluxes of positive and negative magnetic field and between the leading and trailing parts of the sunspot group. The results will be summarised in Section~\ref{conclusion}, where some studies seemingly contradicting our results will also be discussed. 

    	\section{Source data}\label{data}
    
    In this study, for each specified AR, we used data from the Helioseismic and Magnetic Imager (HMI) of the Solar Dynamics Observatory (SDO), which are stored at and available from the Joint Science Operations Center (JSOC; \url{http://jsoc.stanford.edu}). The area to consider was chosen automatically as a Spaceweather HMI Active Region Patch \citep[SHARP; see][]{Bobra_etal_2014} centred at the magnetic-flux-weighted centroid of the given AR. The magnetic-field-vector data were remapped to a Lambert cylindrical equal-area projection (CEA). The continuum optical images were taken for the same CEA-remapped area. We assumed the horizontal velocity fields to be determined with a satisfactory accuracy by tracking the motion of brightness elements (granules) from series of such images using a modified local-correlation-tracking (LCT) technique \citep{Getling_Buchnev_2010}. The images were preliminary subjected to the procedure of Fourier subsonic filtration \citep{Title_etal_1989} with a cutoff phase velocity of 4 km\,s$^{-1}$. They were taken for tracking with a cadence of 135 s and magnified by doubling the number of pixels in both horizontal dimensions (the original pixel size being 0.5~arcsec $\approx 366$~km). The total tracking interval was 1~h~52~m~30~s or 3~h~45~m.

\begin{figure}   
	\flushleft{
    	\includegraphics[width=0.432\textwidth,bb=0 15 680 335,clip]{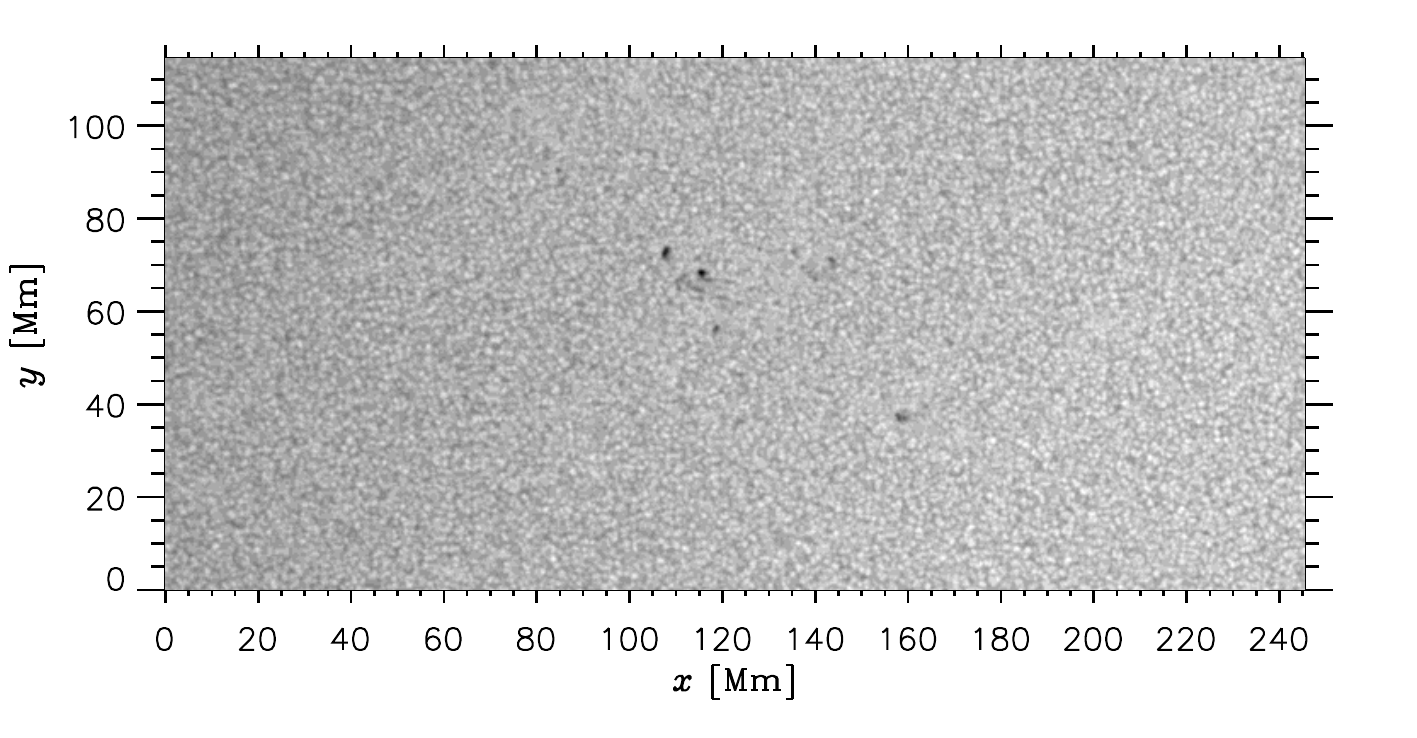}\hspace{0.748cm}		
		\includegraphics[width=0.432\textwidth,bb=0 15 680 335,clip]{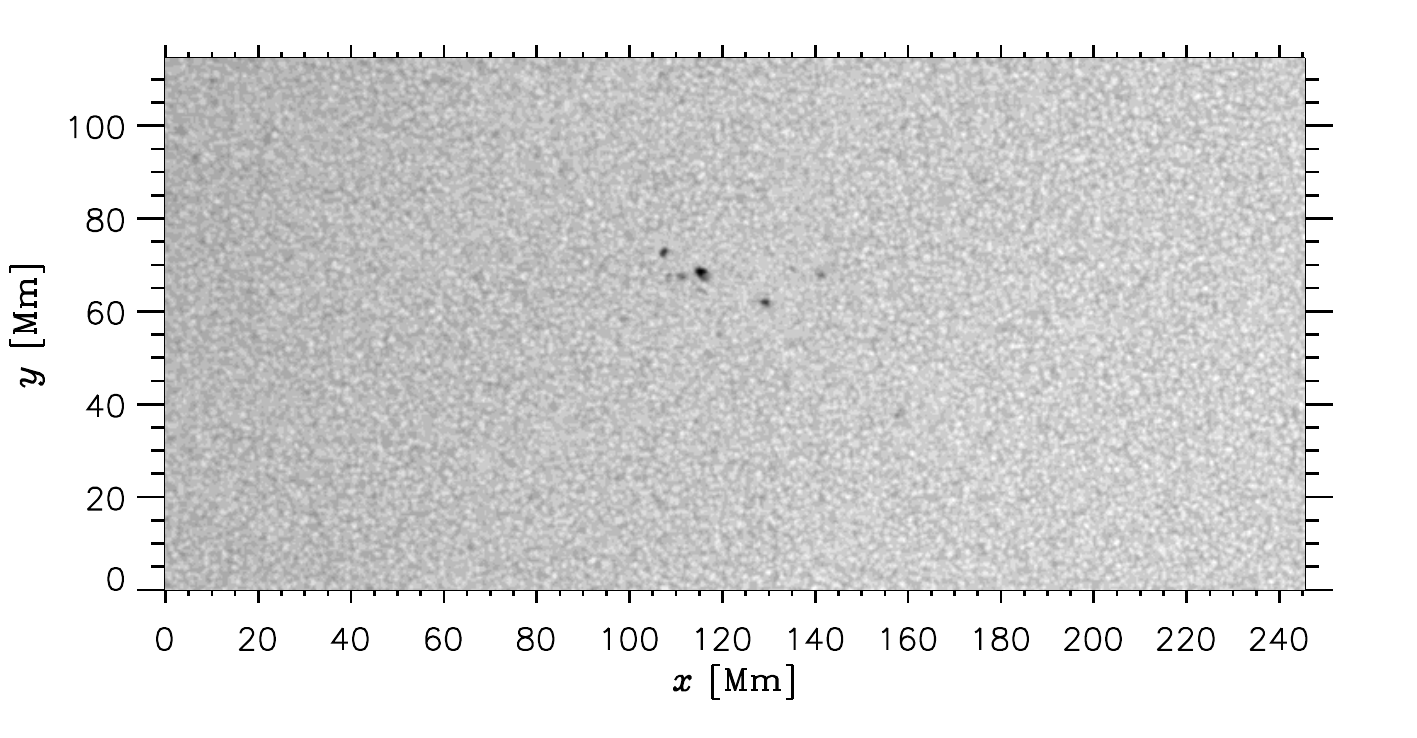}\\
		\includegraphics[width=0.432\textwidth,bb=0 15 680 335,clip]{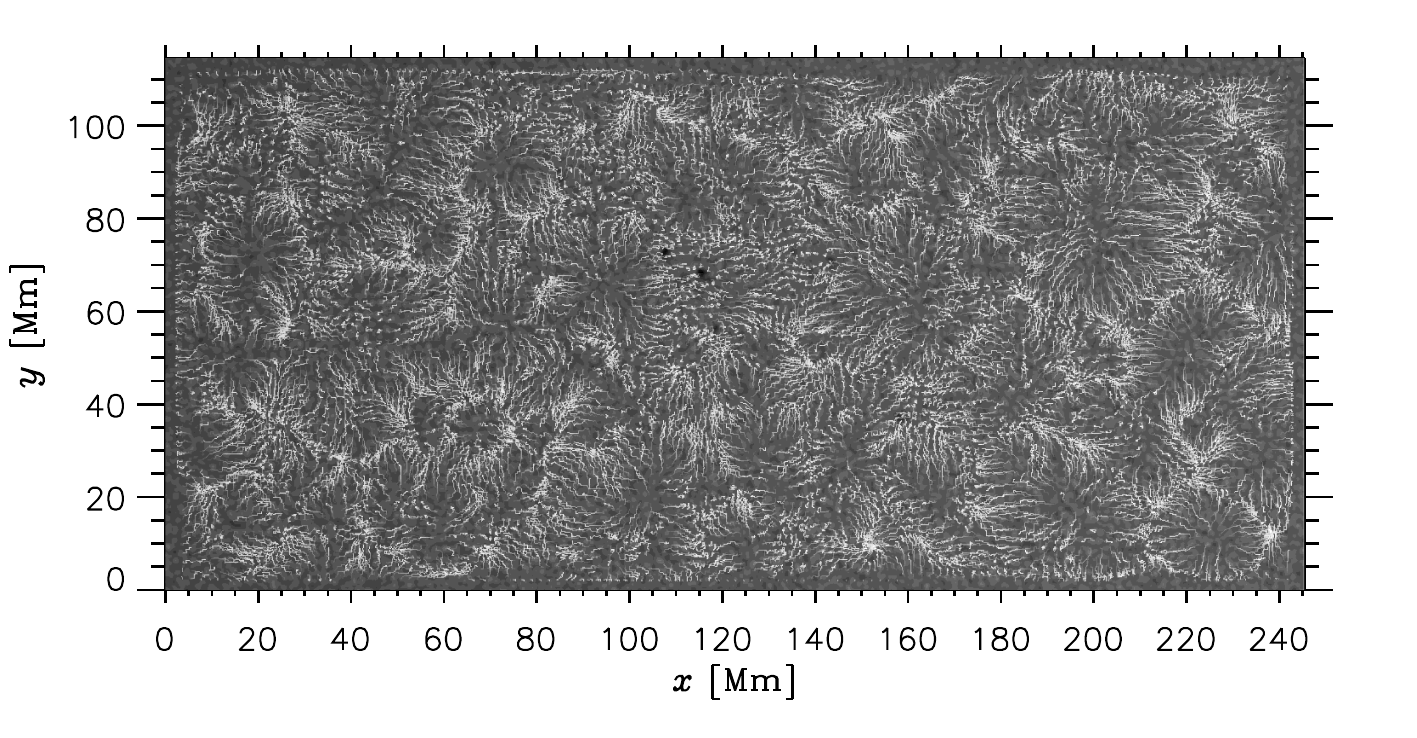}\hspace{0.748cm}
		\includegraphics[width=0.432\textwidth,bb=0 15 680 335,clip]{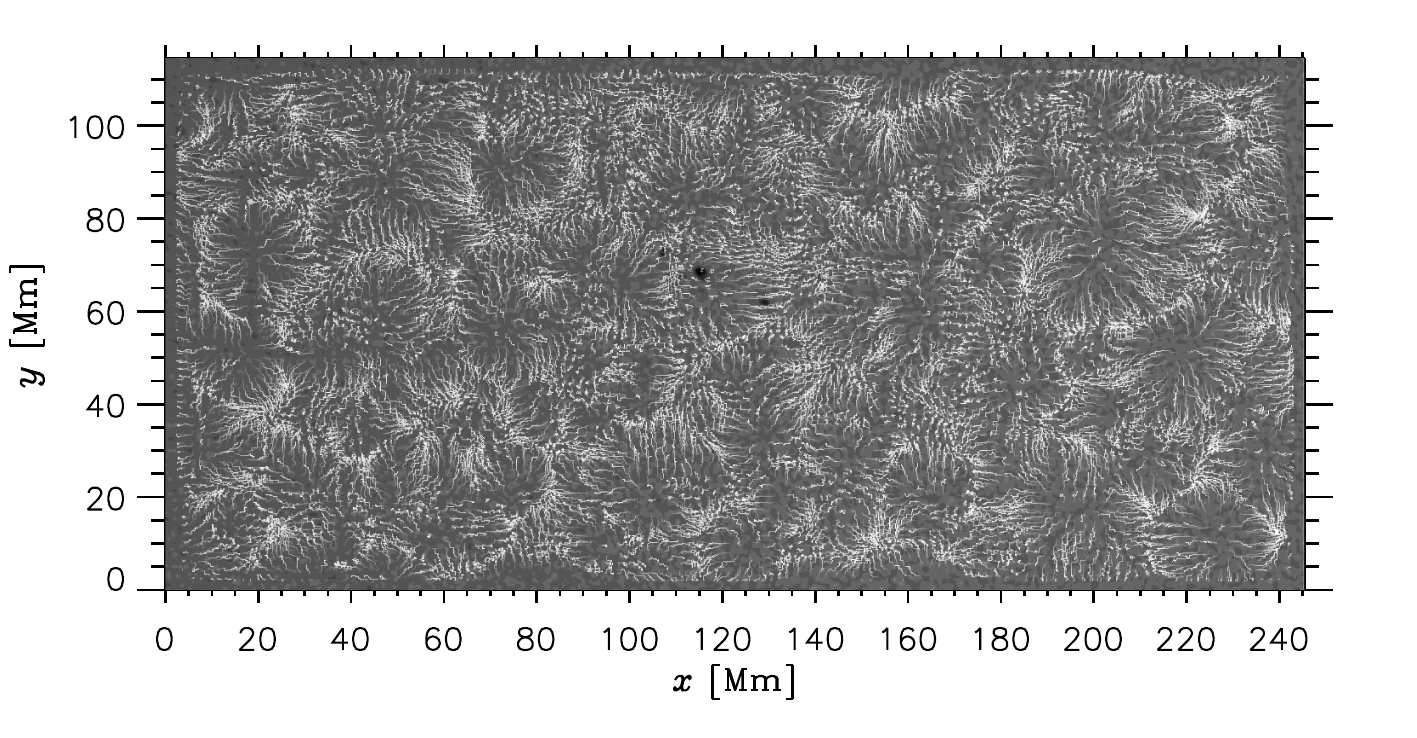}\\
		\includegraphics[width=0.495\textwidth,bb=15 0 793 335,clip]{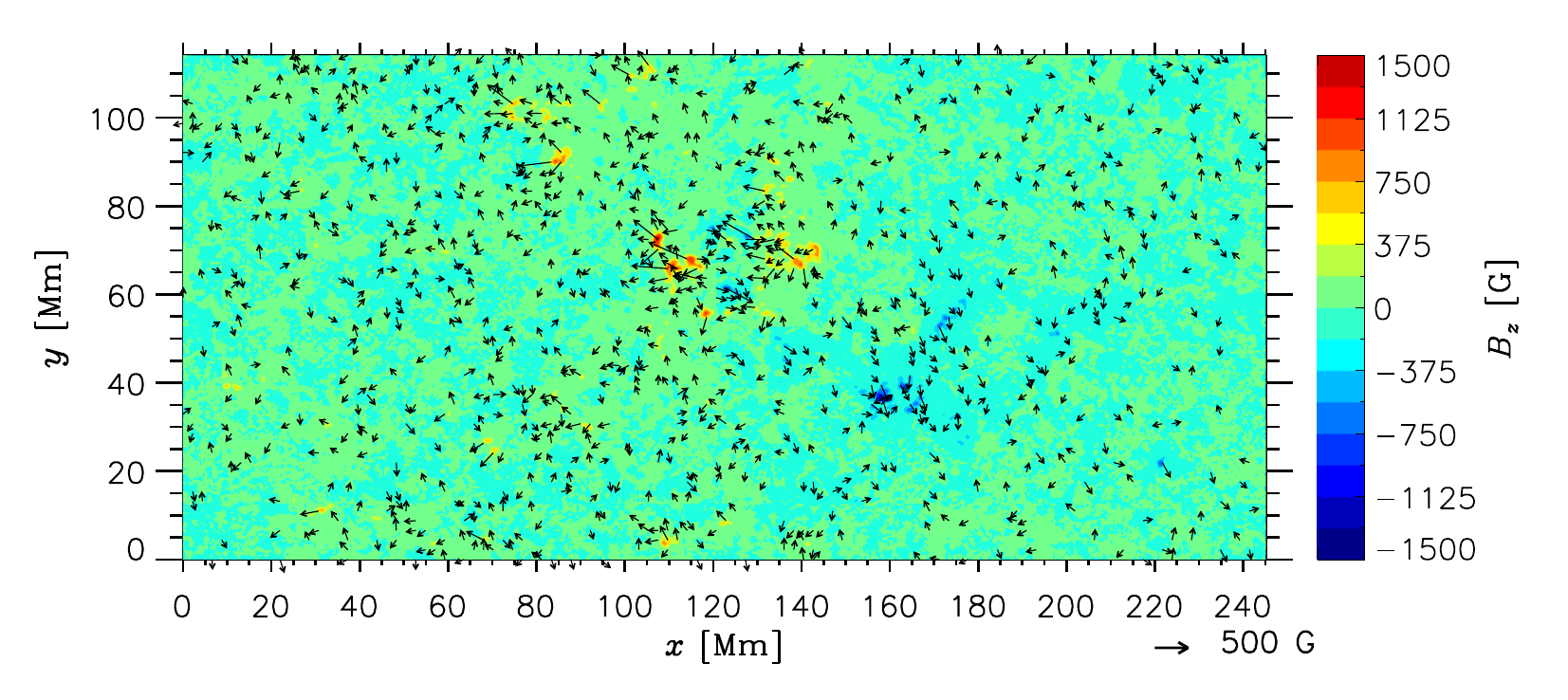}
		\includegraphics[width=0.495\textwidth,bb=15 0 793 335,clip]{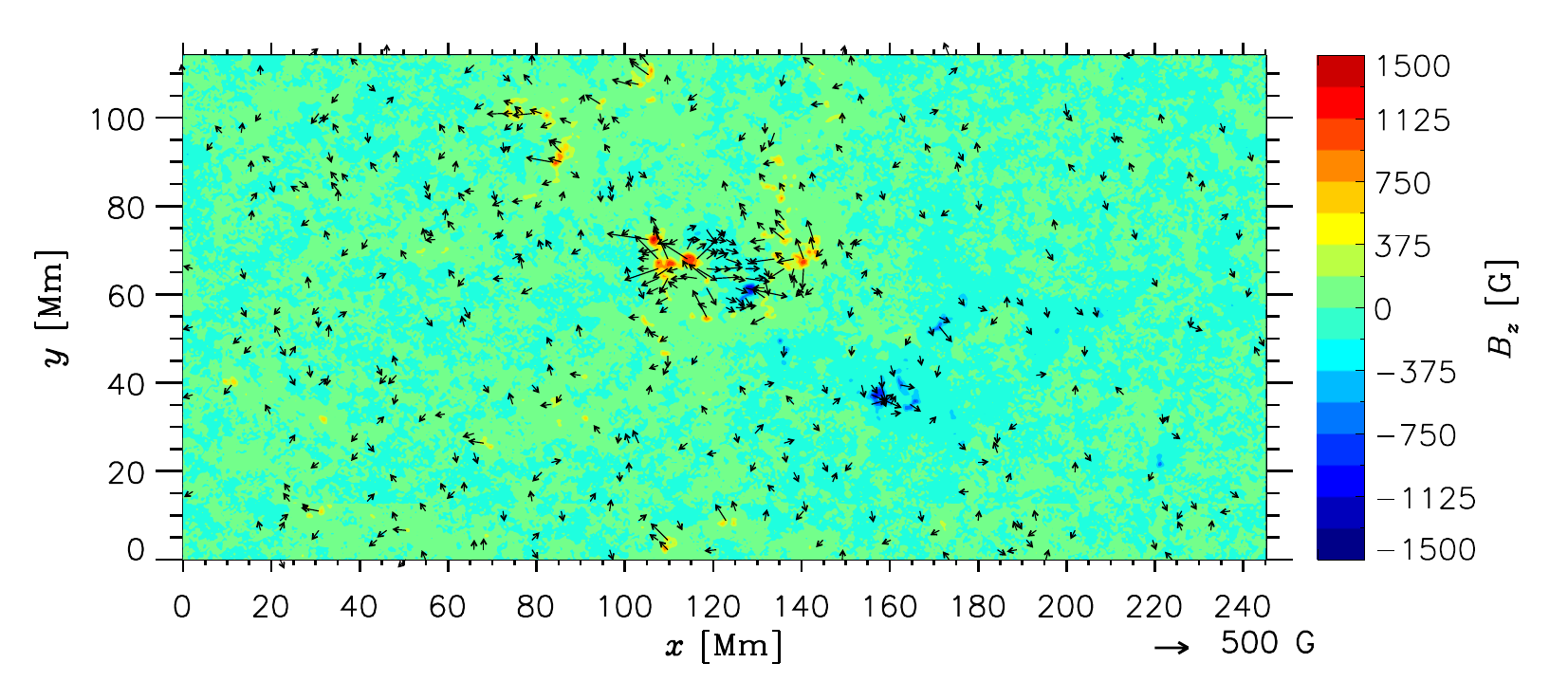}\\[-10pt]
		{\tiny \hspace{1.8cm}  2015 Jan 06, 07:21 \hspace{4.4cm} 2015 Jan 06, 09:13}\\
		\includegraphics[width=0.432\textwidth,bb=0 15 680 335,clip]{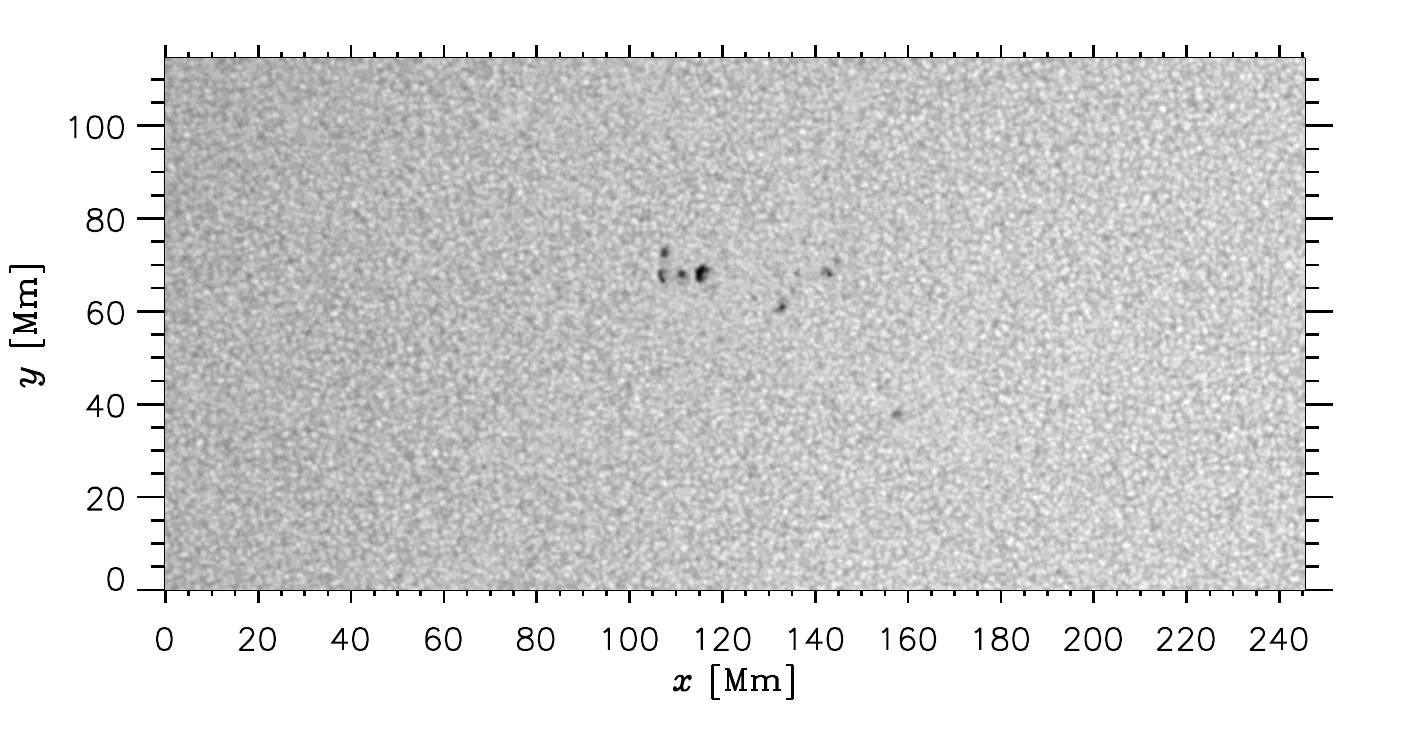}\hspace{0.748cm}		
		\includegraphics[width=0.432\textwidth,bb=0 15 680 335,clip]{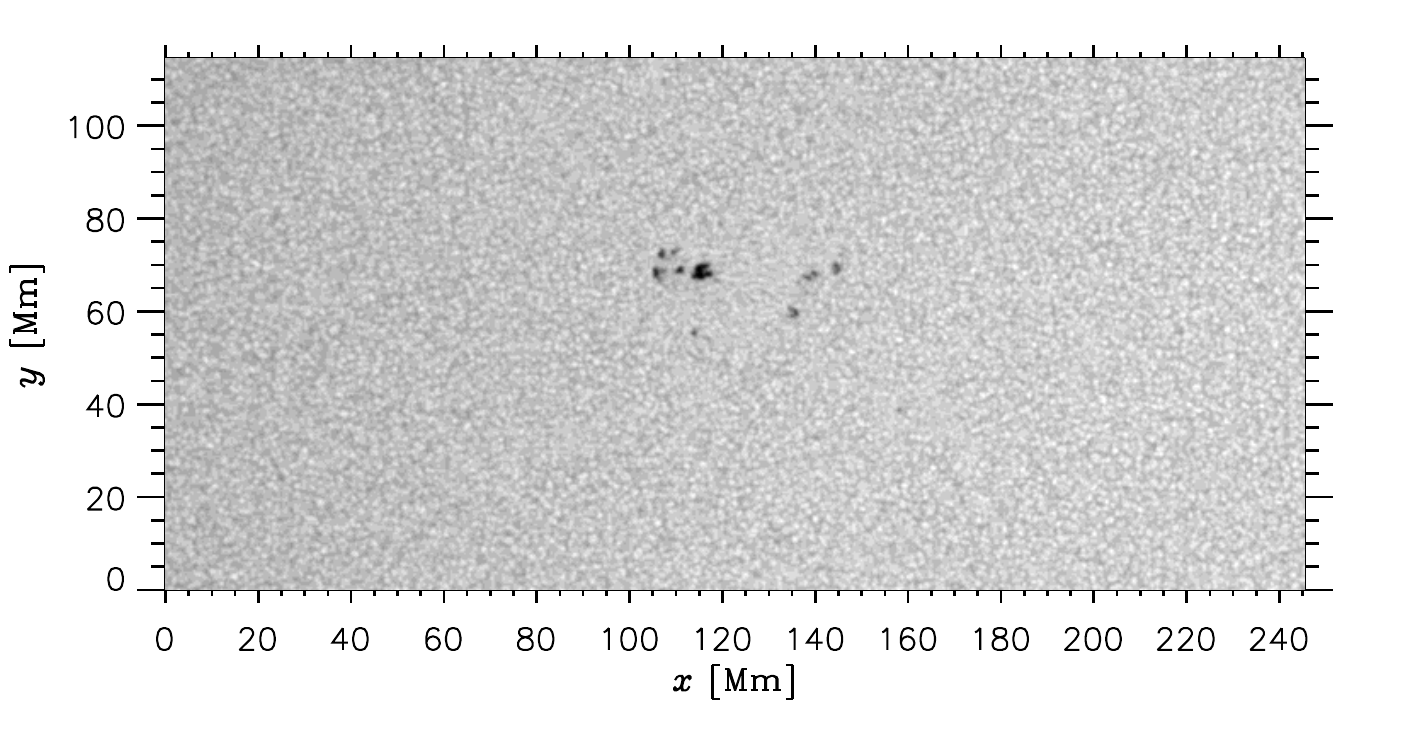}\\
		\includegraphics[width=0.432\textwidth,bb=0 15 680 335,clip]{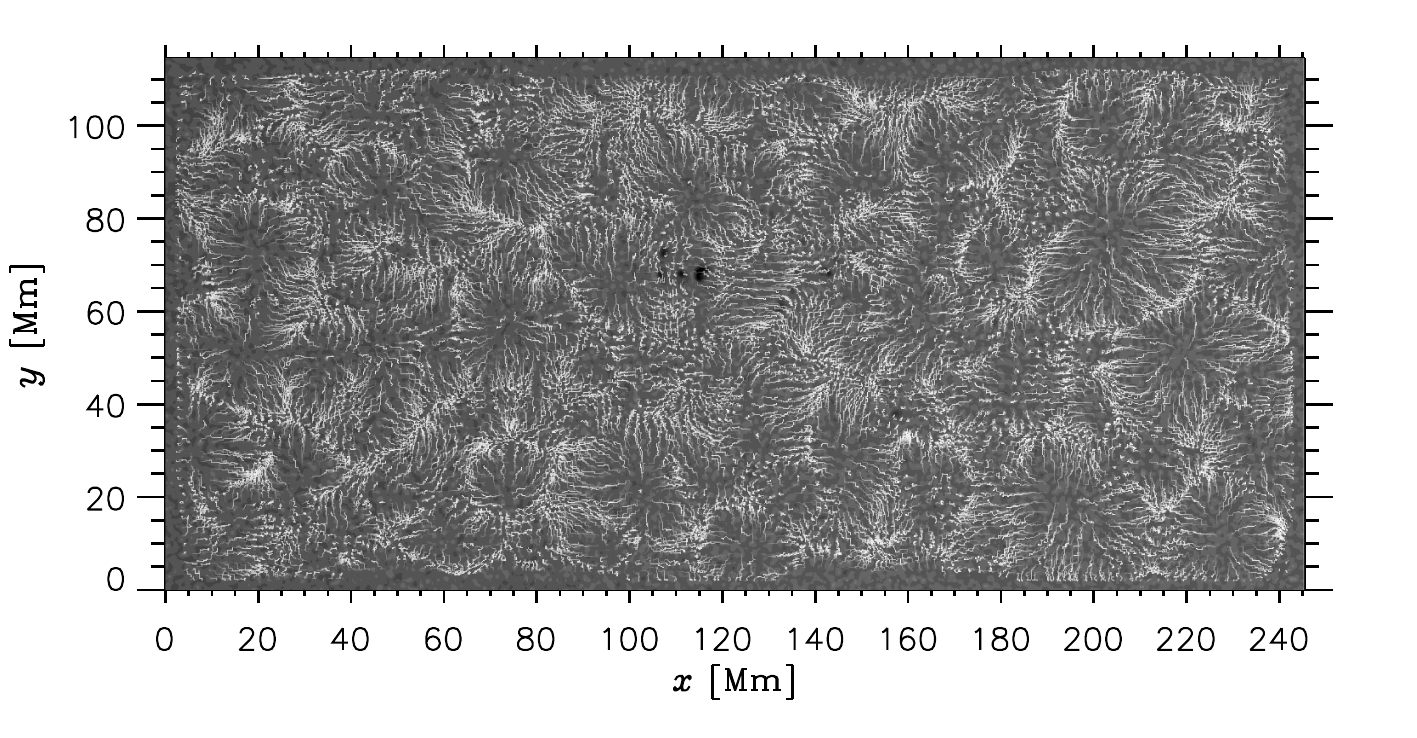}\hspace{0.748cm}
		\includegraphics[width=0.432\textwidth,bb=0 15 680 335,clip]{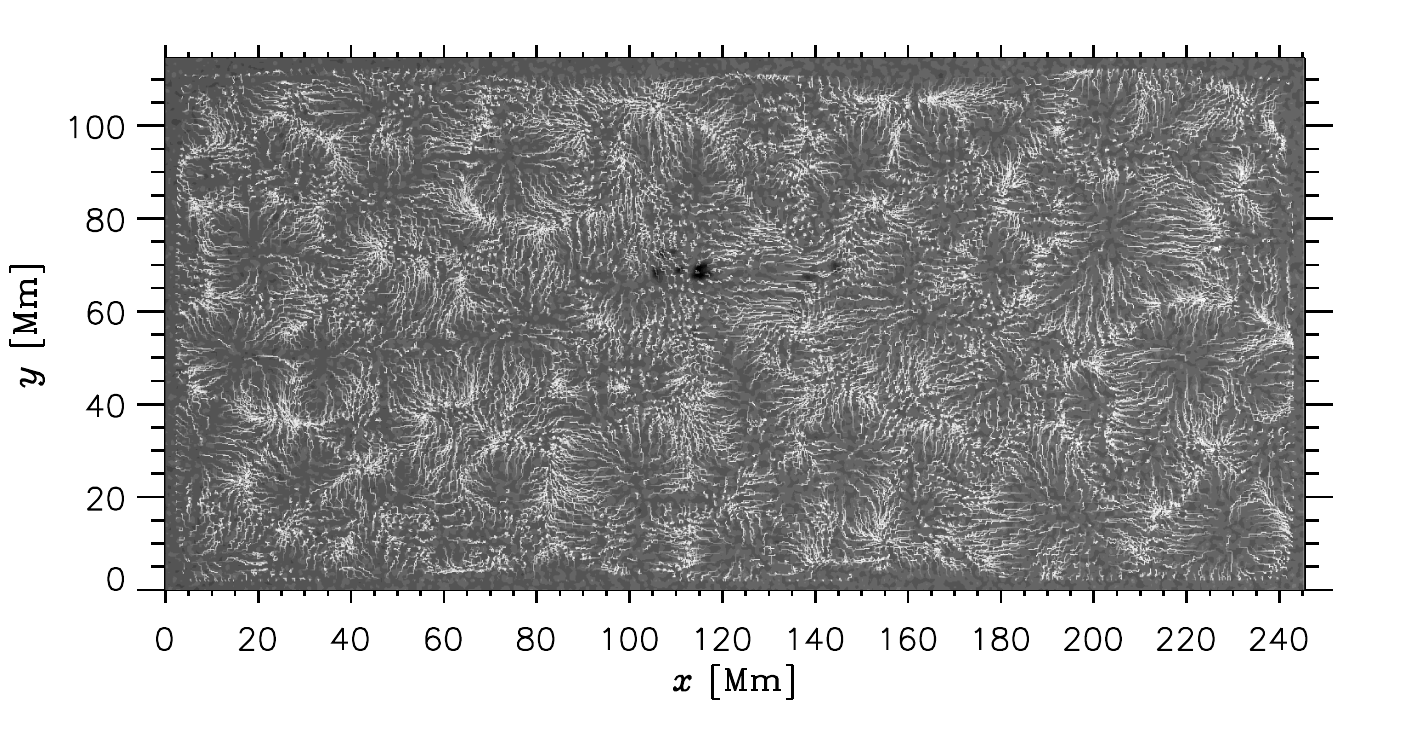}\\
		\includegraphics[width=0.495\textwidth,bb=15 0 793 335,clip]{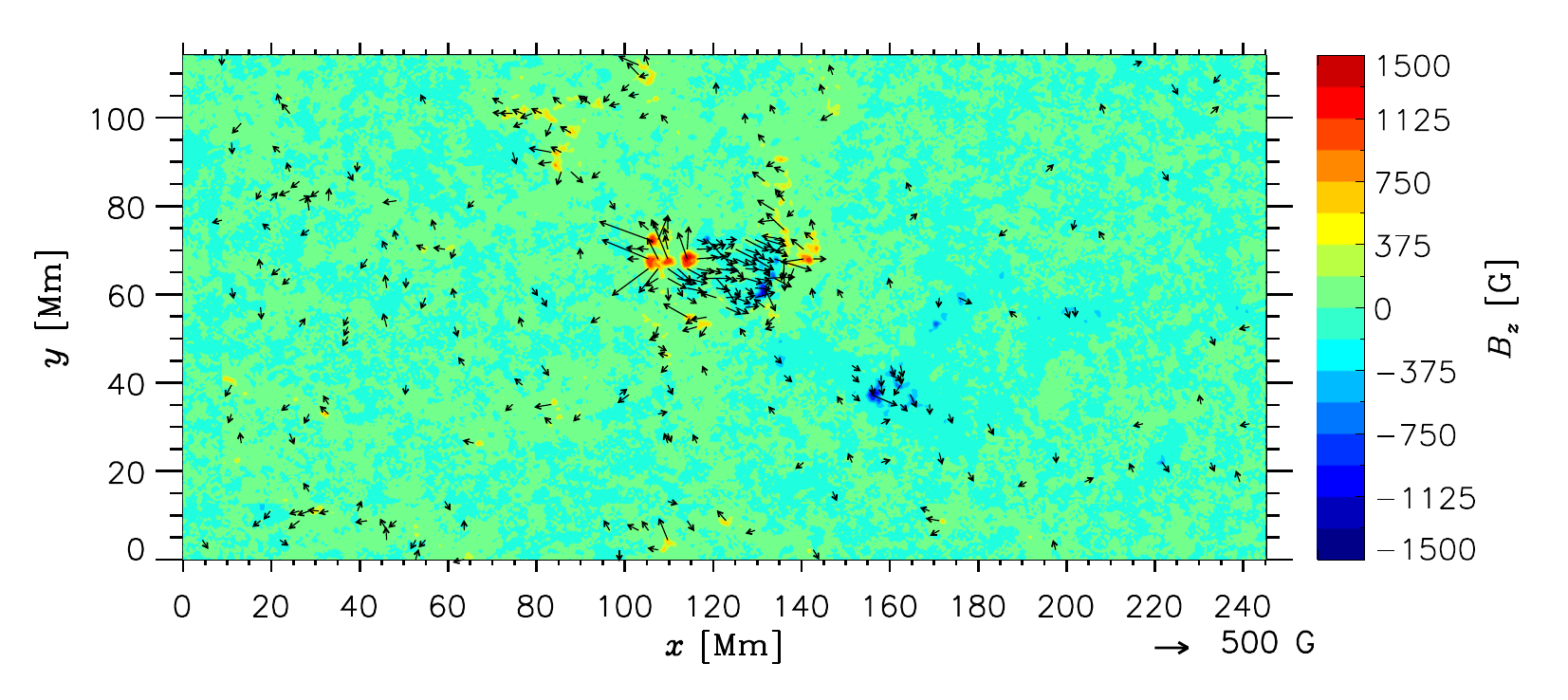}
		\includegraphics[width=0.495\textwidth,bb=15 0 793 335,clip]{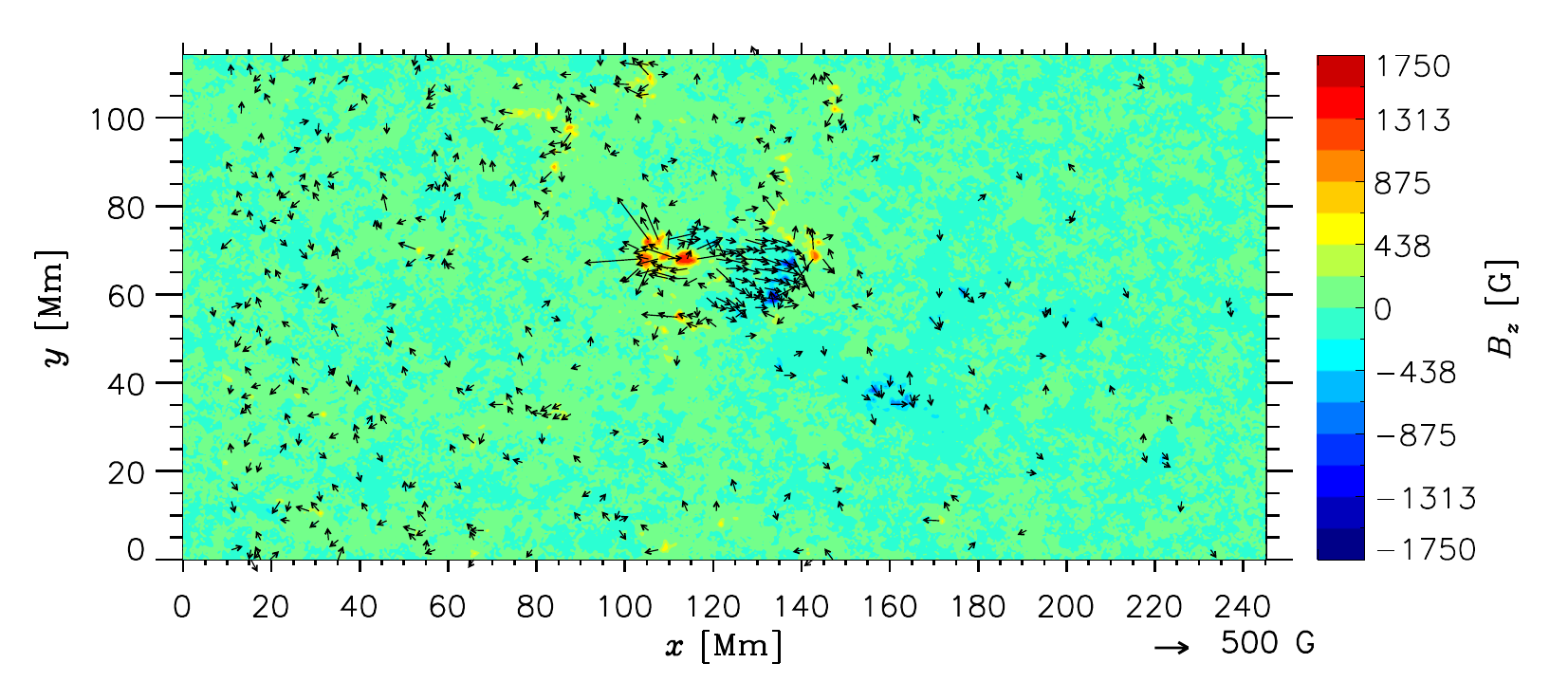}\\[-10pt]
		{\tiny \hspace{1.8cm}  2015 Jan 06, 11:06 \hspace{4.4cm} 2015 Jan 06, 12:58}\\
		\includegraphics[width=0.432\textwidth,bb=0 15 680 335,clip]{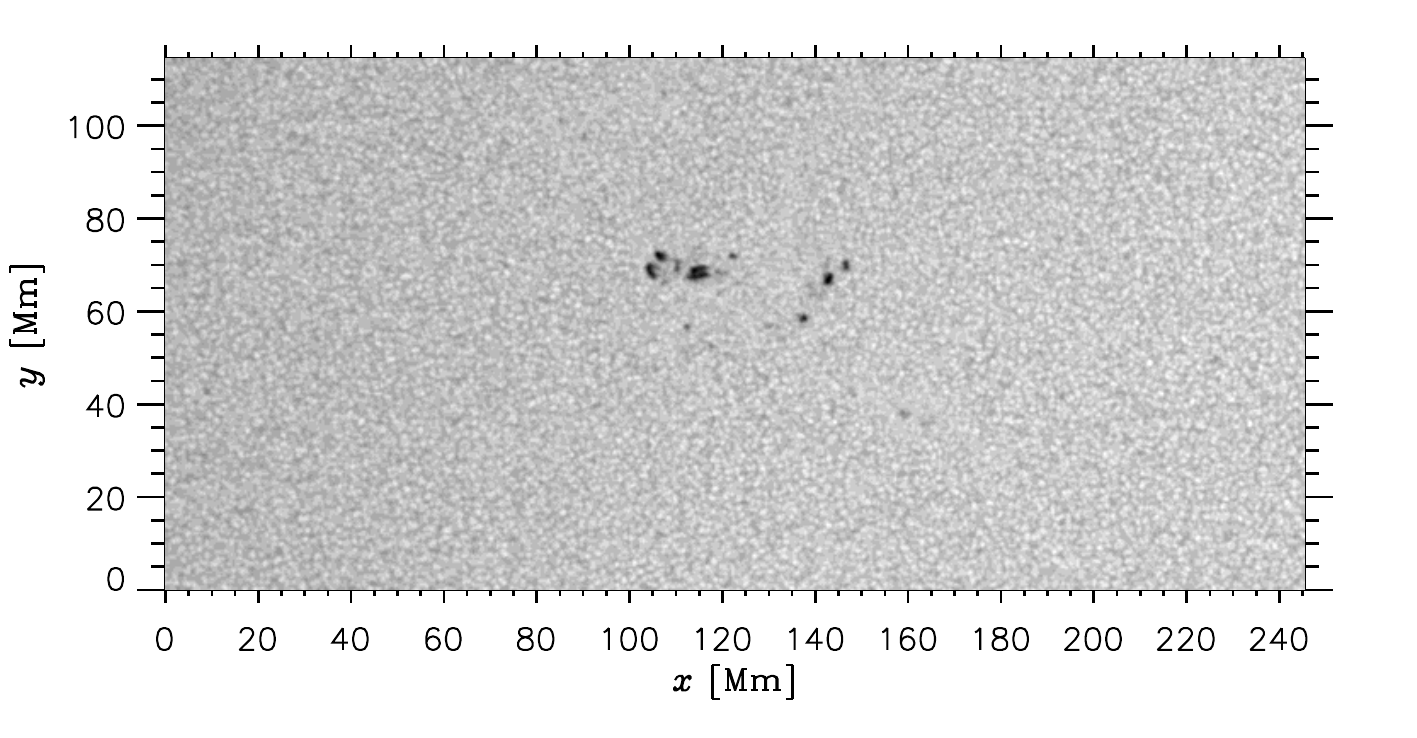}\hspace{0.748cm}		
		\includegraphics[width=0.432\textwidth,bb=0 15 680 335,clip]{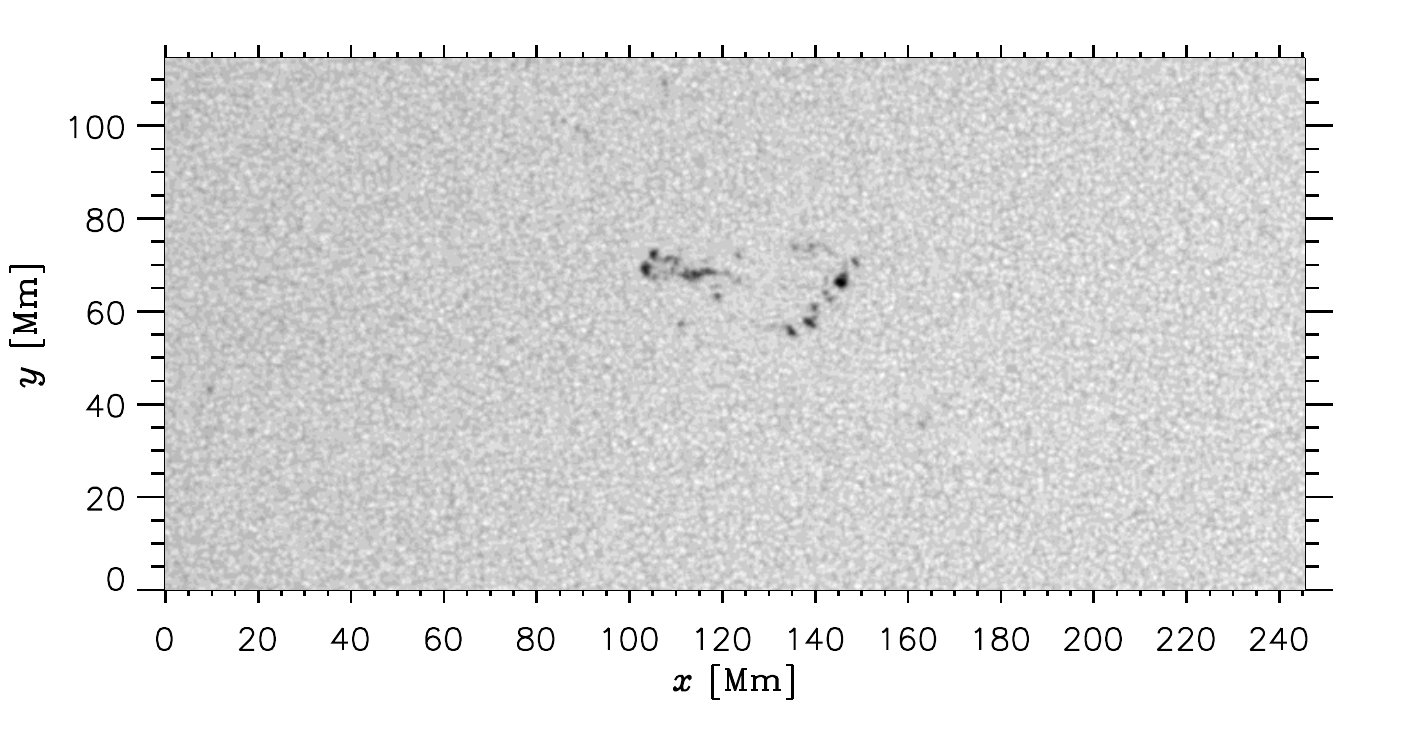}\\
		\includegraphics[width=0.432\textwidth,bb=0 15 680 335,clip]{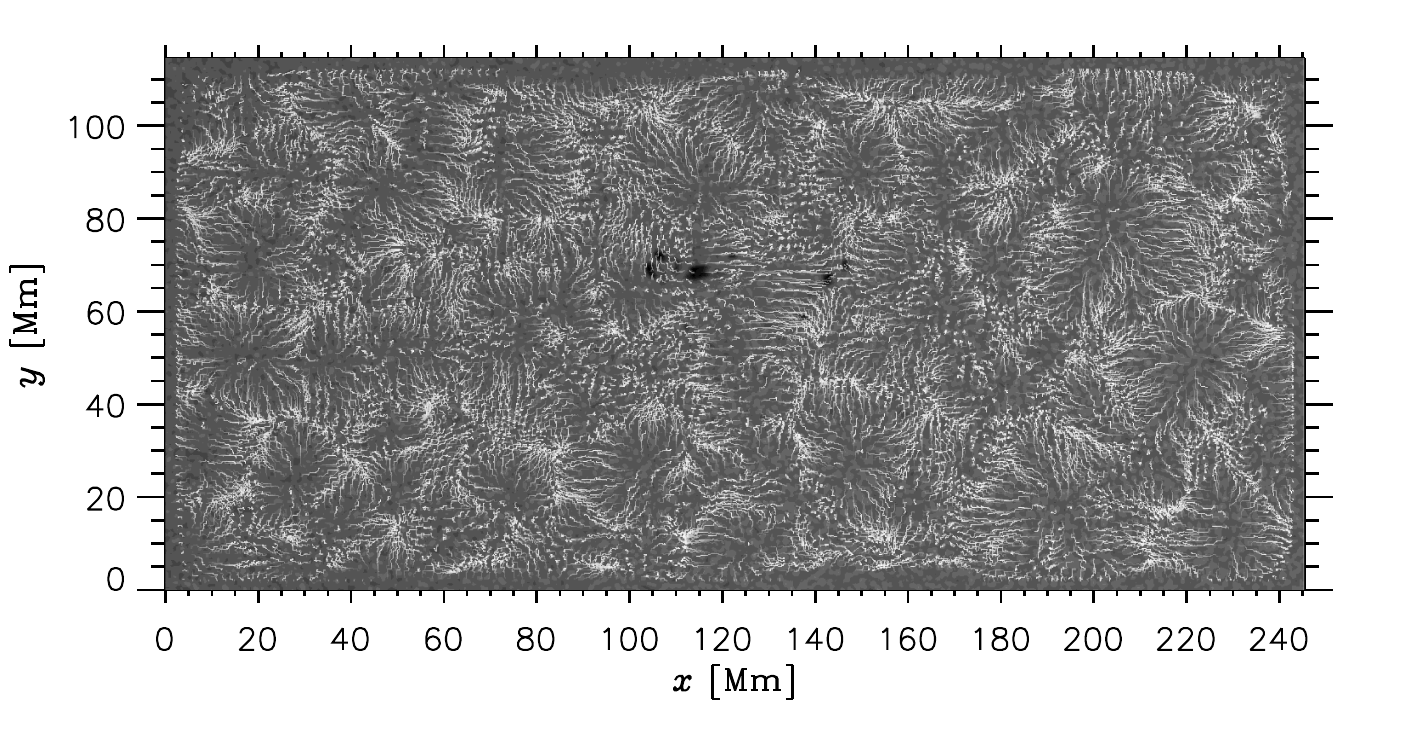}\hspace{0.748cm}
		\includegraphics[width=0.432\textwidth,bb=0 15 680 335,clip]{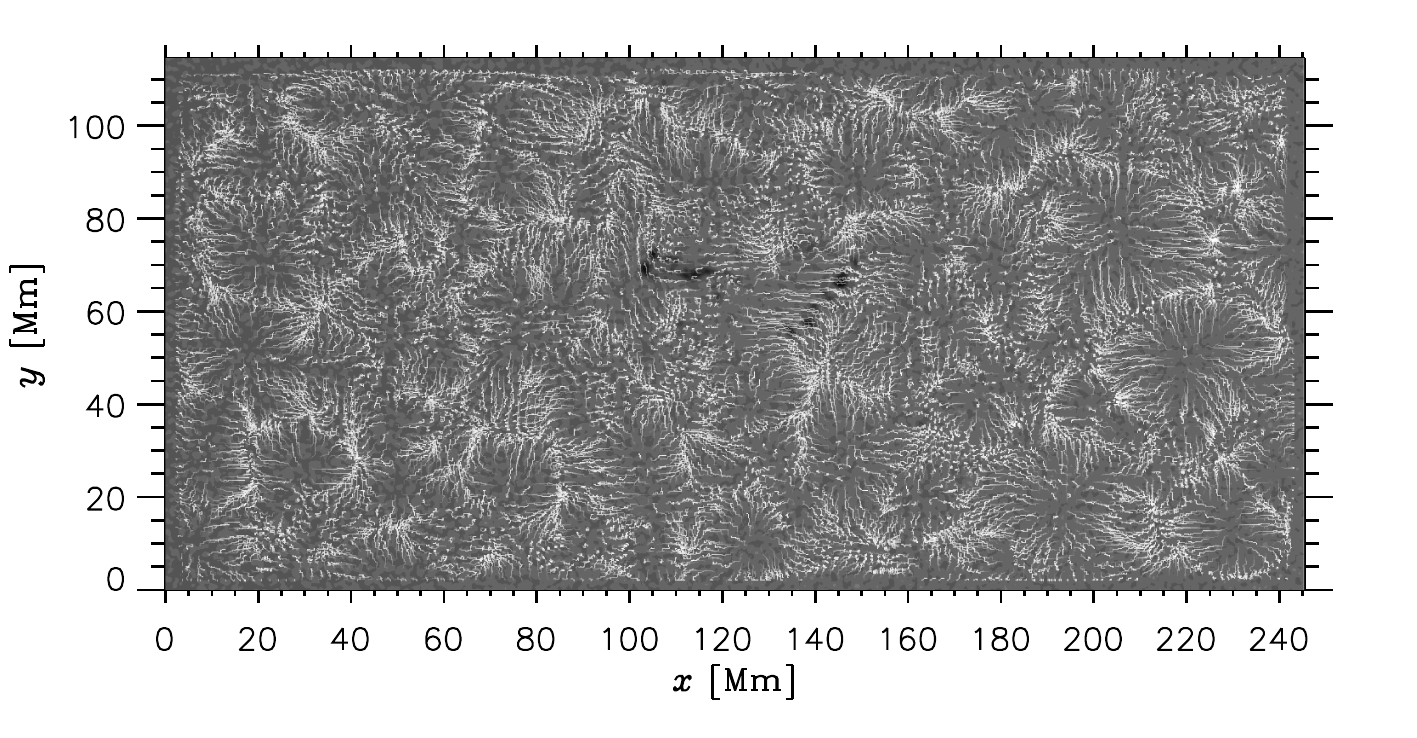}\\
		\includegraphics[width=0.495\textwidth,bb=15 0 793 335,clip]{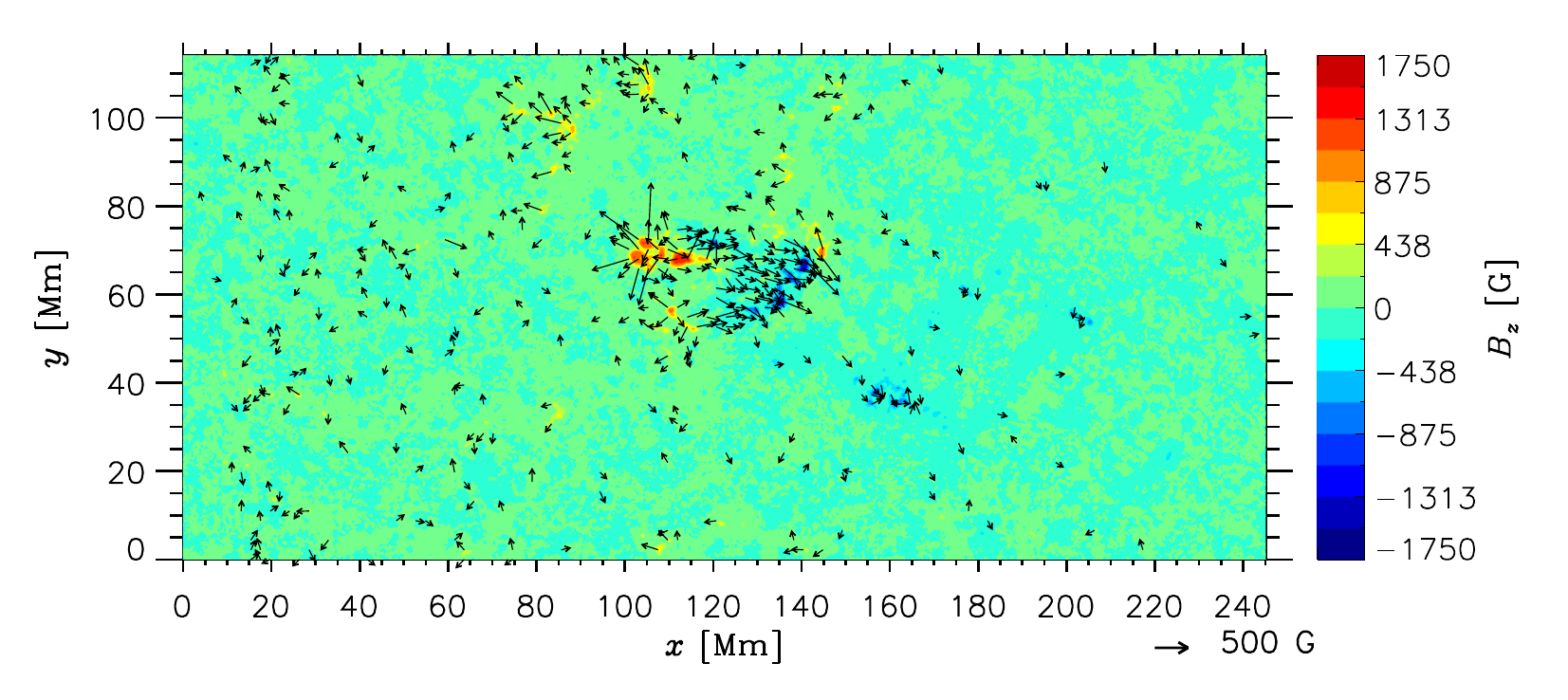}
		\includegraphics[width=0.495\textwidth,bb=15 0 793 335,clip]{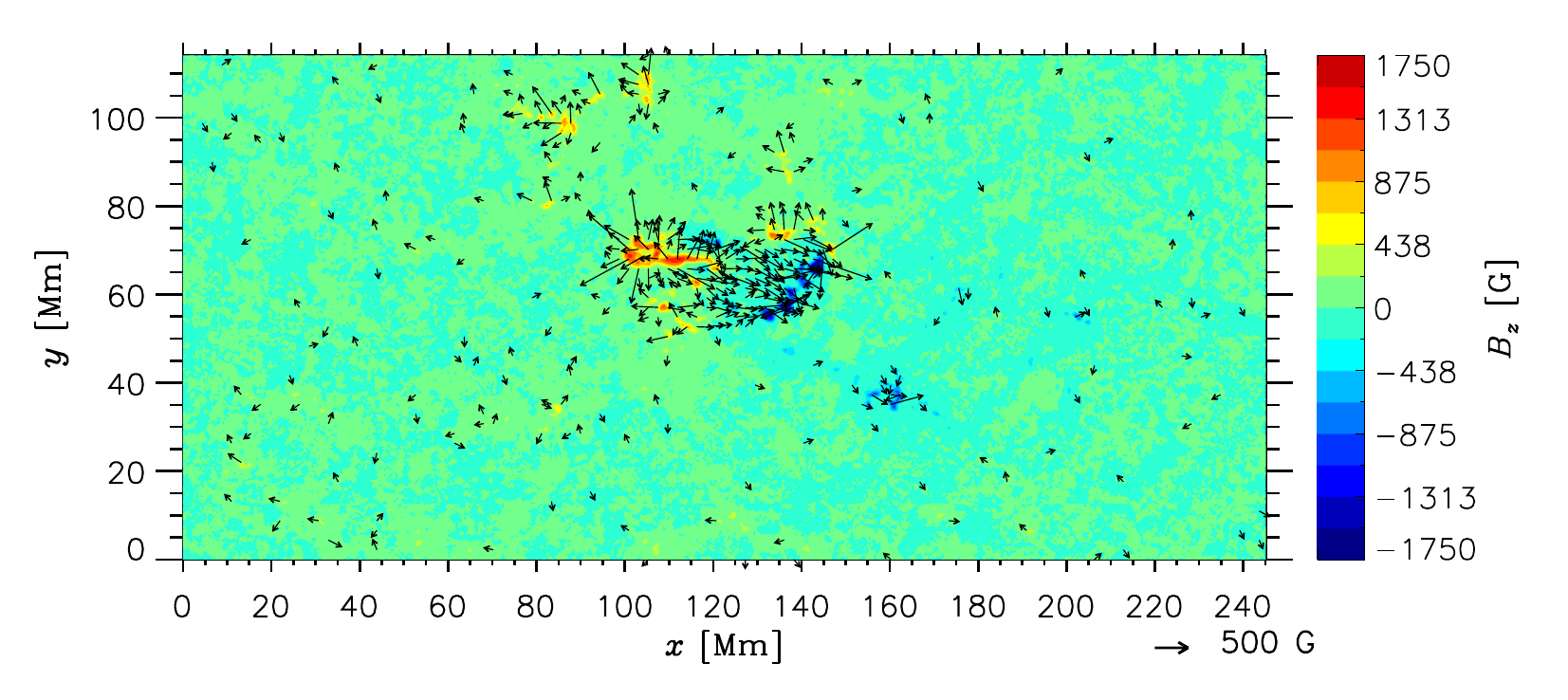}\\[-10pt]
		{\tiny \hspace{1.8cm}  2015 Jan 06, 14:51 \hspace{4.4cm} 2015 Jan 06, 16:43}
	}\caption{For caption see page~\pageref{F1}.}
\end{figure}
\addtocounter{figure}{-1}
\begin{figure}
	\flushleft{
		\includegraphics[width=0.432\textwidth,bb=0 15 680 335,clip]{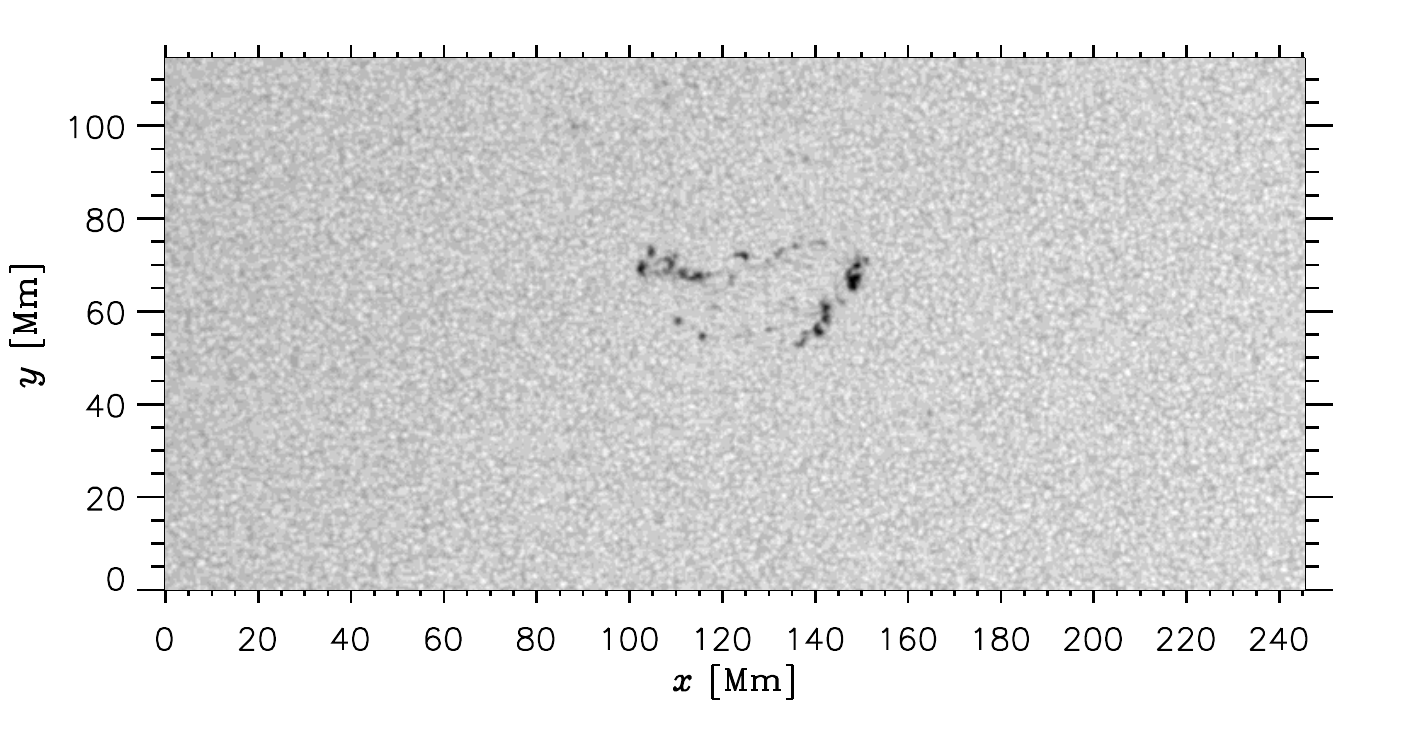}\hspace{0.748cm}	
		\includegraphics[width=0.432\textwidth,bb=0 15 680 335,clip]{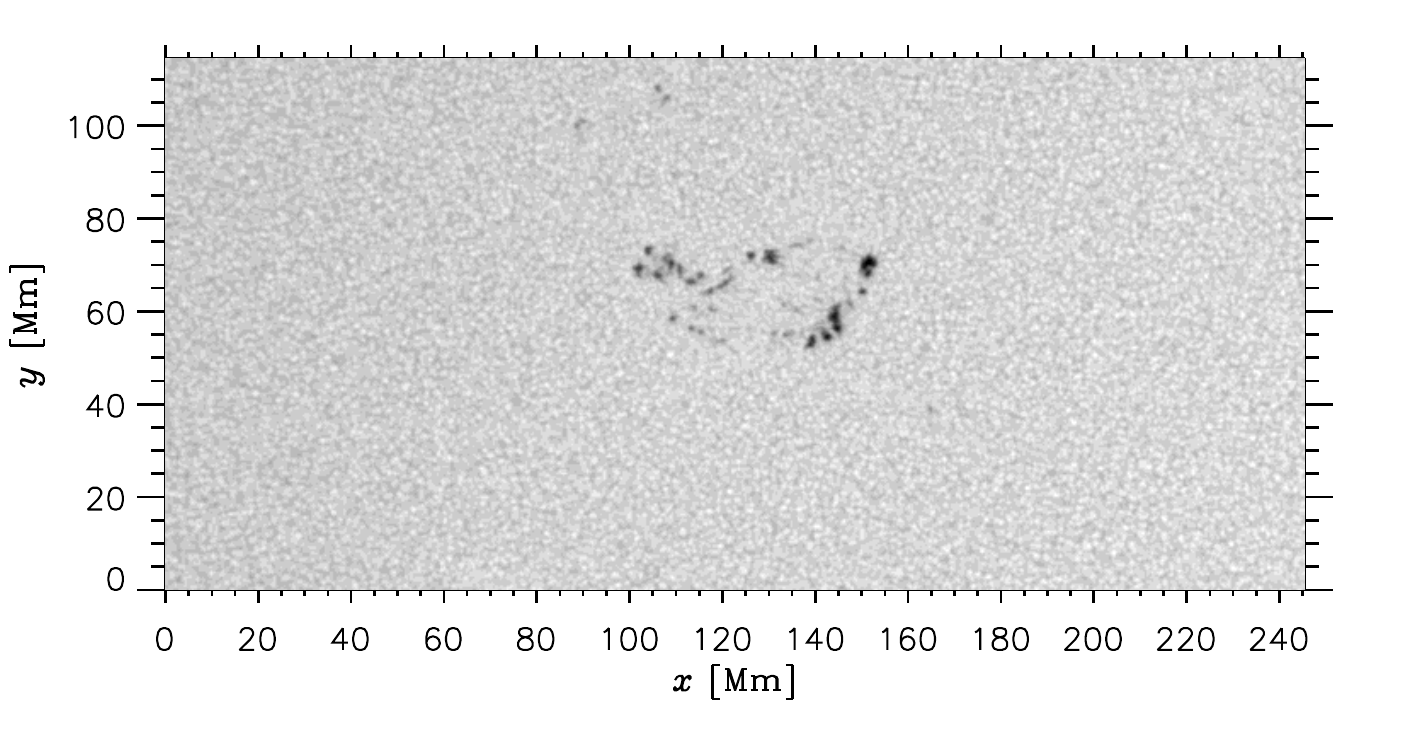}\\
		\includegraphics[width=0.432\textwidth,bb=0 15 680 335,clip]{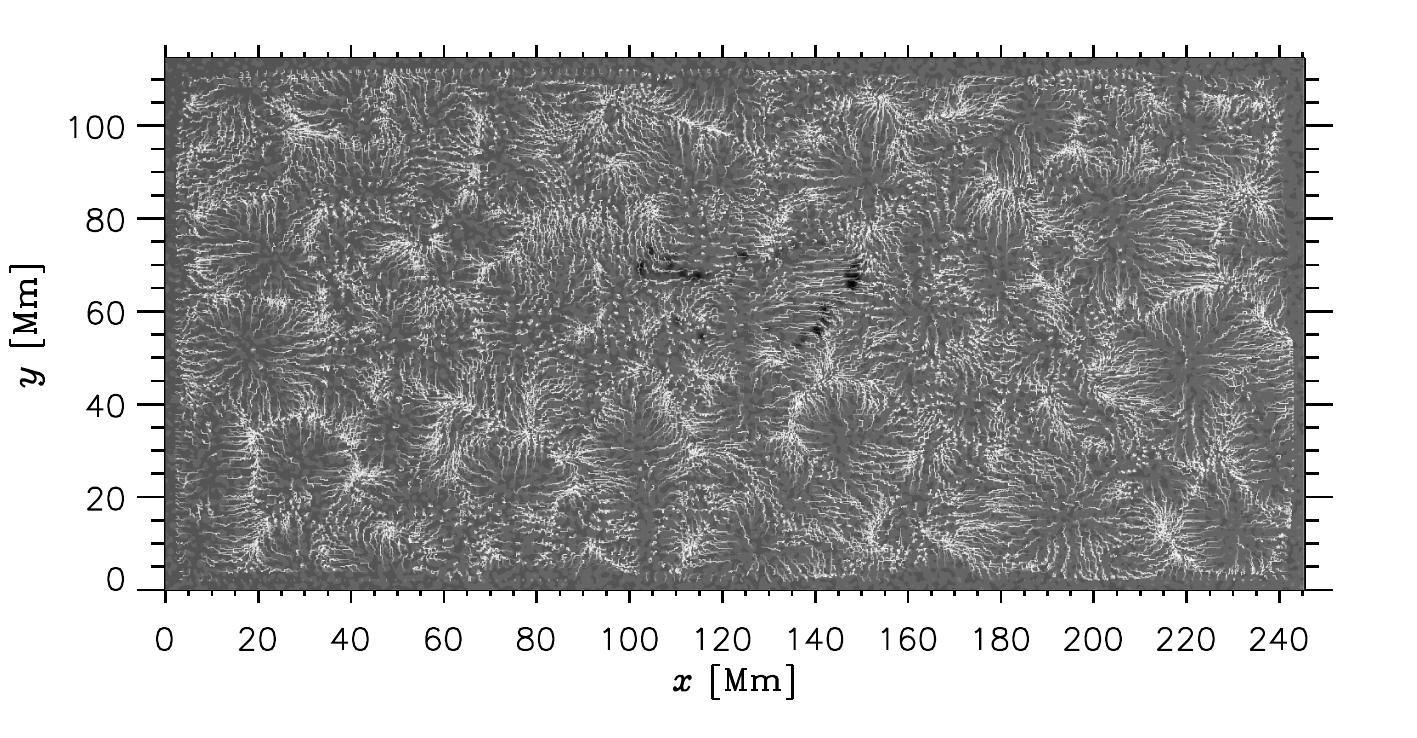}\hspace{0.748cm}
		\includegraphics[width=0.432\textwidth,bb=0 15 680 335,clip]{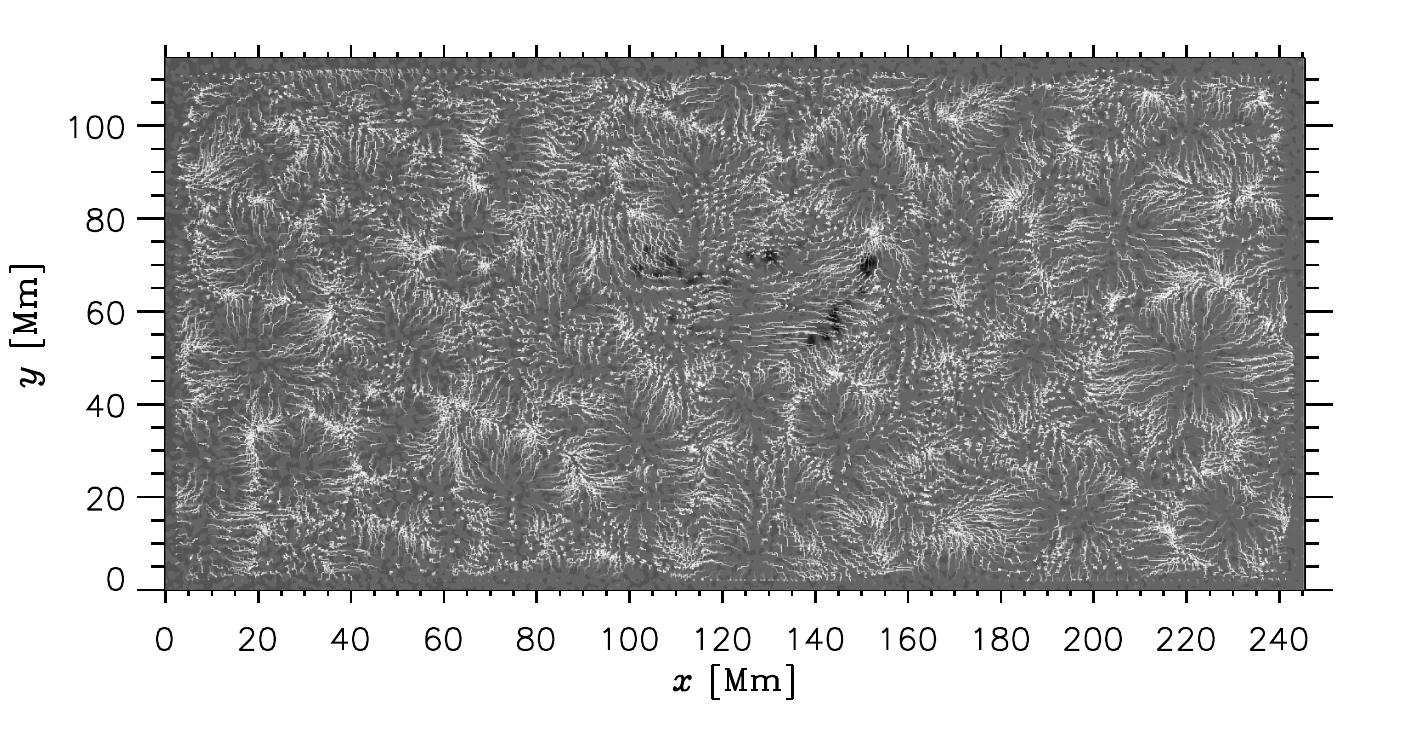}\\
		\includegraphics[width=0.495\textwidth,bb=15 0 793 335,clip]{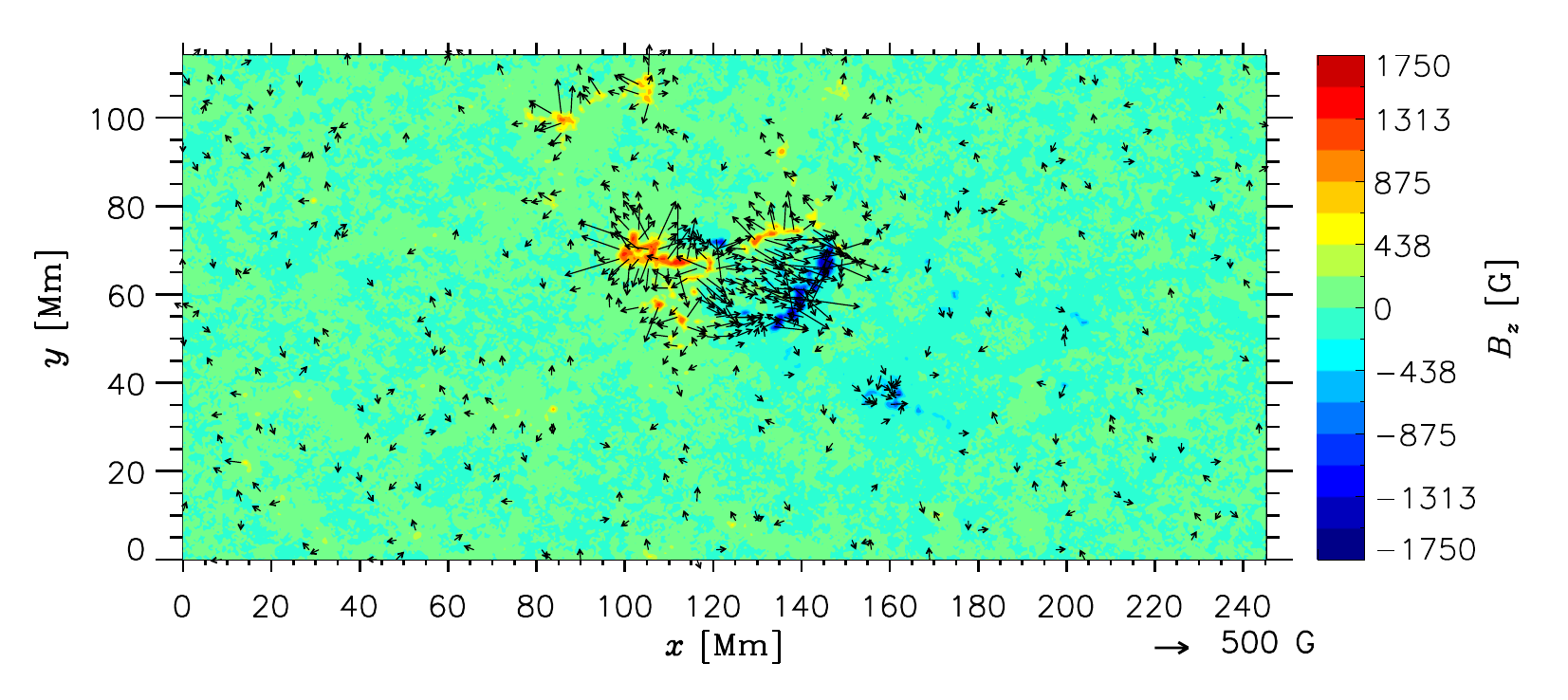}
		\includegraphics[width=0.495\textwidth,bb=15 0 793 335,clip]{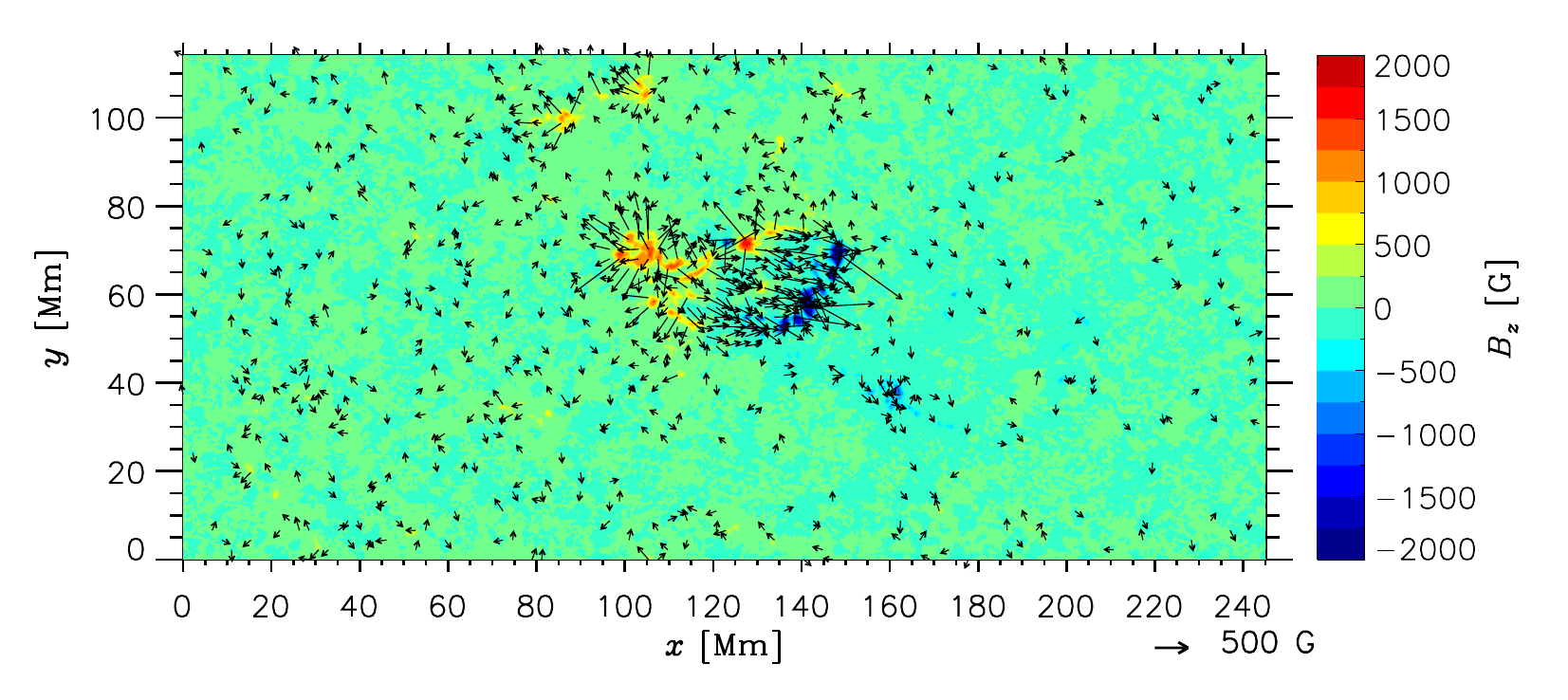}}\\[-10pt]
        {\tiny \hspace{1.8cm}  2015 Jan 06, 18:40 \hspace{4.4cm} 2015 Jan 06, 20:40}\\
					
		\includegraphics[width=0.432\textwidth,bb=0 15 680 335,clip]{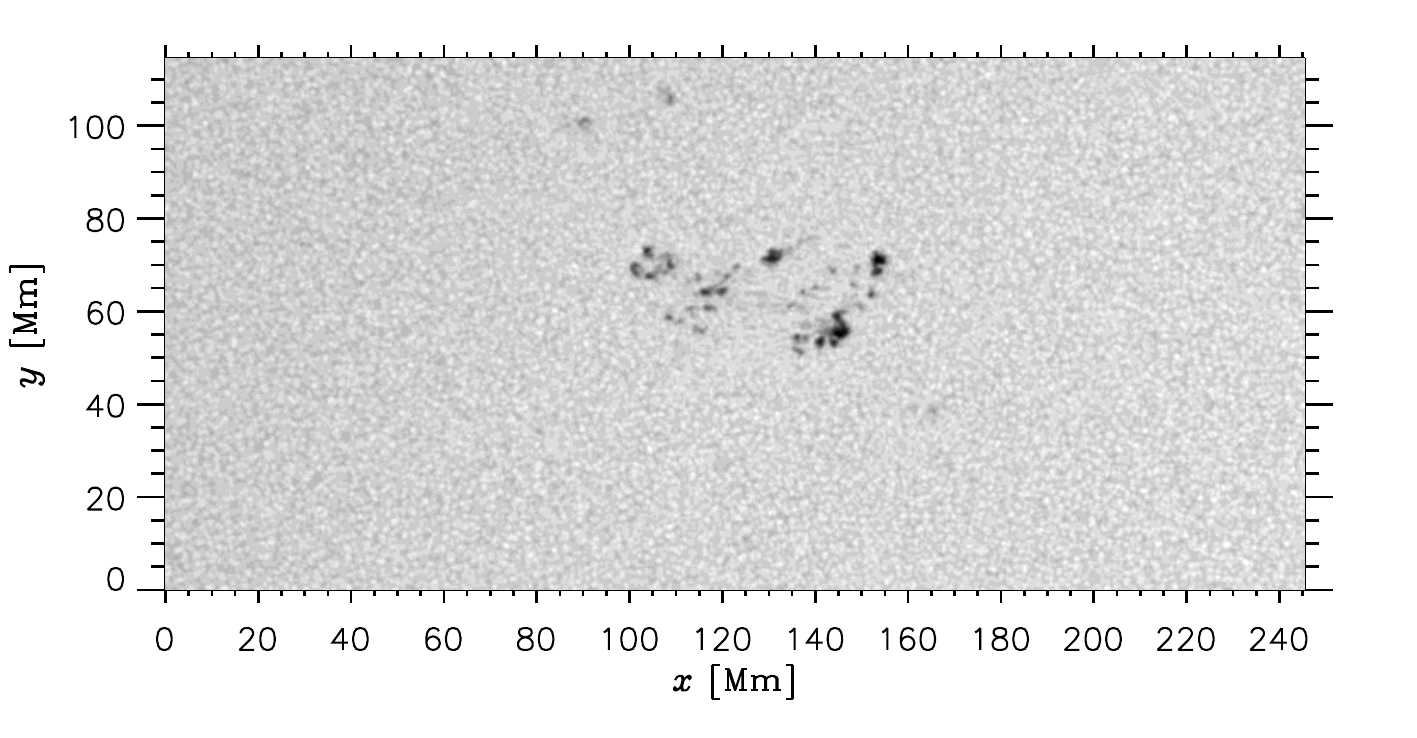}\hspace{0.748cm}	
		\includegraphics[width=0.432\textwidth,bb=0 15 680 335,clip]{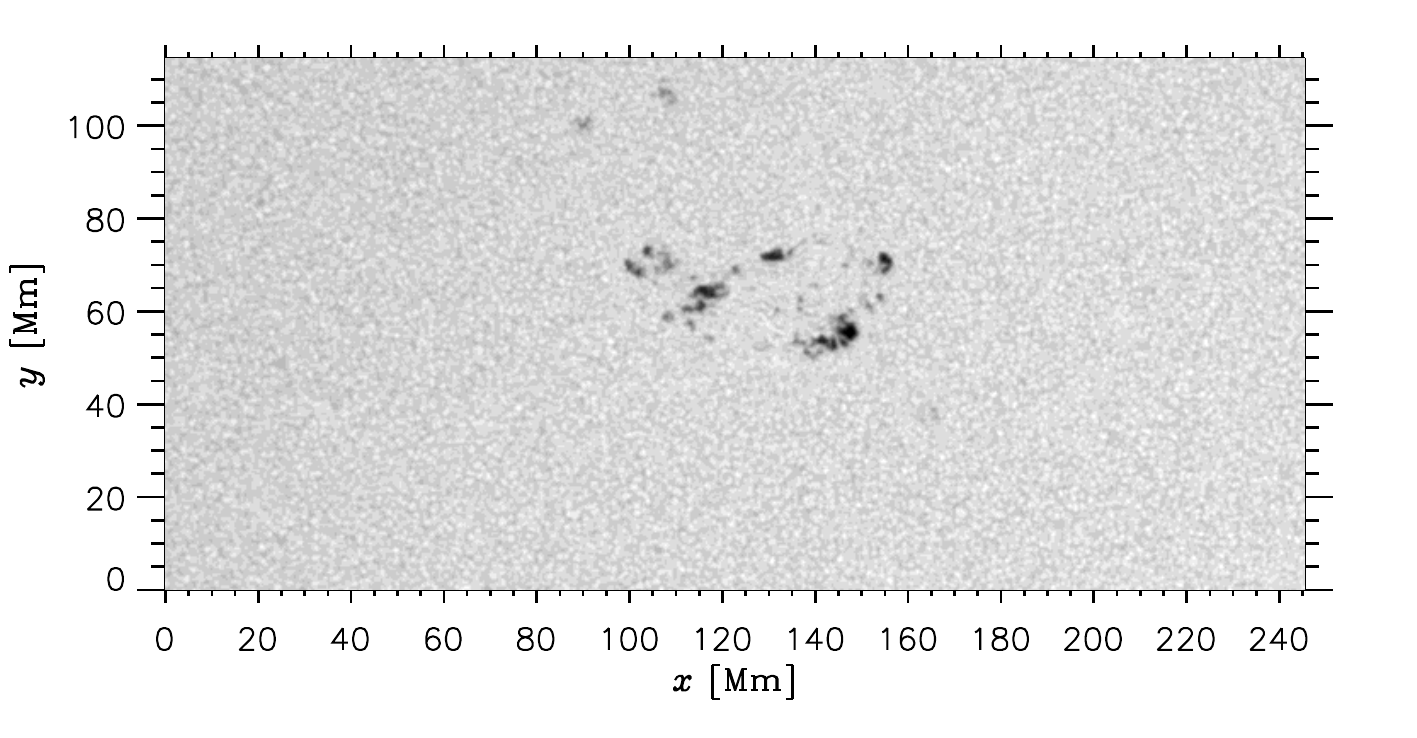}\\
		\includegraphics[width=0.432\textwidth,bb=0 15 680 335,clip]{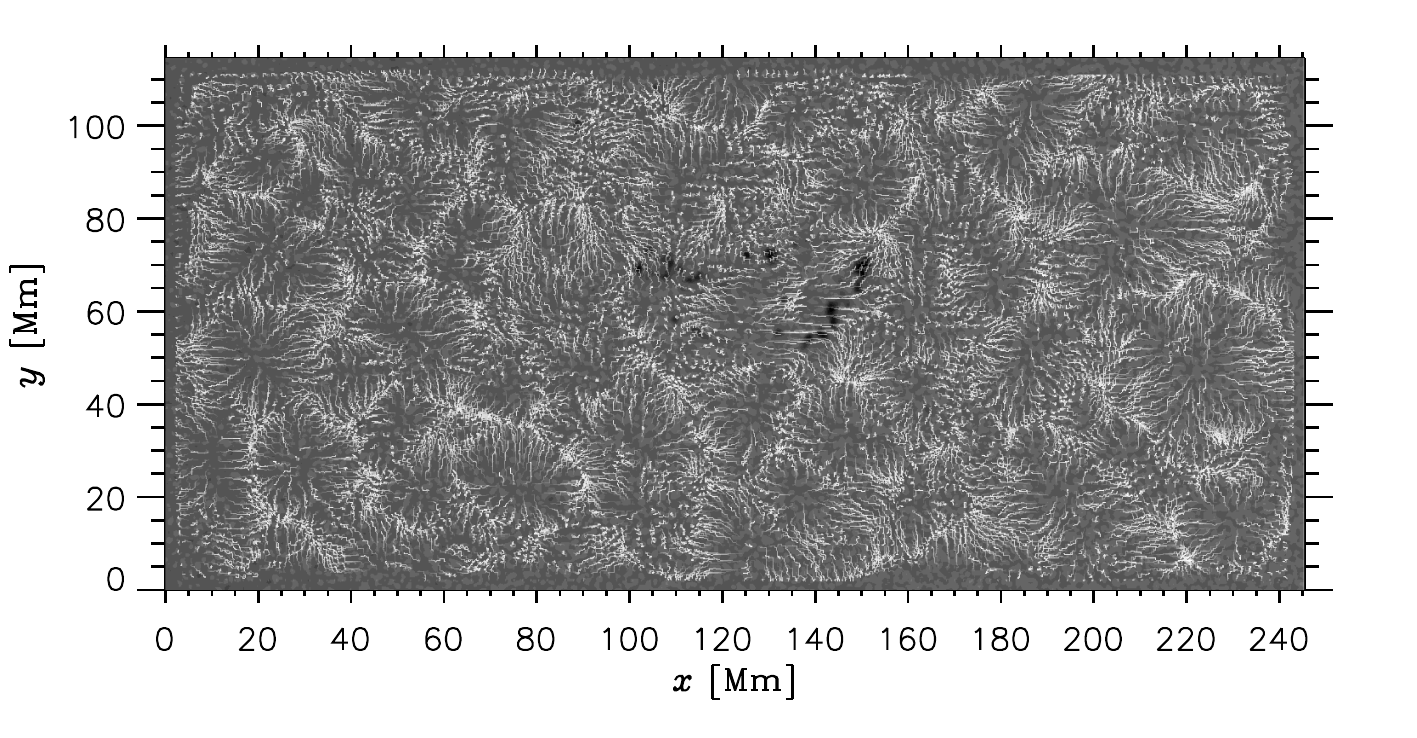}\hspace{0.748cm}
		\includegraphics[width=0.432\textwidth,bb=0 15 680 335,clip]{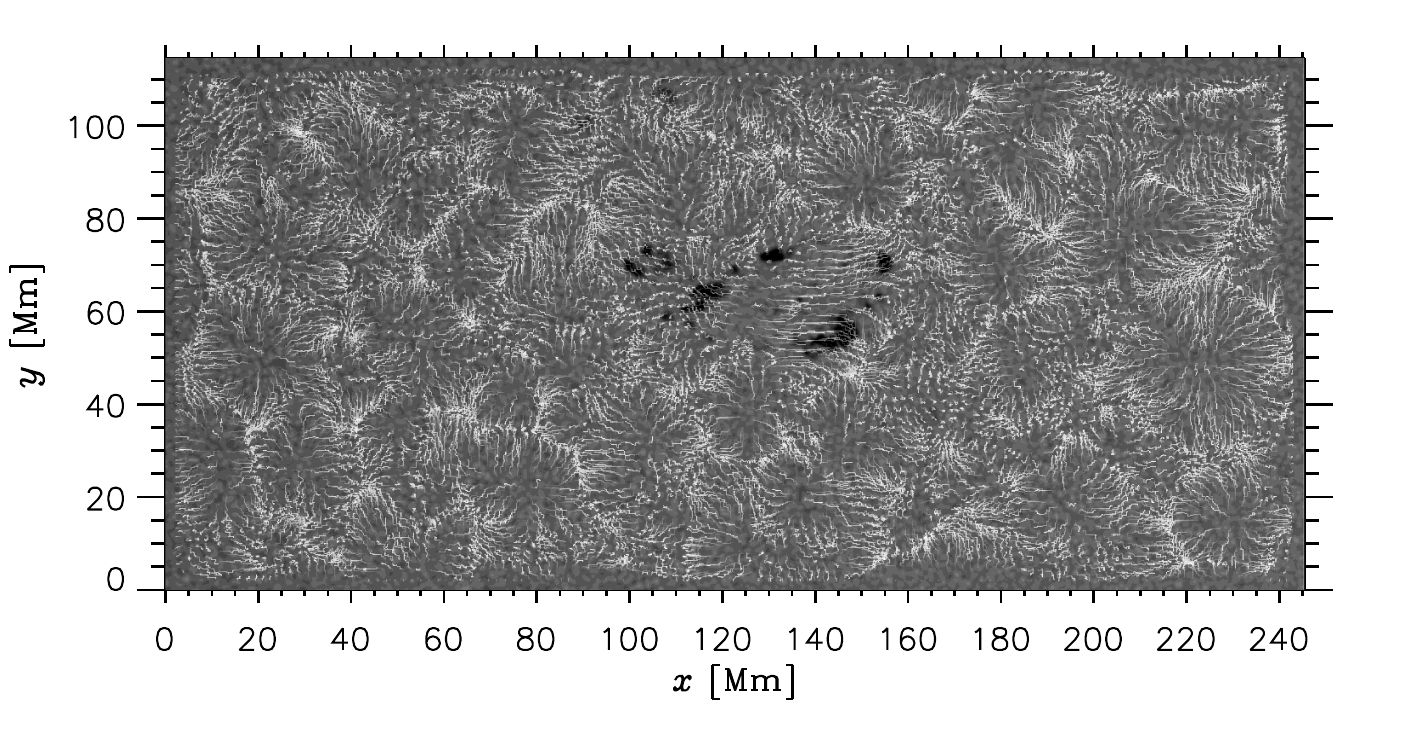}\\
		\includegraphics[width=0.495\textwidth,bb=15 0 793 335,clip]{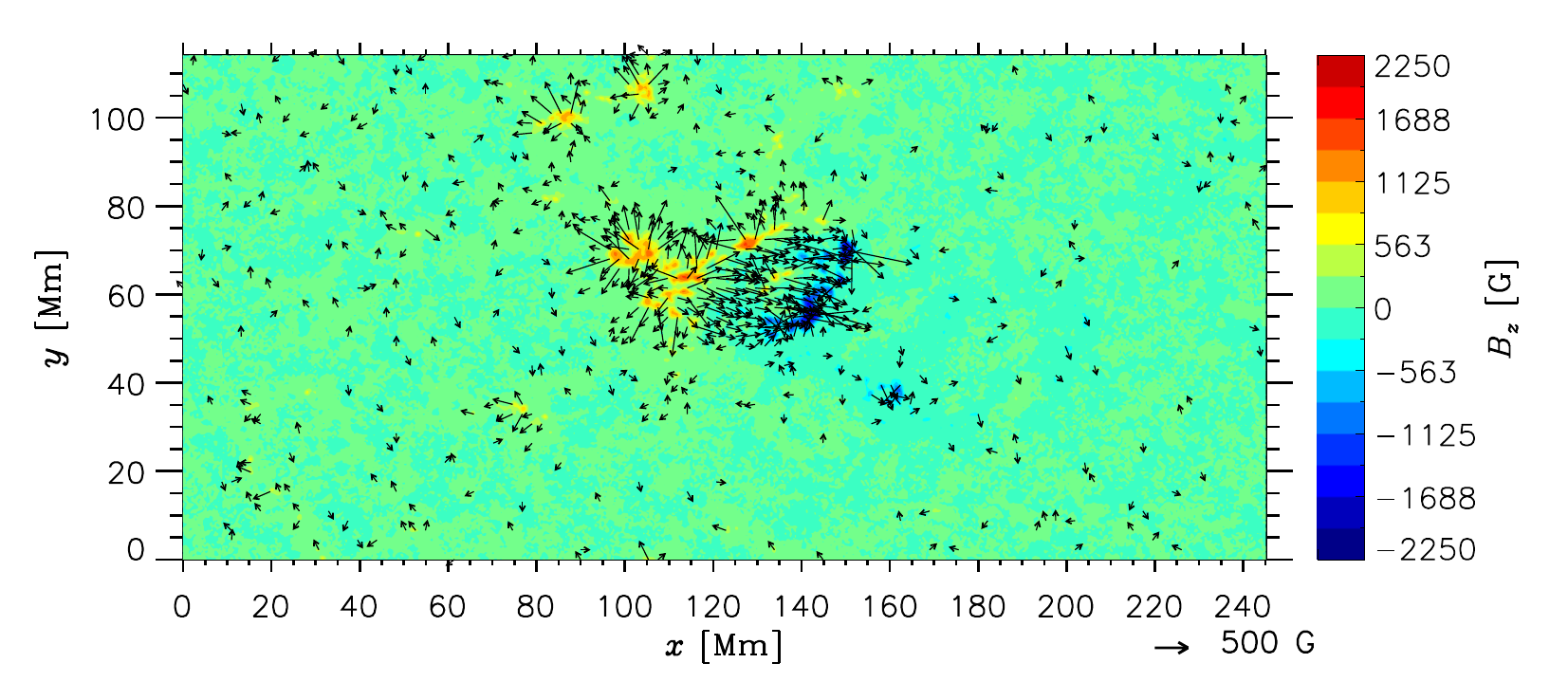}
		\includegraphics[width=0.495\textwidth,bb=15 0 793 335,clip]{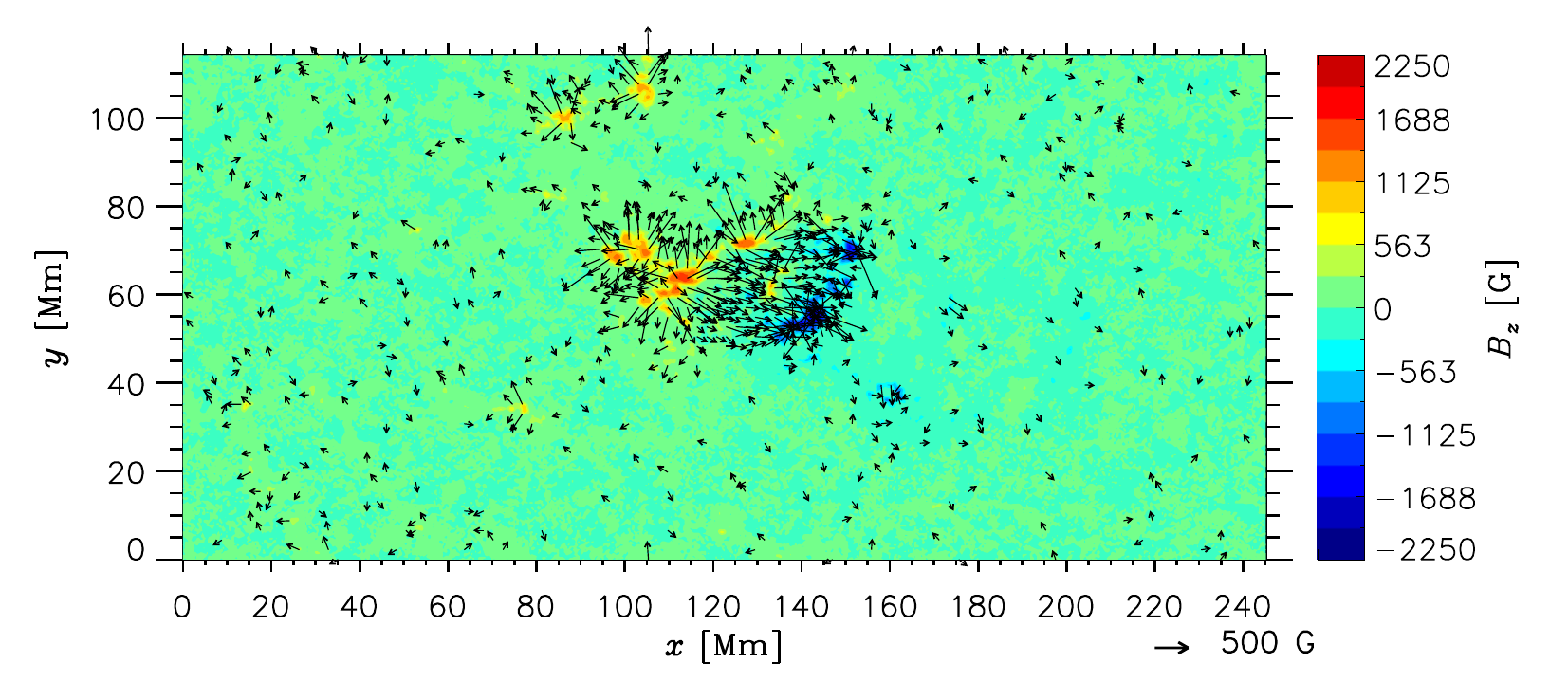}\\[-10pt]
		{\tiny \hspace{1.8cm}  2015 Jan 06, 22:32  \hspace{4.4cm}  2015 Jan 07, 00:25}\\
		
		\includegraphics[width=0.432\textwidth,bb=0 15 680 335,clip]{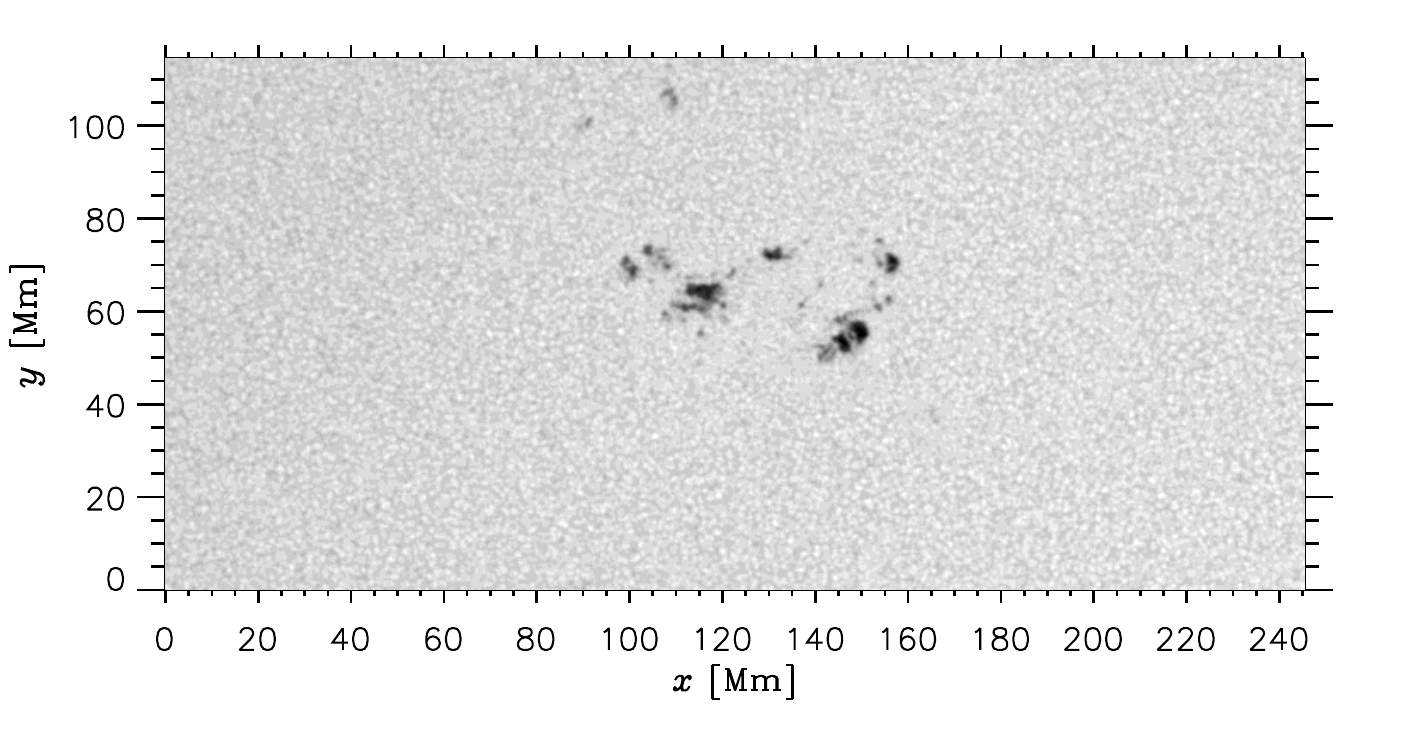}\hspace{0.748cm}
		\includegraphics[width=0.432\textwidth,bb=0 15 680 335,clip]{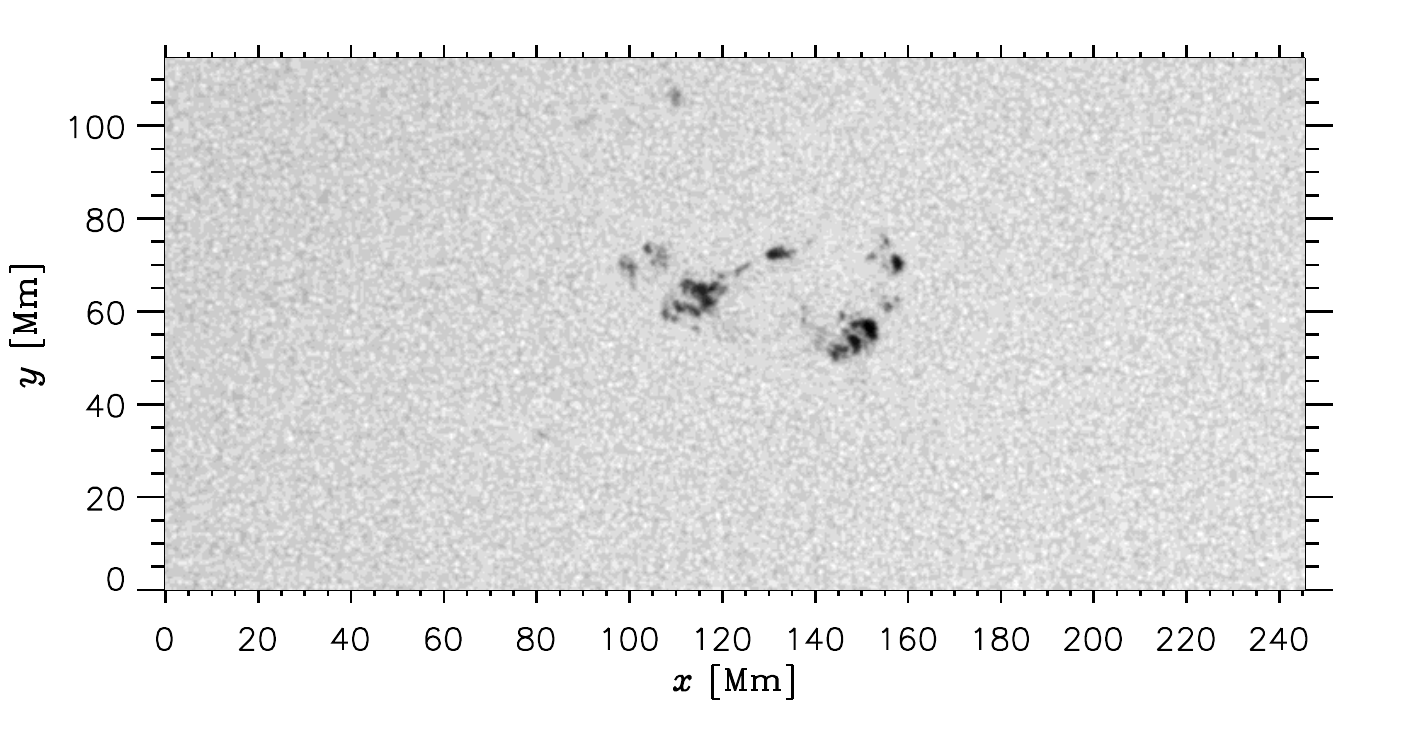}\\
		\includegraphics[width=0.432\textwidth,bb=0 15 680 335,clip]{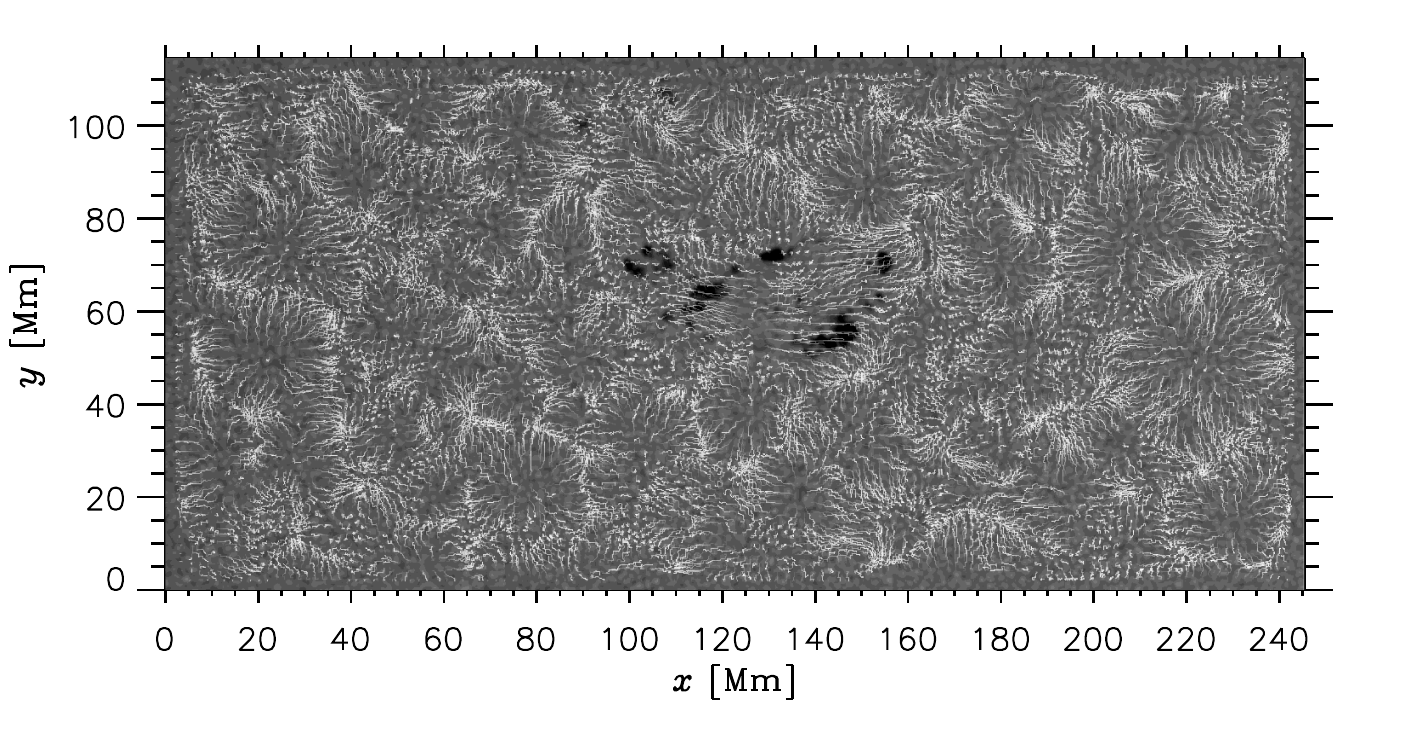}\hspace{0.748cm}
		\includegraphics[width=0.432\textwidth,bb=0 15 680 335,clip]{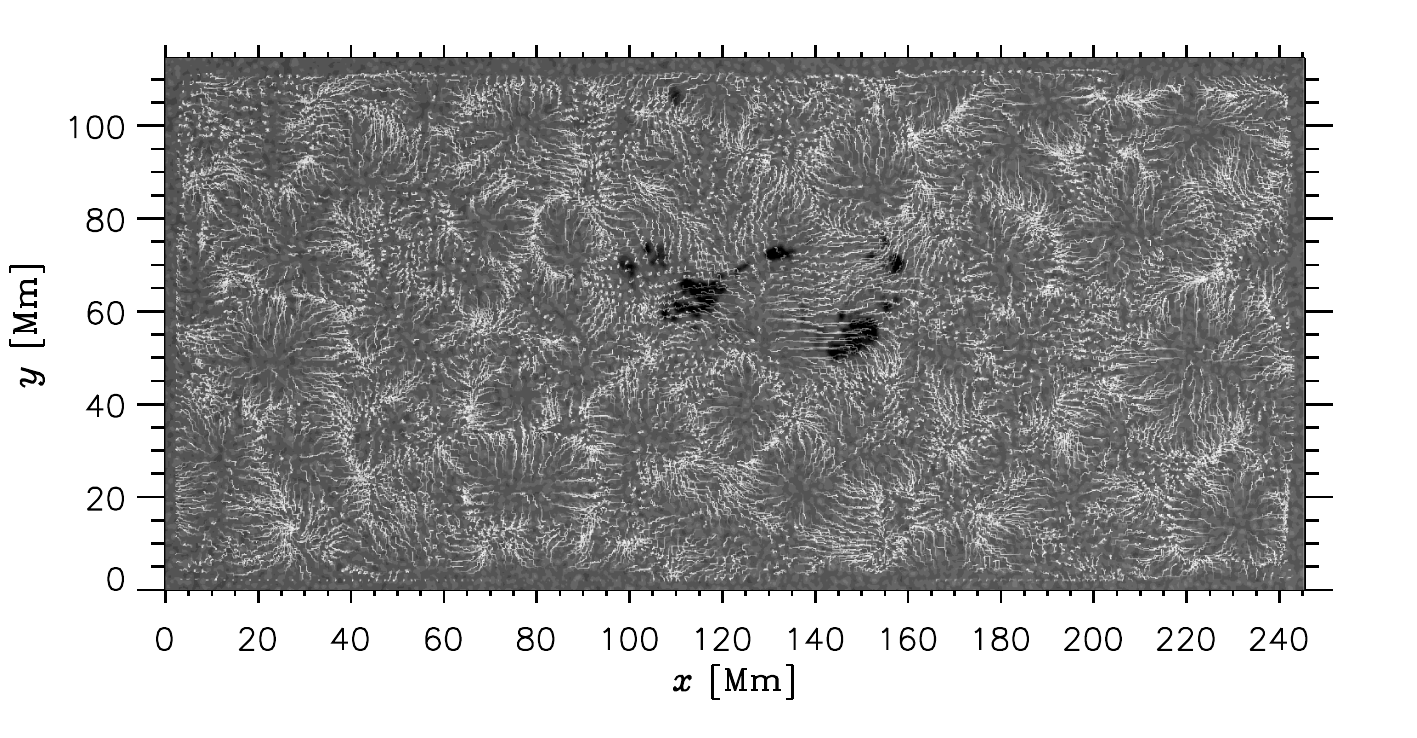}\\
		\includegraphics[width=0.495\textwidth,bb=15 0 793 335,clip]{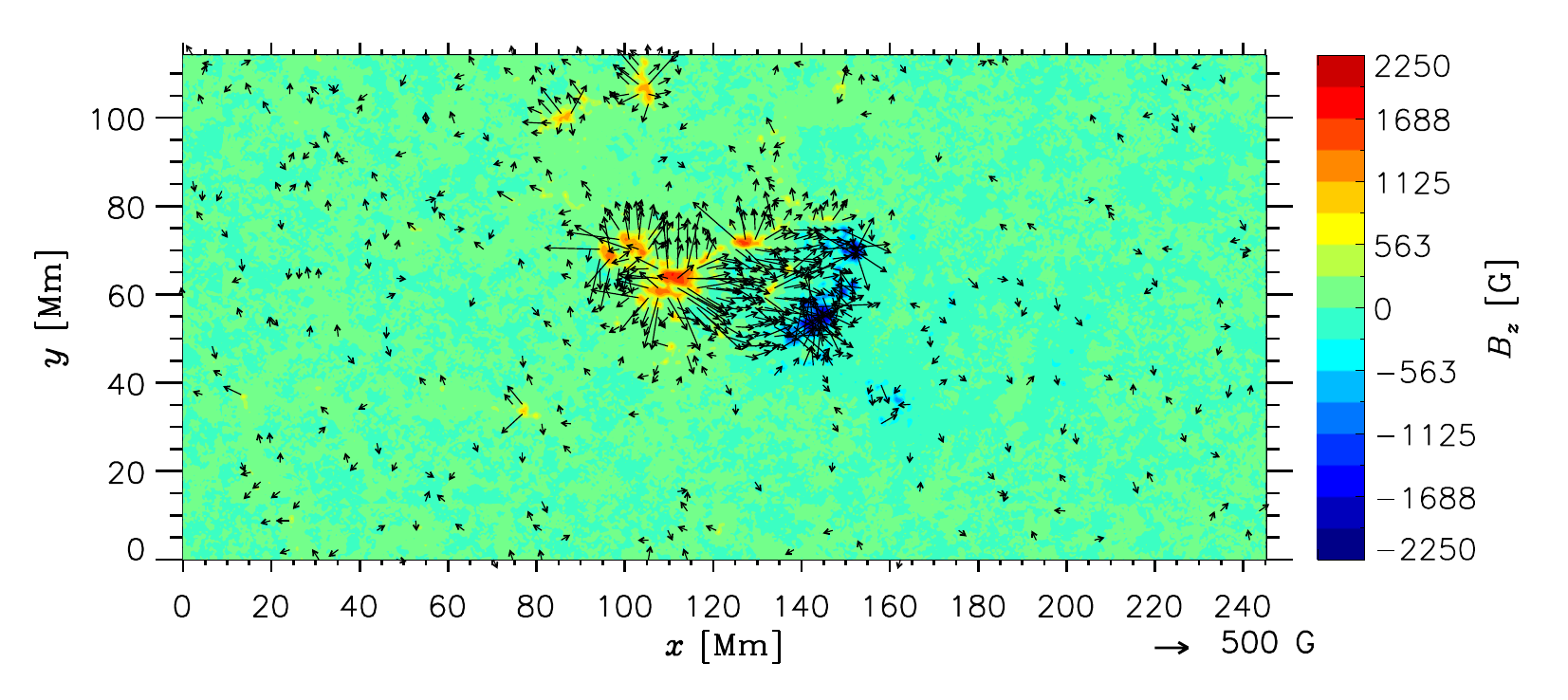}
		\includegraphics[width=0.495\textwidth,bb=15 0 793 335,clip]{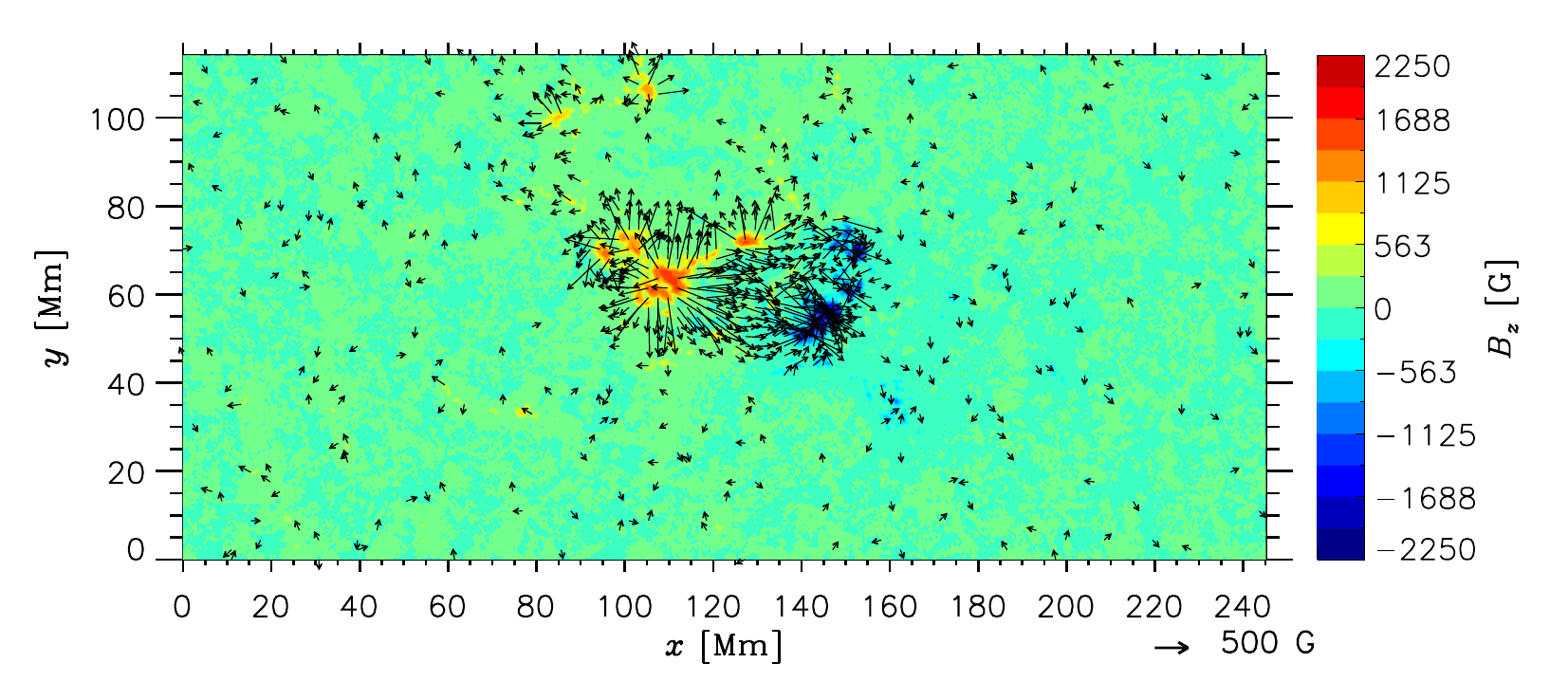}\\[-10pt]
        {\tiny \hspace{1.8cm}  2015 Jan 07, 02:17  \hspace{4.4cm}  2015 Jan 07, 04:10}
        \caption{(continued). For caption see page~\pageref{F1}.}
	\end{figure}
		
	\addtocounter{figure}{-1}
	\begin{figure}
		\flushleft{
			\includegraphics[width=0.432\textwidth,bb=0 15 680 335,clip]{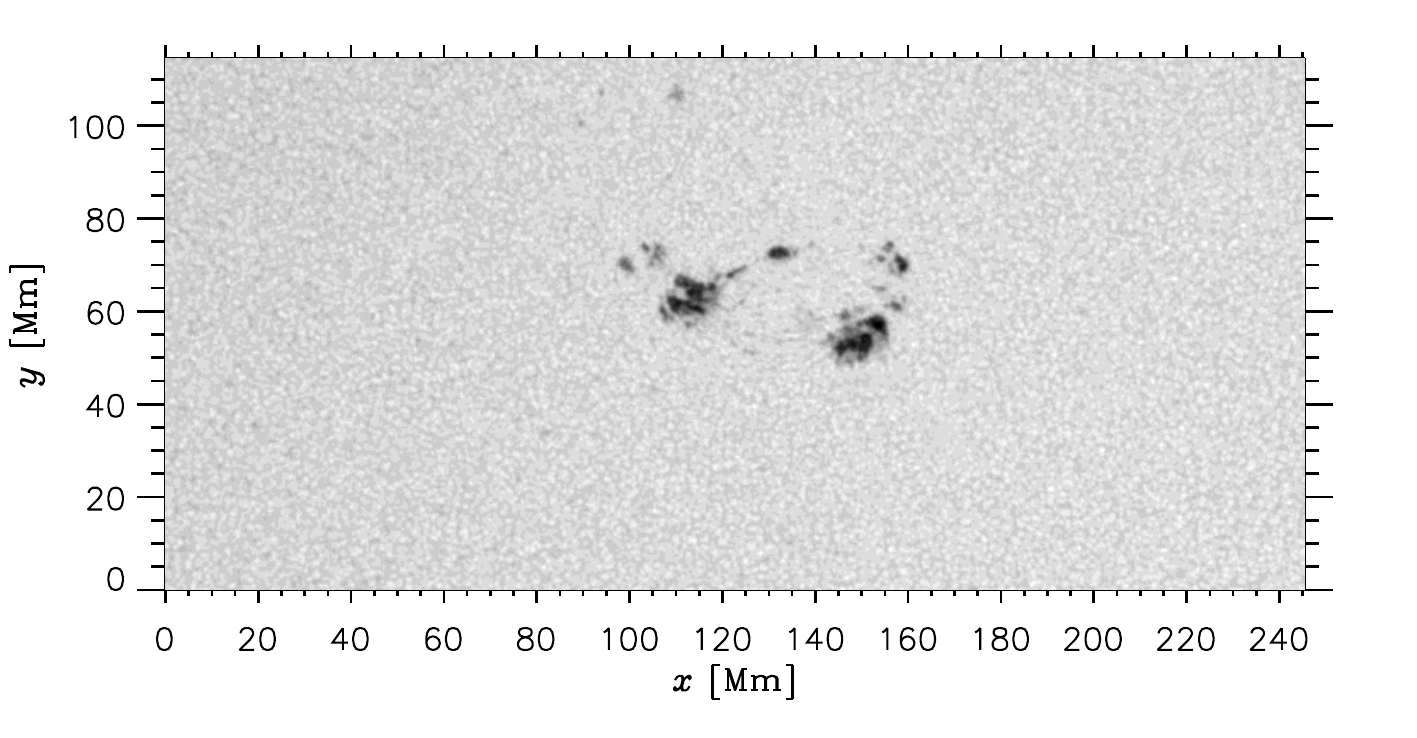}\hspace{0.748cm}
			\includegraphics[width=0.432\textwidth,bb=0 15 680 335,clip]{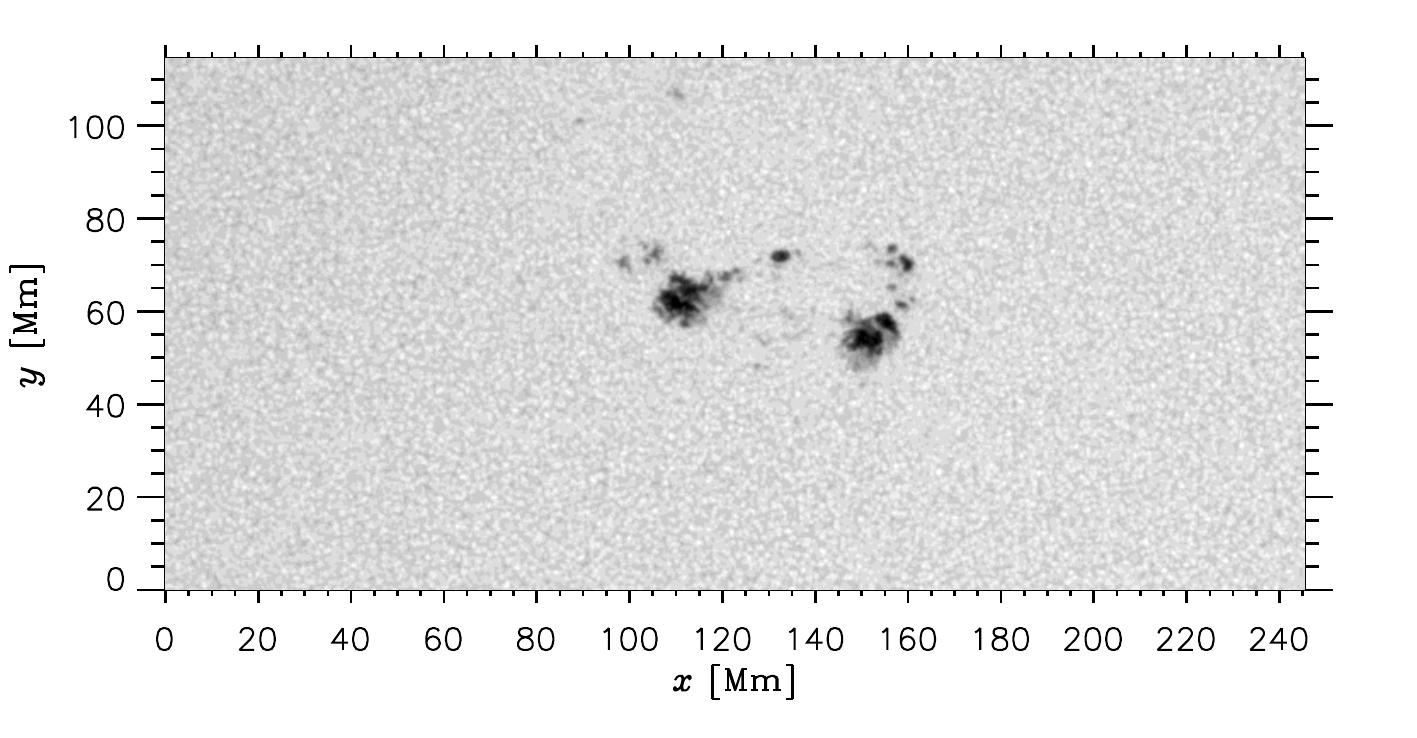}\\
			\includegraphics[width=0.432\textwidth,bb=0 15 680 335,clip]{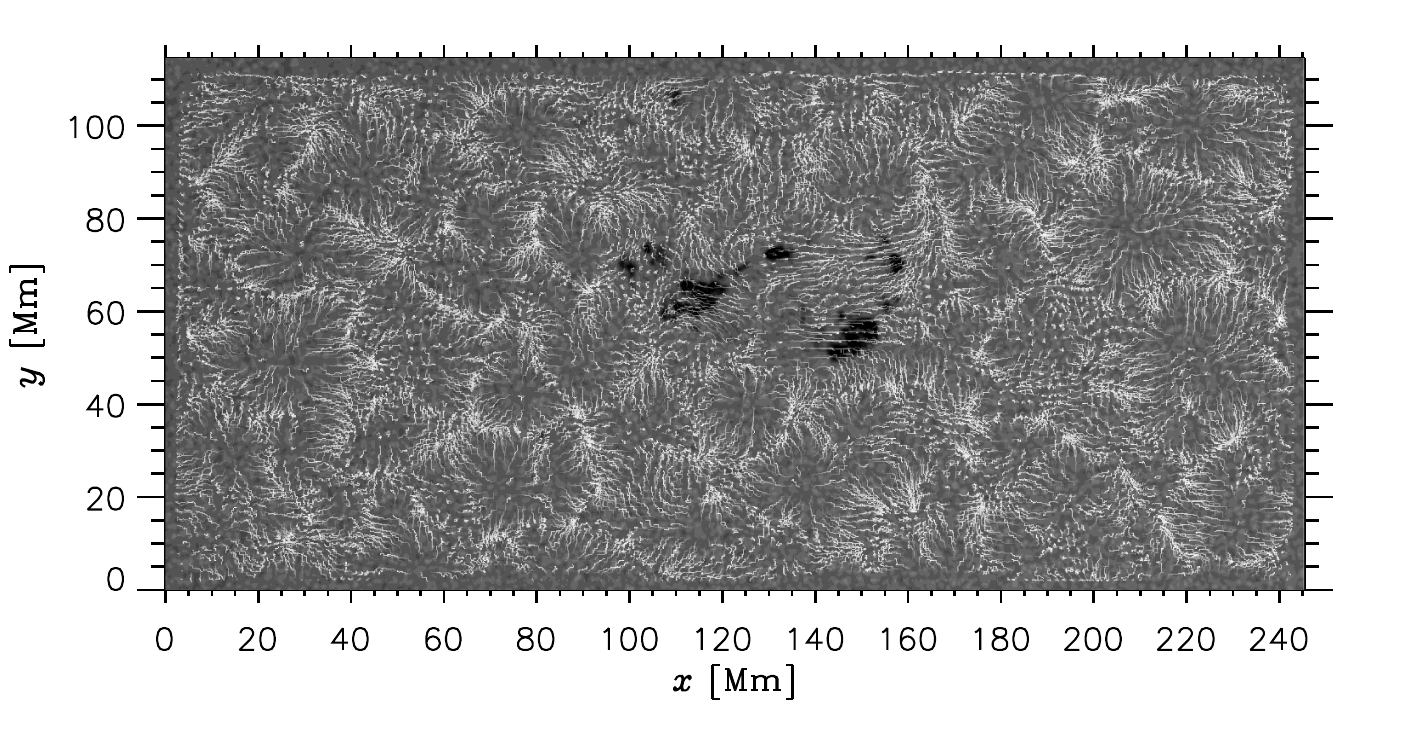}\hspace{0.748cm}
			\includegraphics[width=0.432\textwidth,bb=0 15 680 335,clip]{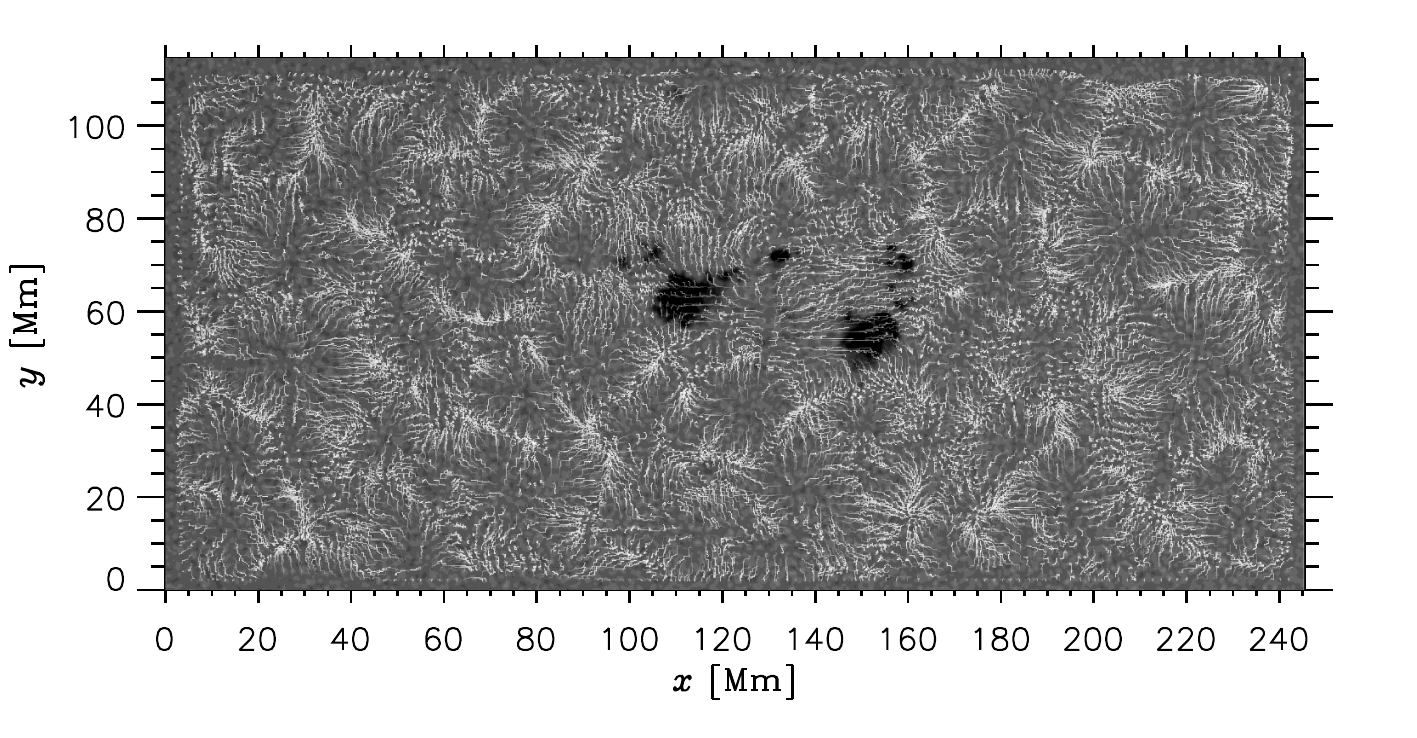}\\
			\includegraphics[width=0.495\textwidth,bb=15 0 793 335,clip]{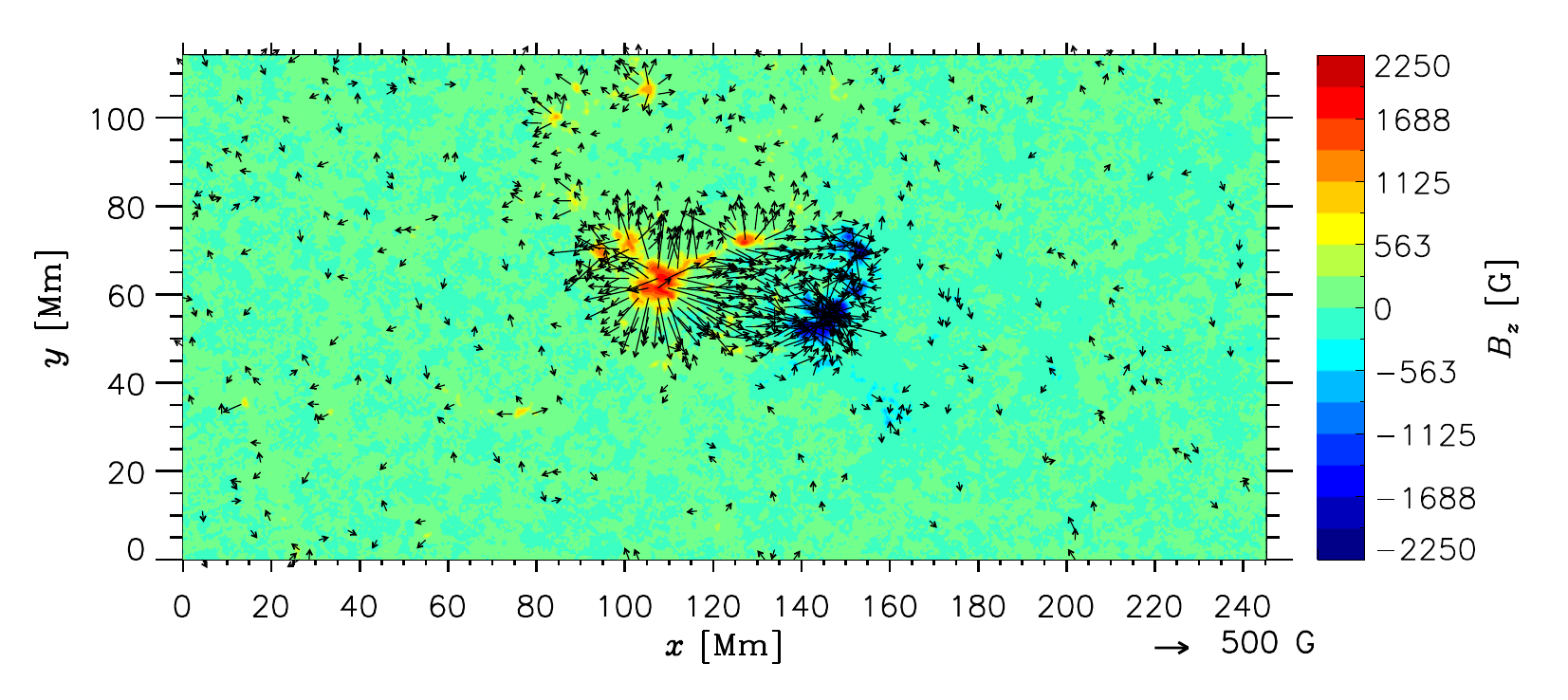}
			\includegraphics[width=0.495\textwidth,bb=15 0 793 335,clip]{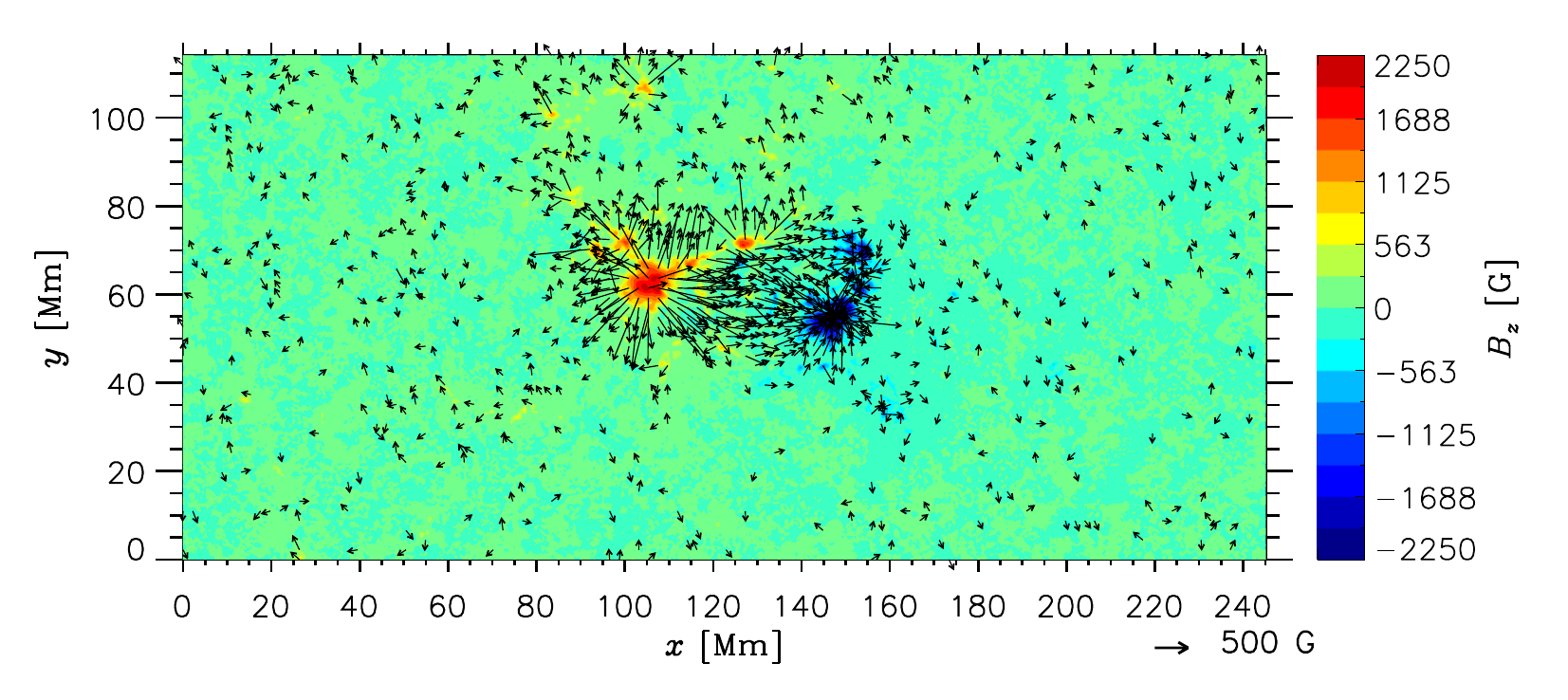}\\[-10pt]
            {\tiny \hspace{1.8cm}  2015 Jan 07, 06:04  \hspace{4.4cm}  2015 Jan 07, 07:57}
		}	\caption{(continued). The origin of AR 12257: images in continuum, maps of tracks overlaid with continuum images and full-vector maps of magnetic field for times indicated below each group of three panels.}\label{Fig12257}\label{F1}
	  \end{figure}		

    To analyse the formation of the ARs, we compare series of their optical images, LCT-derived velocity fields represented by the trajectories of imaginary corks \citep[selected based on the maximum-contrast criterion, see][]{Getling_Buchnev_2010} and full-vector maps of the magnetic field (constructed of its radial, latitudinal and longitudinal components provided by the SHARPs). Each trajectory is shown in the maps in black at the initial time, with gradually increasing  brightness to fully white by the end of tracking. As the trajectories are plotted in this way, their final segments clearly outline the boundaries of supergranules. Below, the cork-trajectory maps will be overlaid with optical images corresponding to the final time of tracking.

    \section{Results}\label{results}

    Before considering particular ARs, let us make some general remarks concerning the flow patterns during the formation of pores and sunspots. It would be extremely strange if the supergranulation pattern remained unperturbed in the areas where strong magnetic fields develop. Some changes, possibly considerable, will be inevitable. However, if the rising-tube mechanism were acting, we would observe a strong, multidirectional divergent flow \textit{preceding} the emergence of magnetic elements. This means that the supergranulation pattern should undergo considerable changes \textit{before} the emergence of pores. Specifically, flows should develop that resemble those arising in the uppermost water layer above the emerging body, no matter a submarine or a diver's head.
        
	\subsection{Active region 12257}\label{12257}
    
    The origin of AR 12257 is illustrated by Figure~\ref{Fig12257}, where groups of maps are shown for an approximately one-day period (2015 Jan 6--7) with an interval slightly shorter than 2 hours. Each group consists of a white-light image, a map of cork trajectories overlaid with the same white-light image and a full-vector map of magnetic field. The tracking interval (of duration 1~h~52~m~30~s) ends at the time for which the continuum image is shown. It is this time that is indicated below the group, and the magnetic-field map is taken for a time differing by no more than 3--6 minutes.

    \begin{figure}   
		\flushleft{
			\includegraphics[width=0.432\textwidth,bb=0 15 680 335,clip]{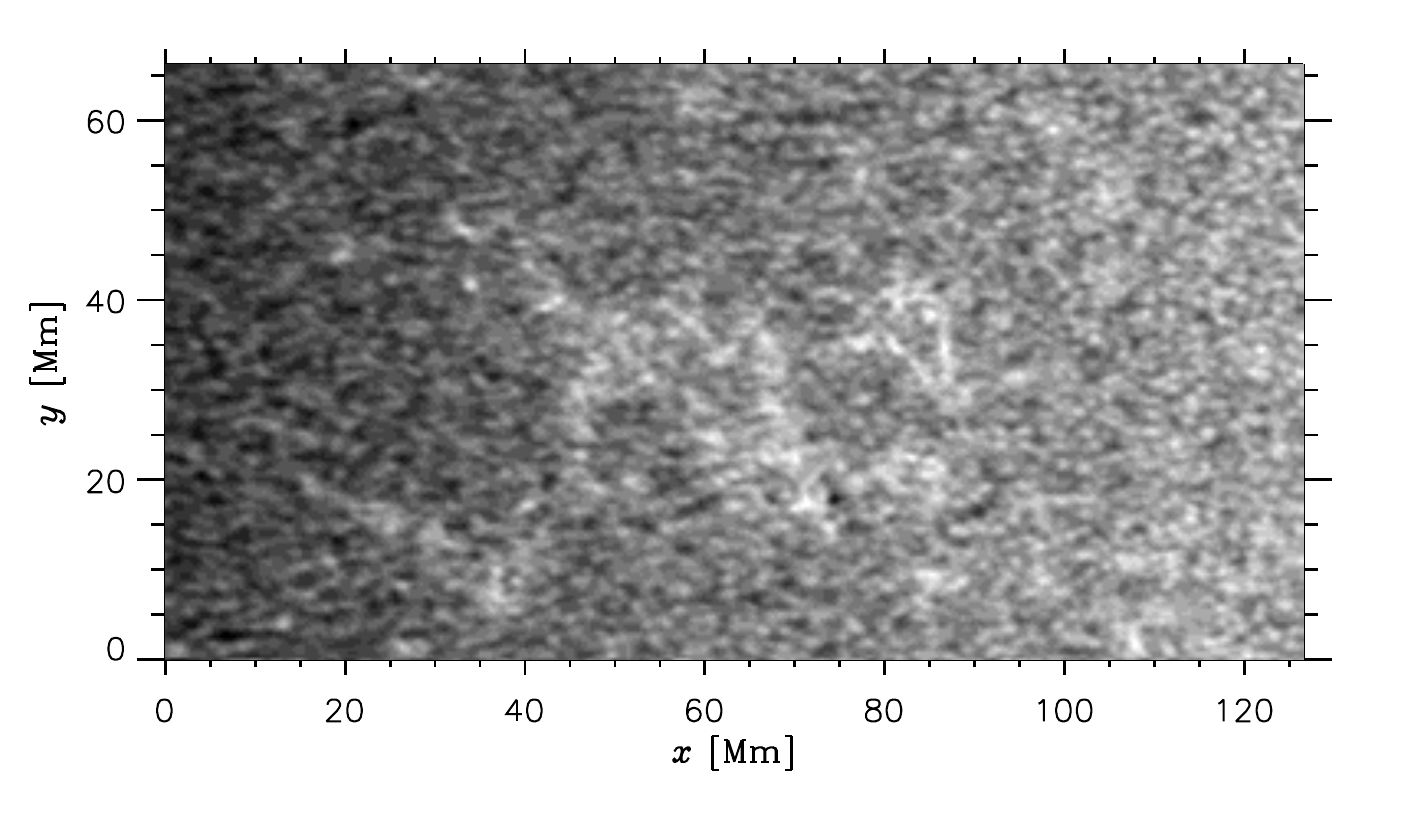}\hspace{0.75cm}		
			\includegraphics[width=0.432\textwidth,bb=0 15 680 335,clip]{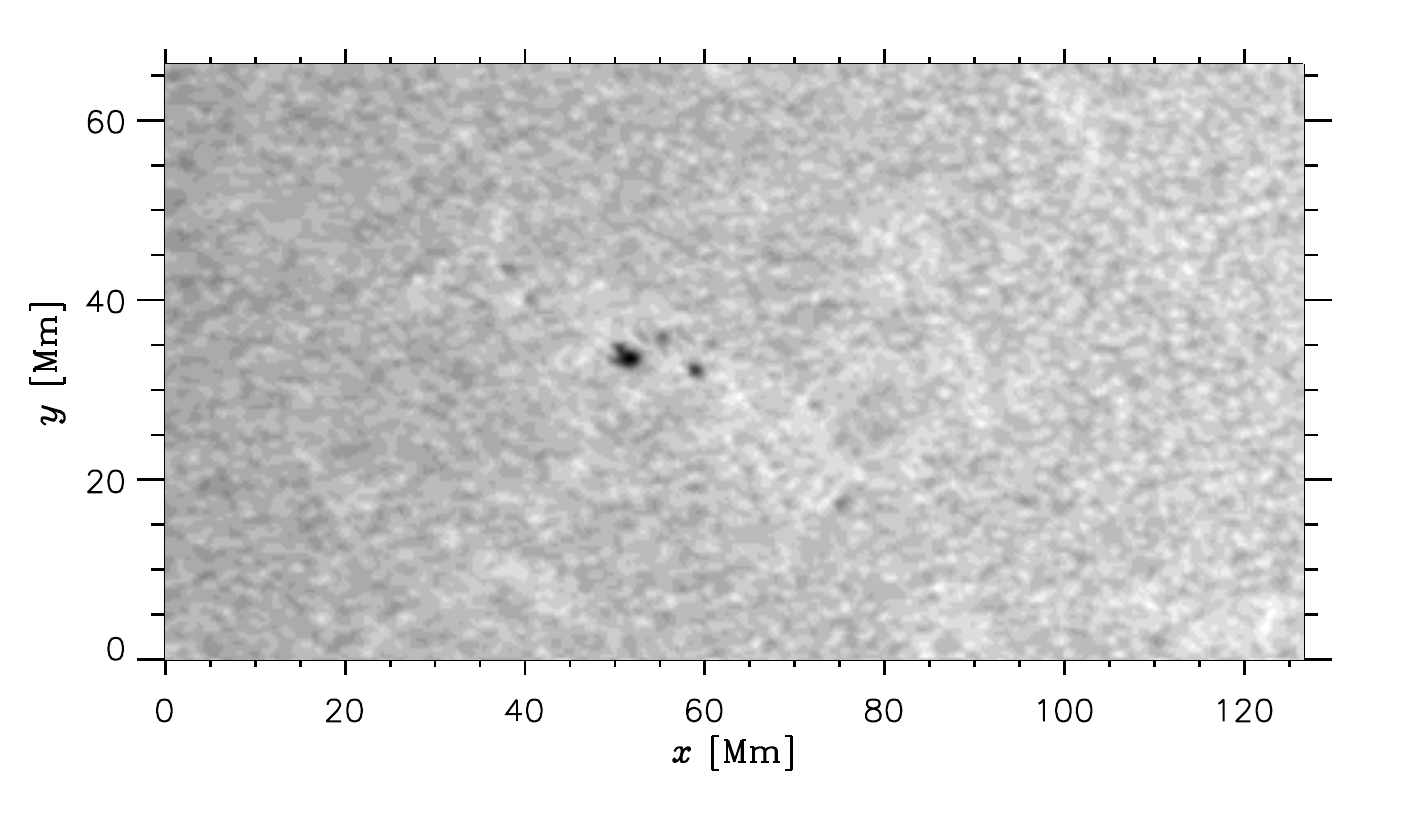}\\
			\includegraphics[width=0.432\textwidth,bb=0 15 680 335,clip]{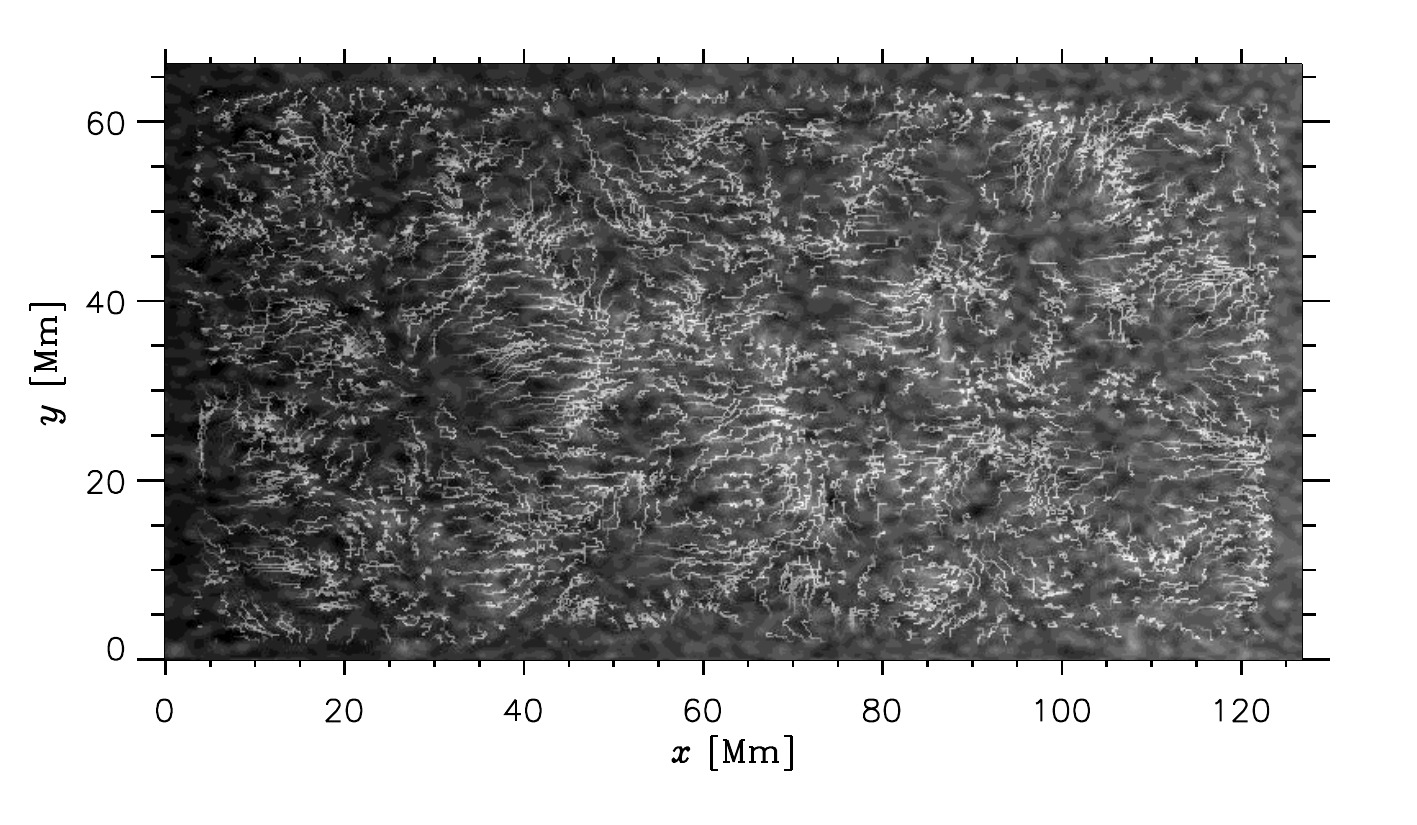}\hspace{0.75cm}
			\includegraphics[width=0.432\textwidth,bb=0 15 680 335,clip]{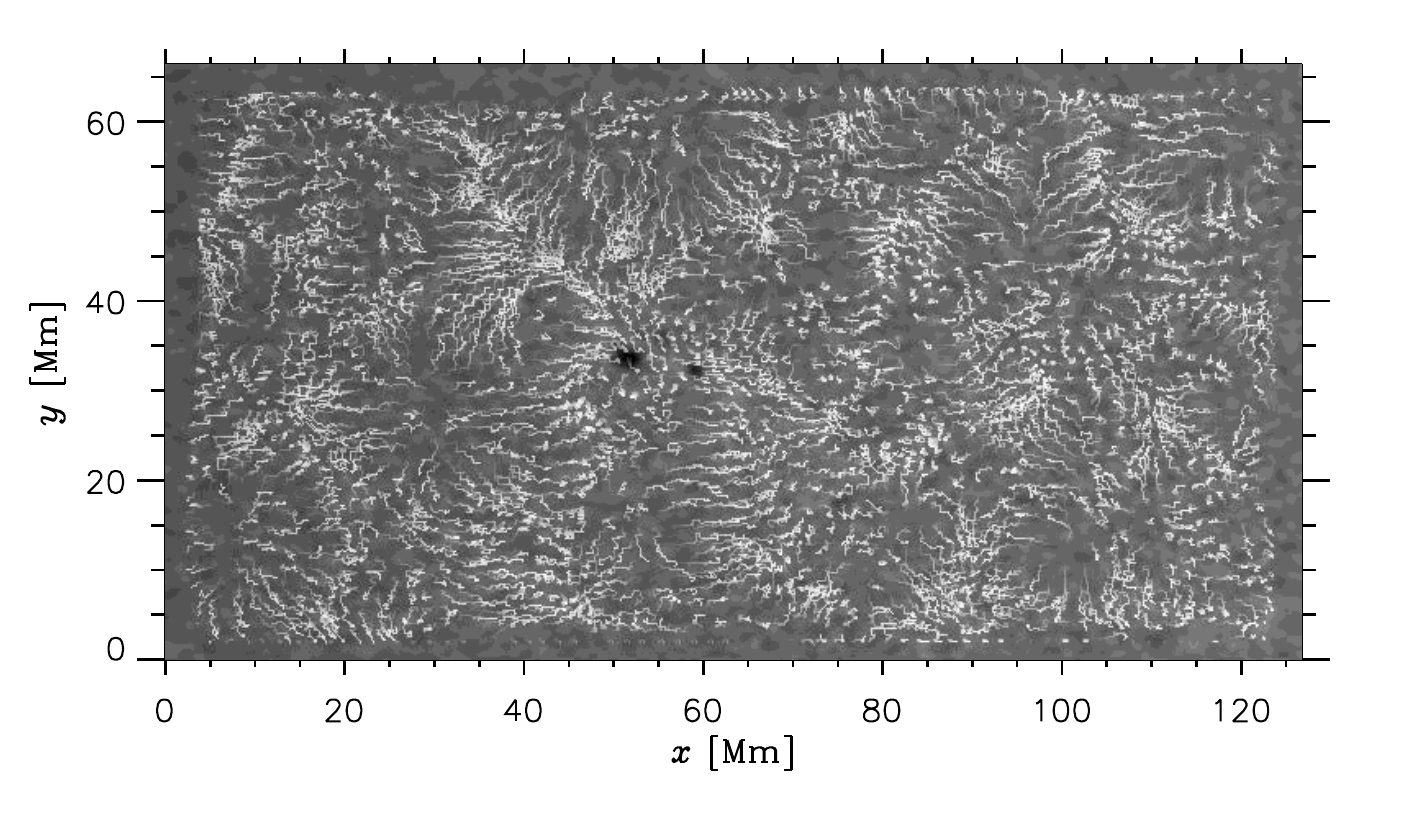}\\		
			\includegraphics[width=0.495\textwidth,bb=15 0 793 335,clip]{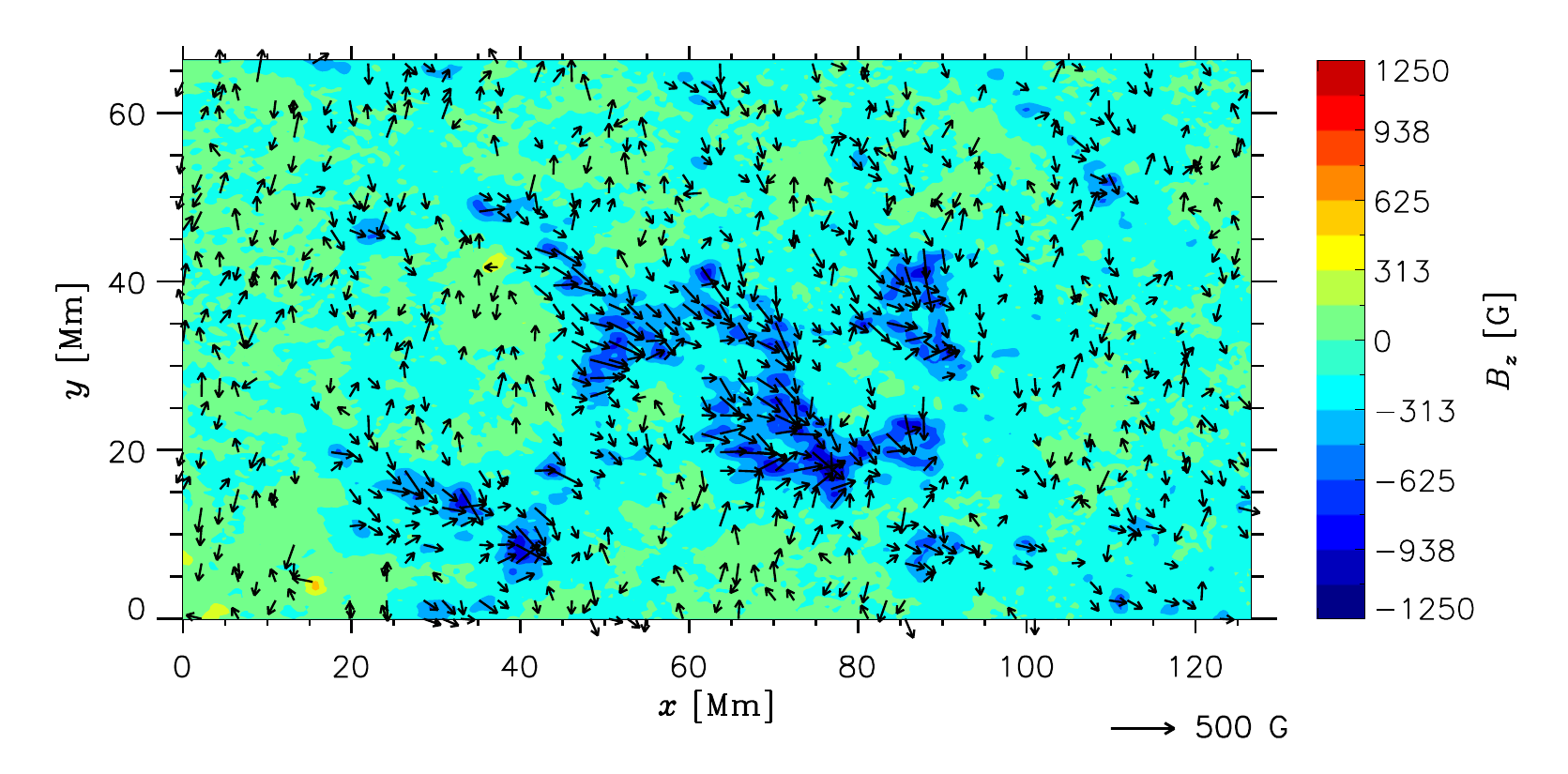}
			\includegraphics[width=0.495\textwidth,bb=15 0 793 335,clip]{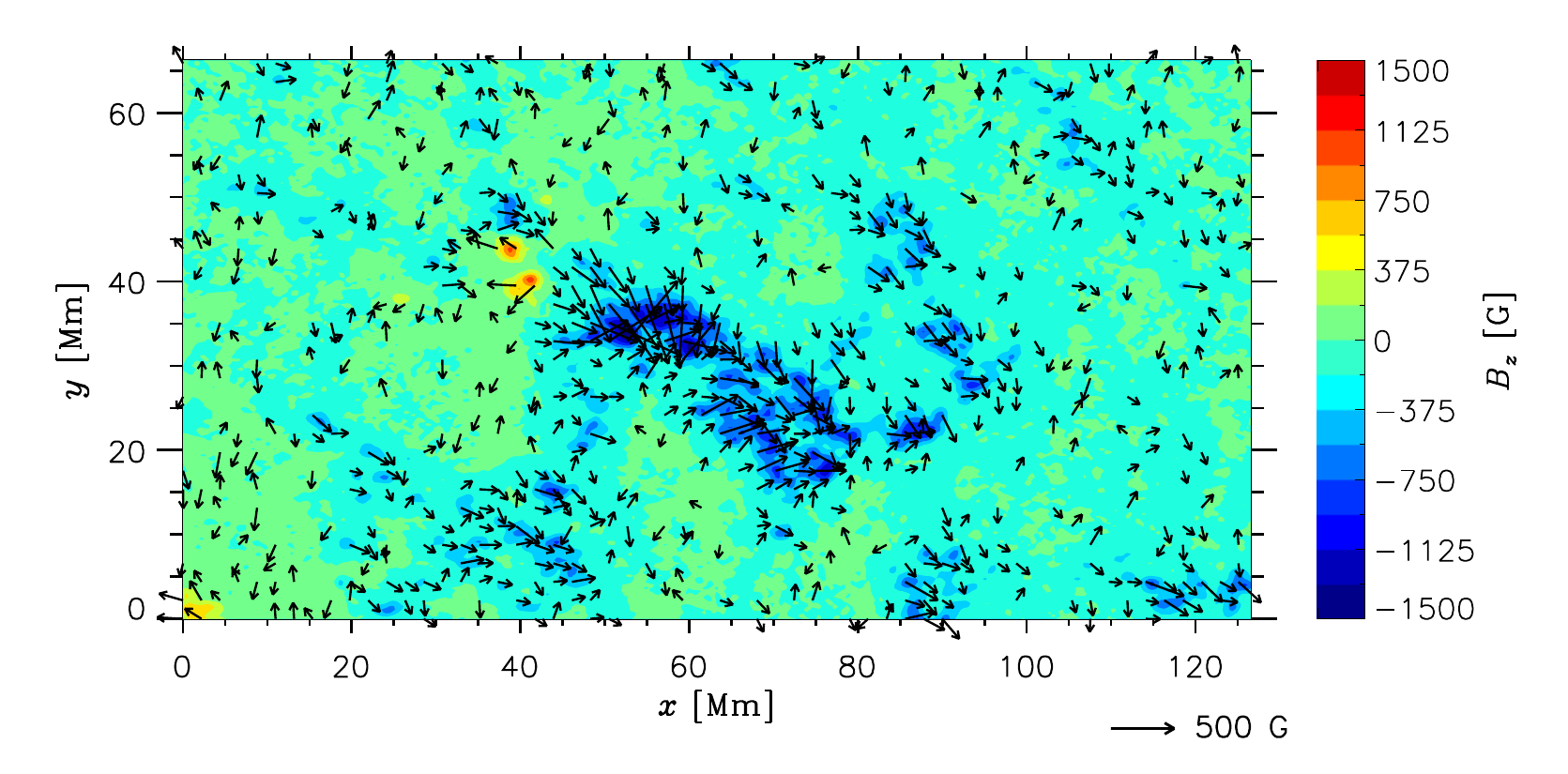}
			}\\[-10pt]
            {\tiny \hspace{1.8cm}  2015 Aug 10, 15:22  \hspace{4.4cm}  2015 Aug 11, 02:51}\\
			\includegraphics[width=0.432\textwidth,bb=0 15 680 335,clip]{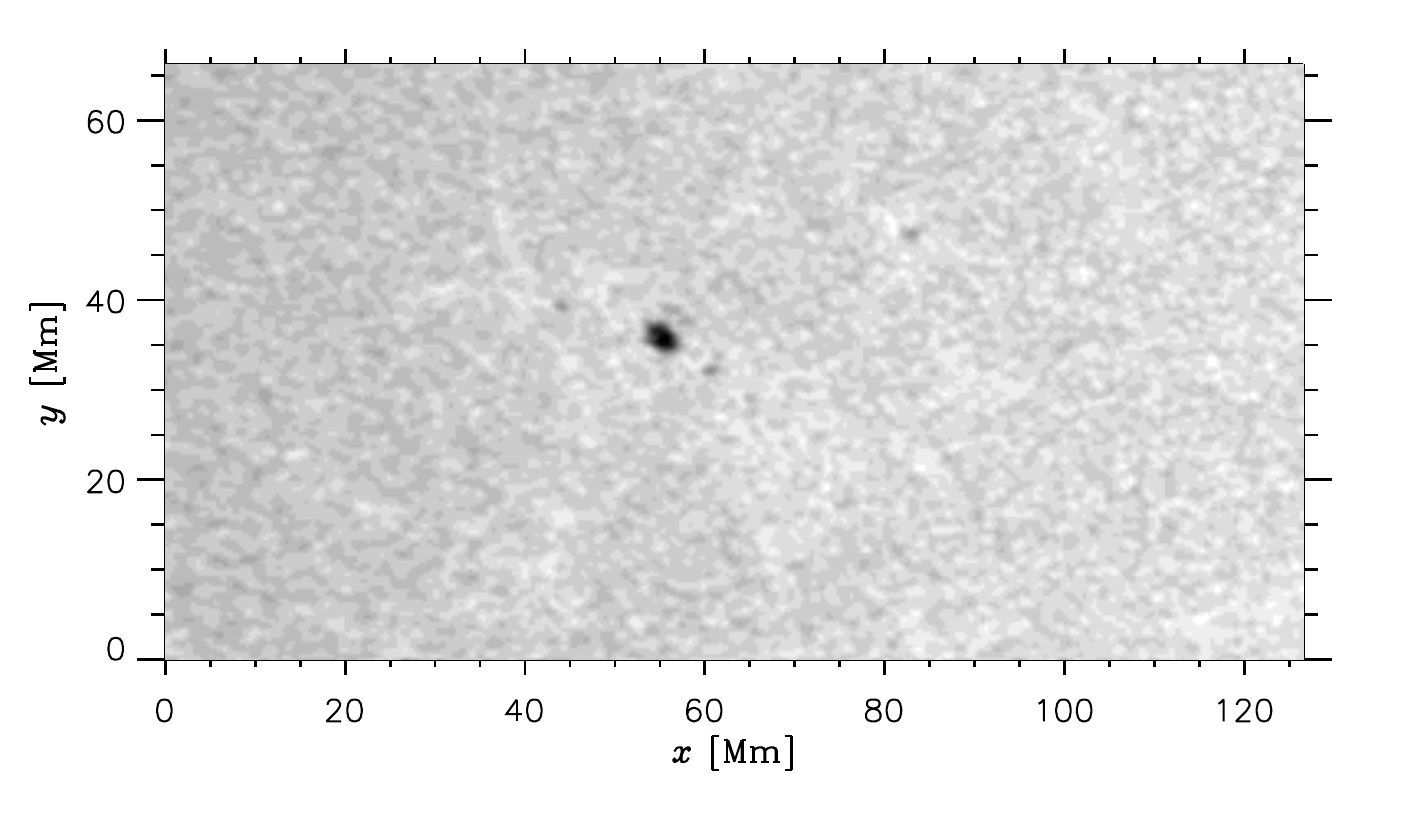}\hspace{0.75cm}		
			\includegraphics[width=0.432\textwidth,bb=0 15 680 335,clip]{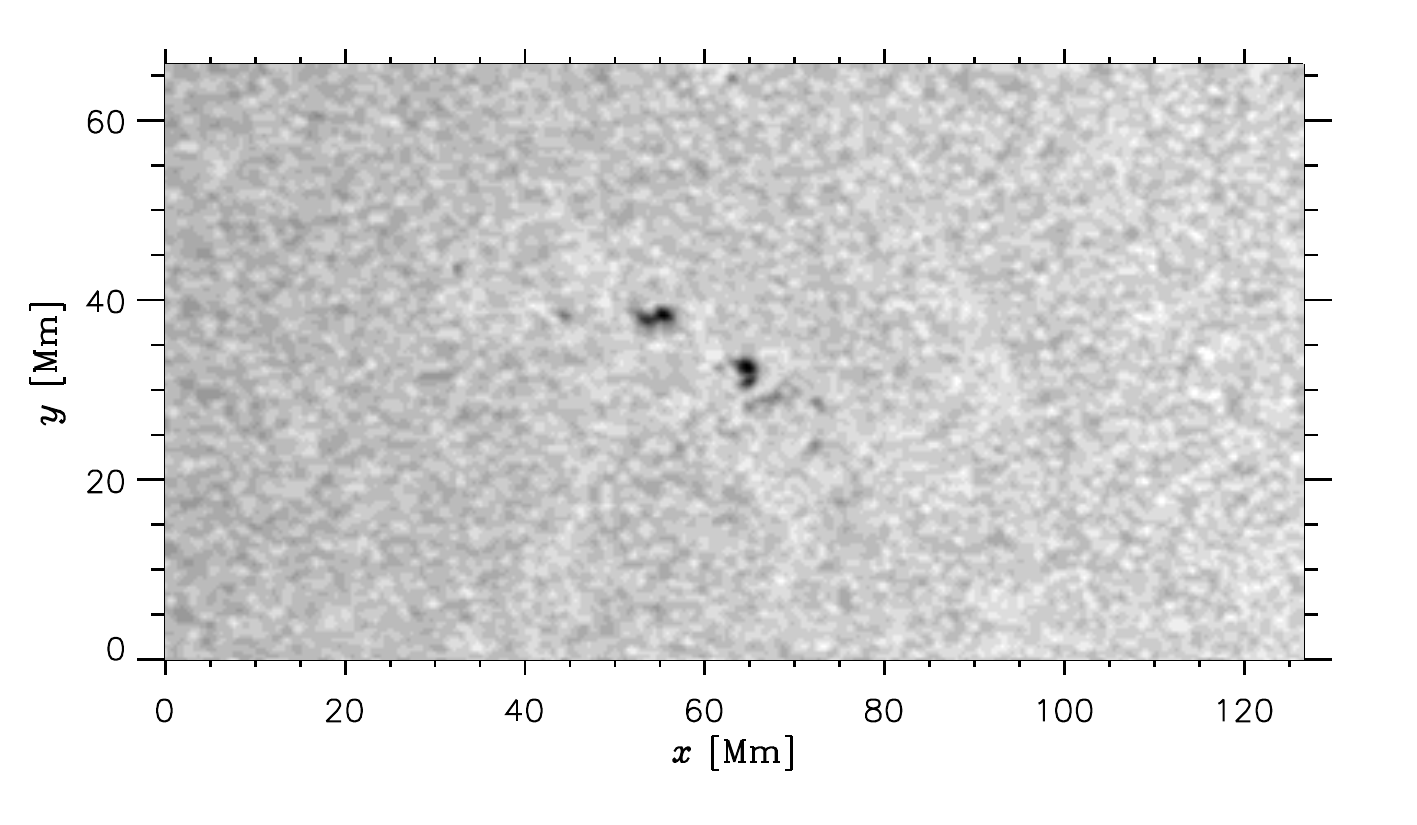}\\
			\includegraphics[width=0.432\textwidth,bb=0 15 680 335,clip]{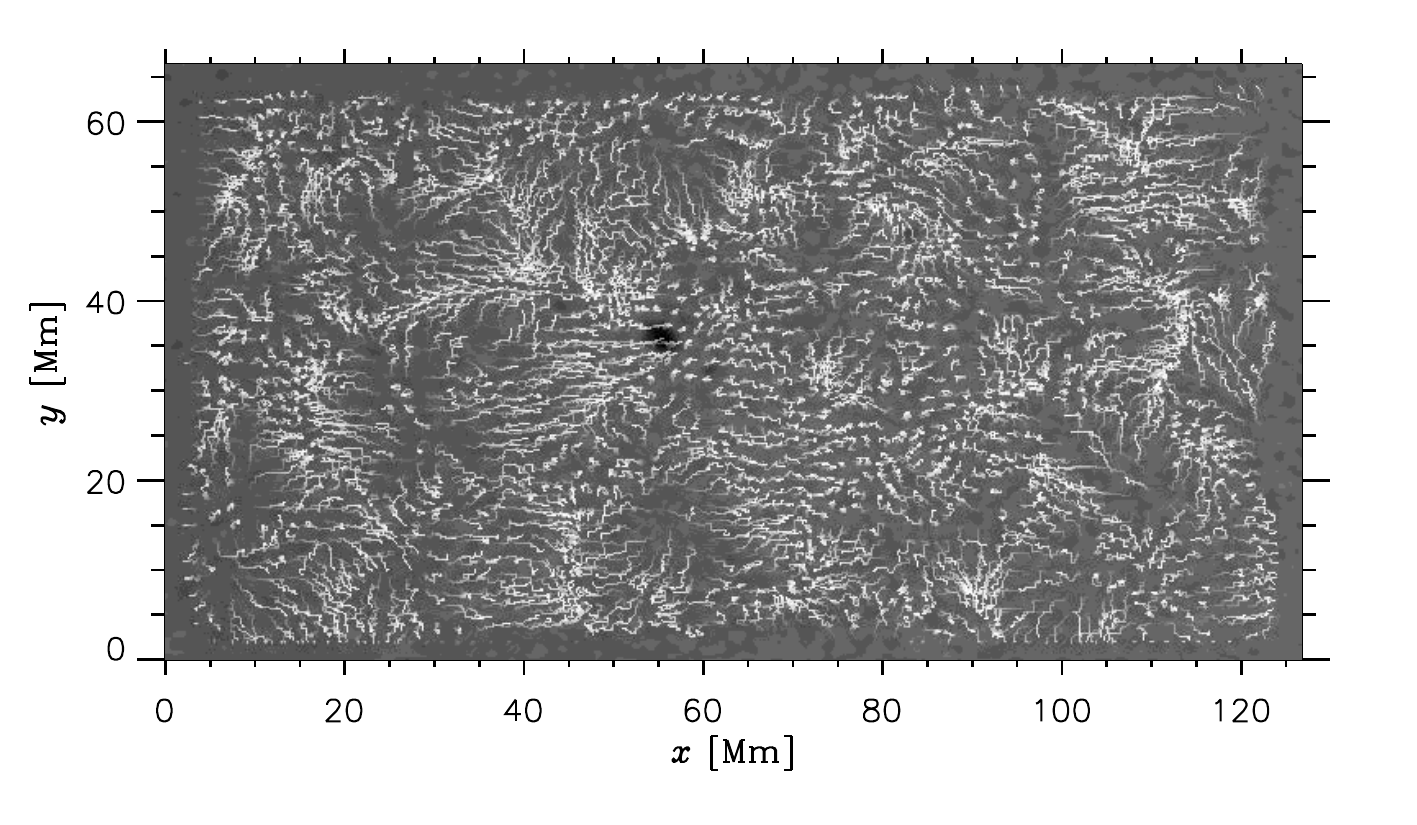}\hspace{0.75cm}
			\includegraphics[width=0.432\textwidth,bb=0 15 680 335,clip]{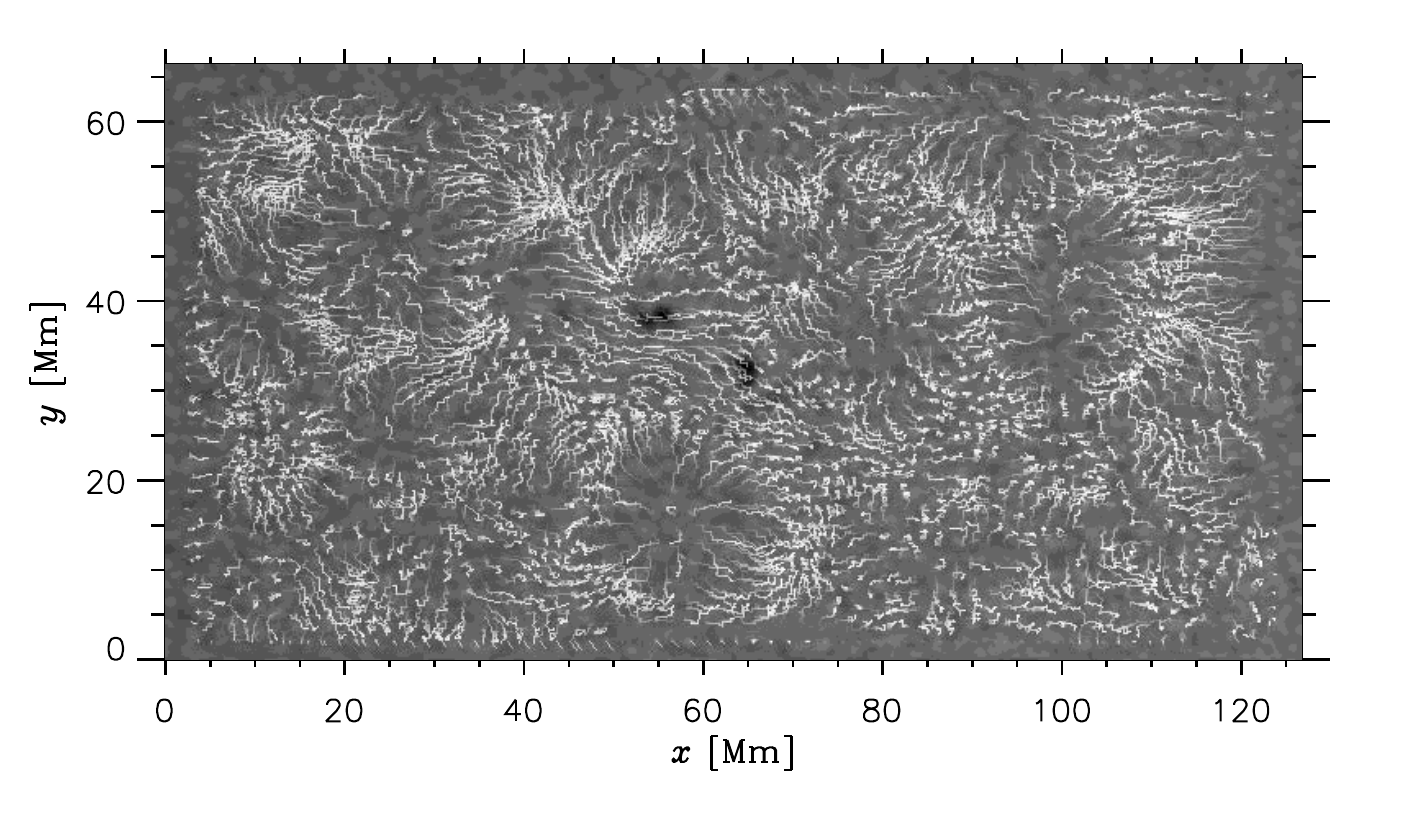}\\		
			\includegraphics[width=0.495\textwidth,bb=15 0 793 335,clip]{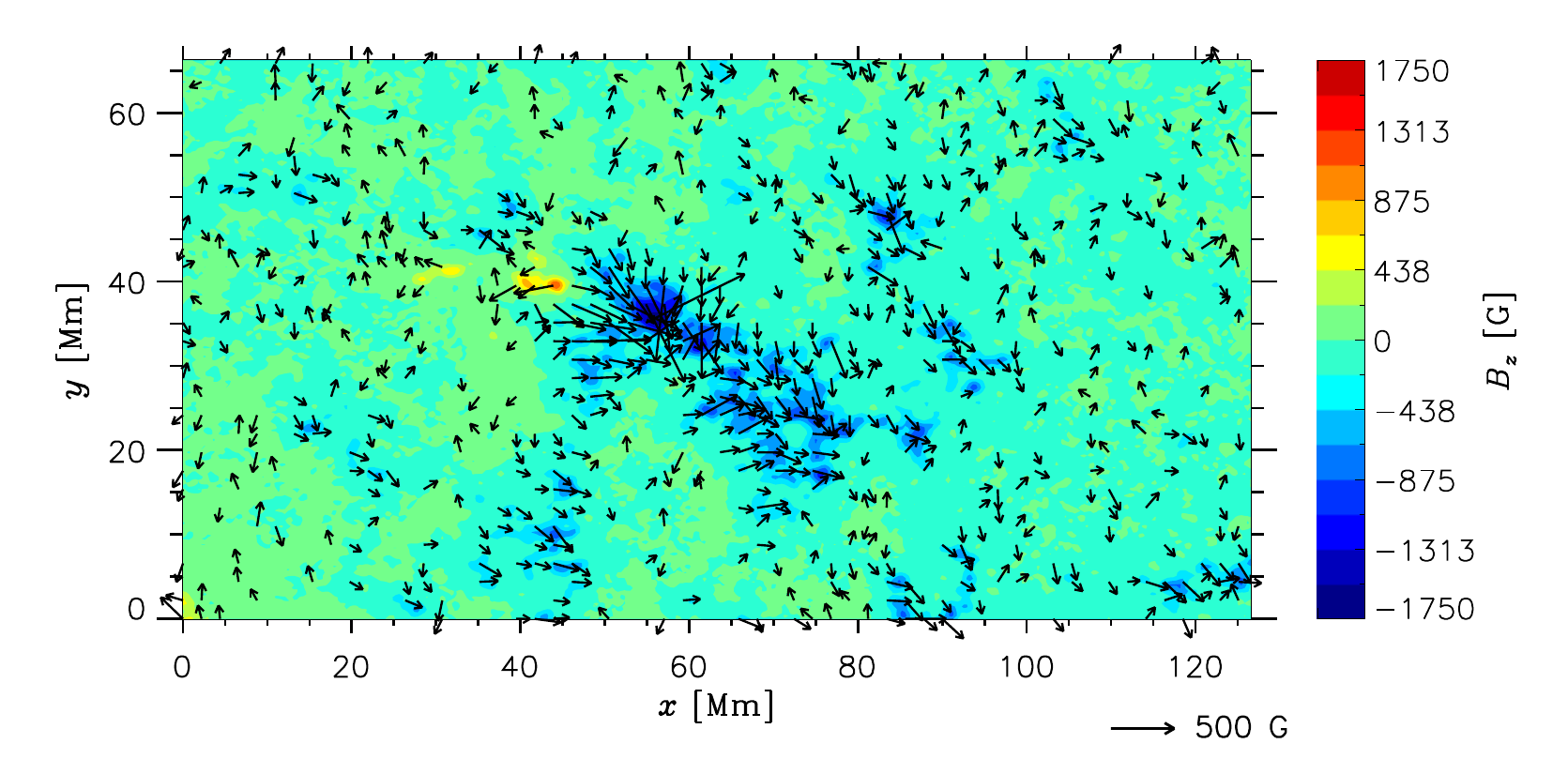}
			\includegraphics[width=0.495\textwidth,bb=15 0 793 335,clip]{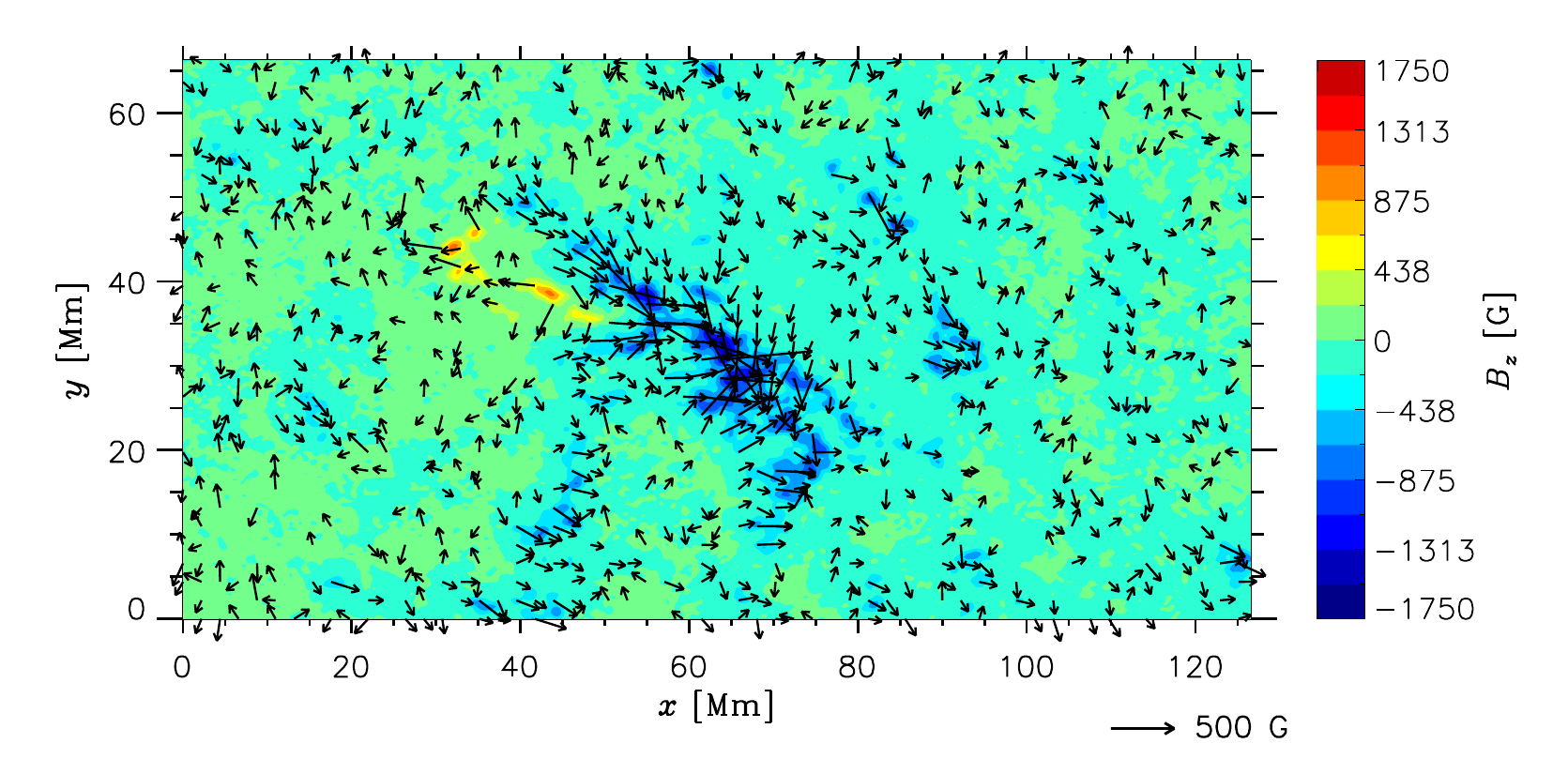}\\[-10pt]
		      {\tiny \hspace{1.8cm}  2015 Aug 11, 06:47  \hspace{4.4cm}  2015 Aug 11, 14:49}\\
			
			\includegraphics[width=0.432\textwidth,bb=0 15 680 335,clip]{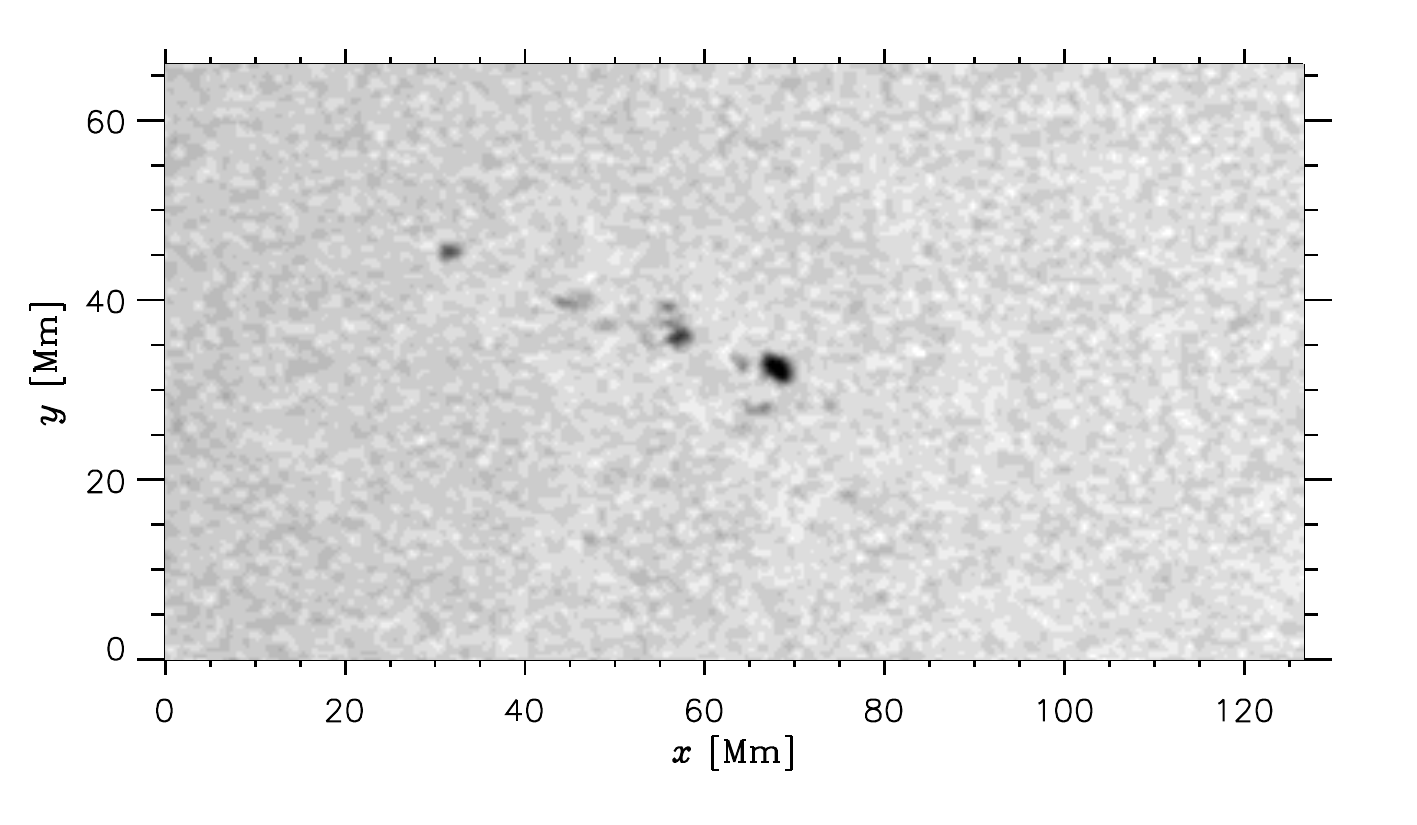}\hspace{0.75cm}		
			\includegraphics[width=0.432\textwidth,bb=0 15 680 335,clip]{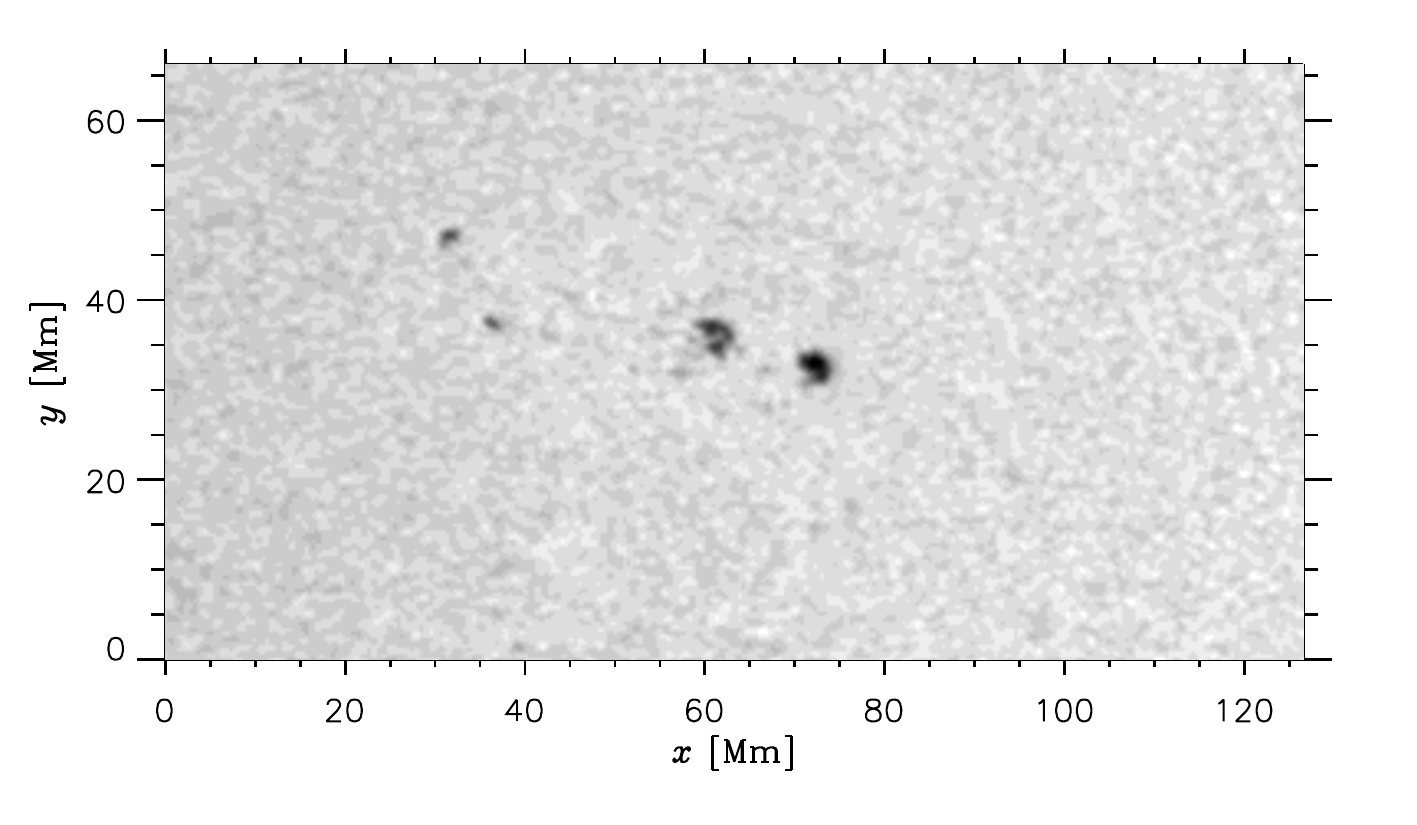}\\
			\includegraphics[width=0.432\textwidth,bb=0 15 680 335,clip]{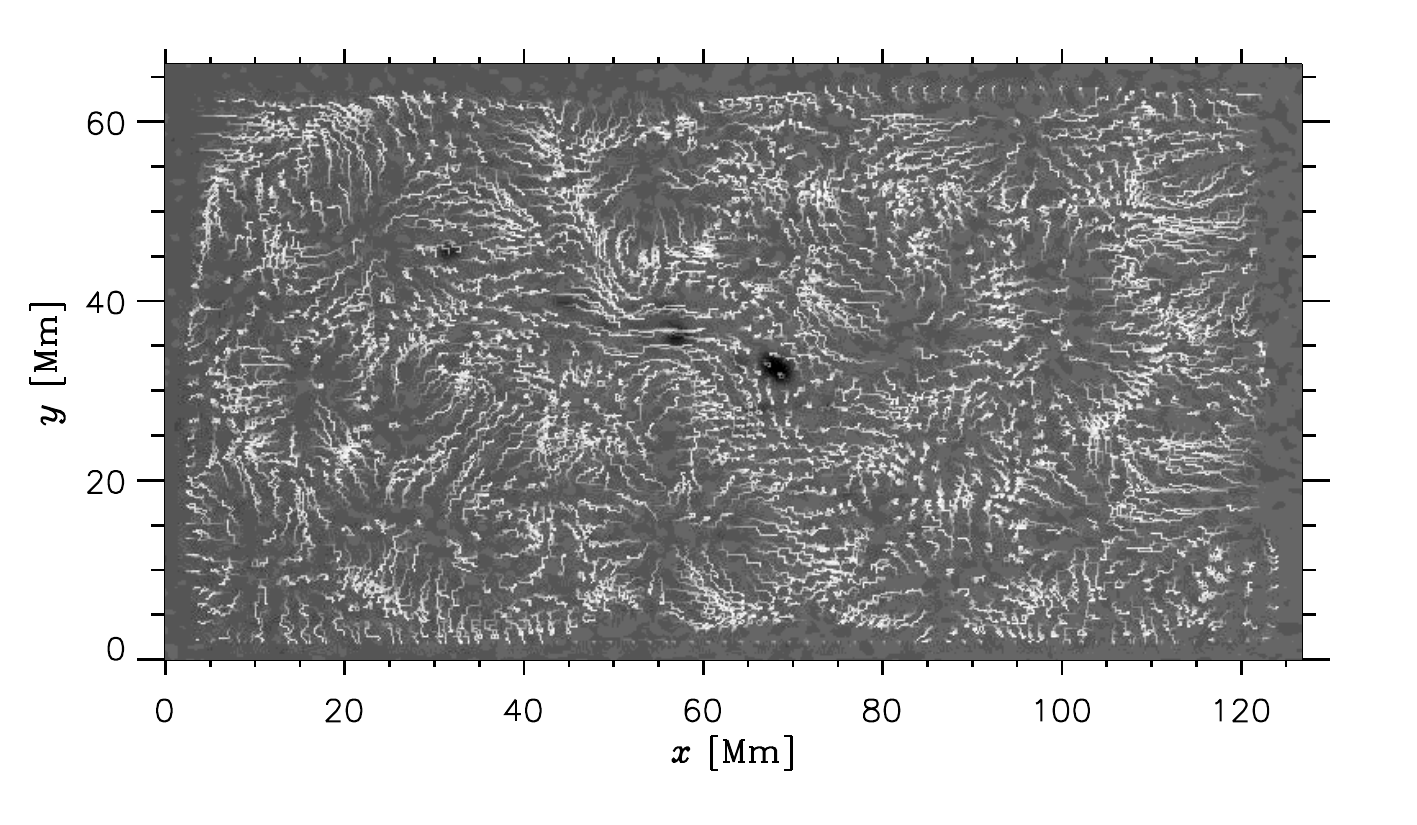}\hspace{0.75cm}
			\includegraphics[width=0.432\textwidth,bb=0 15 680 335,clip]{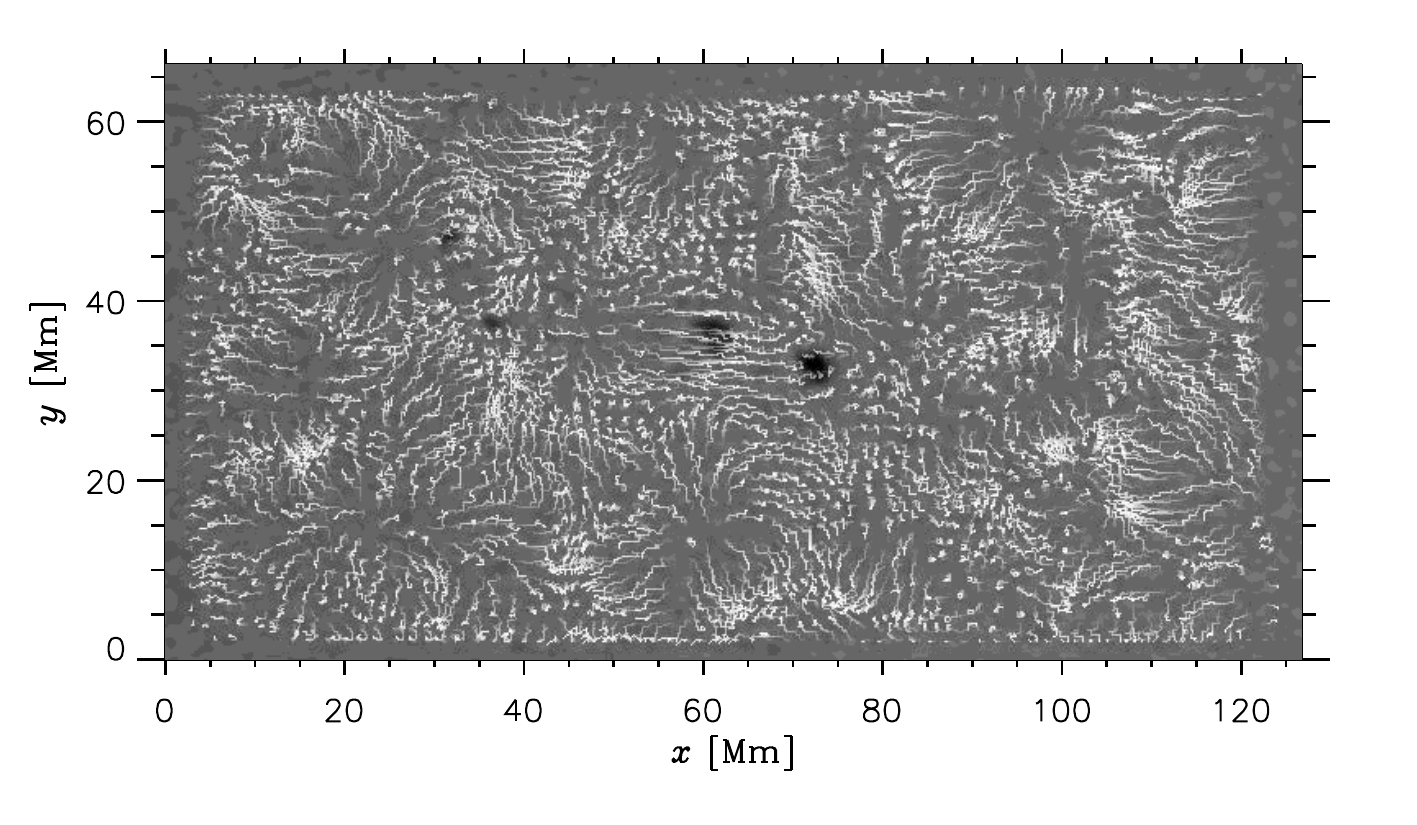}\\		
			\includegraphics[width=0.495\textwidth,bb=15 0 793 335,clip]{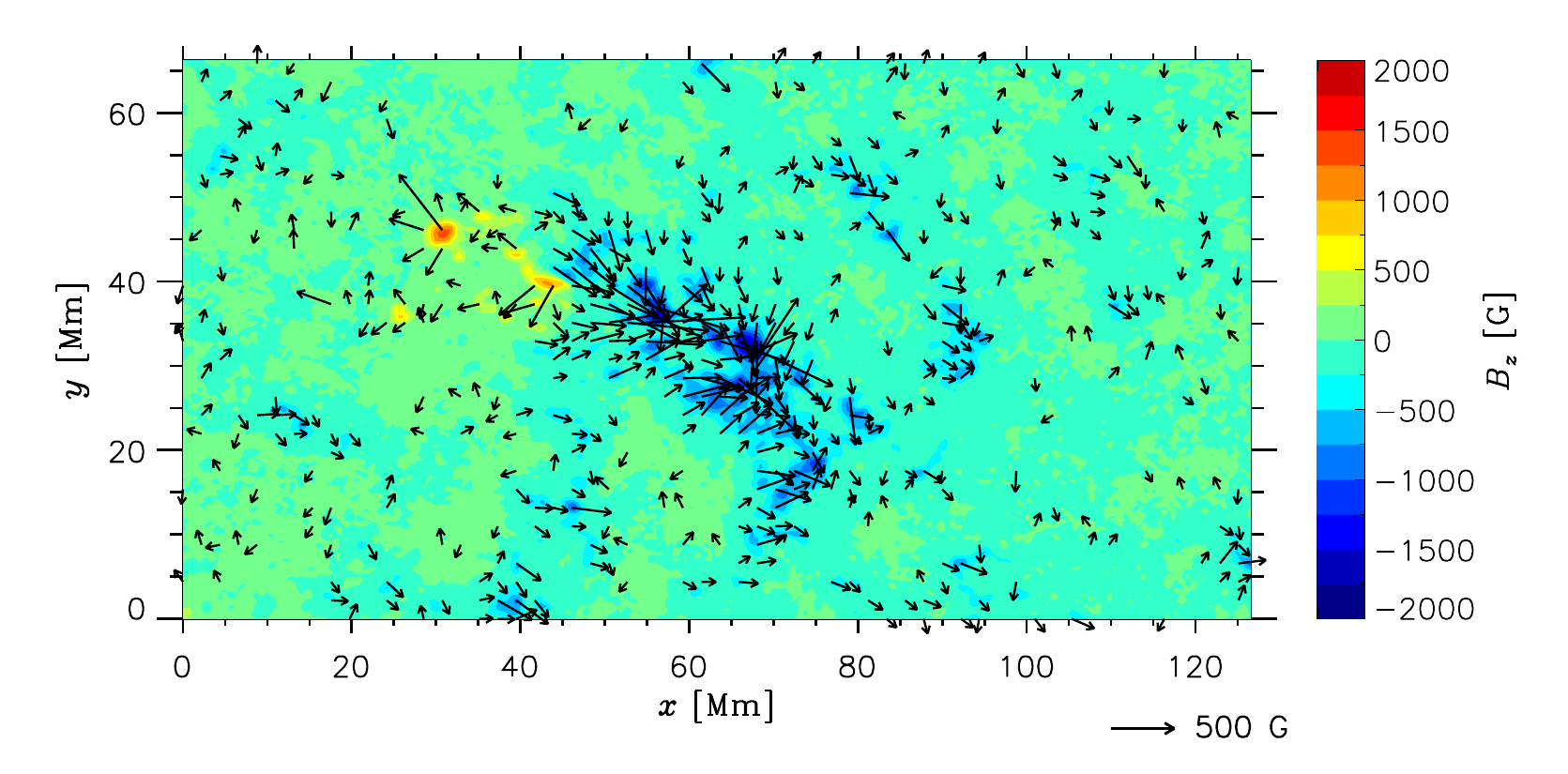}
			\includegraphics[width=0.495\textwidth,bb=15 0 793 335,clip]{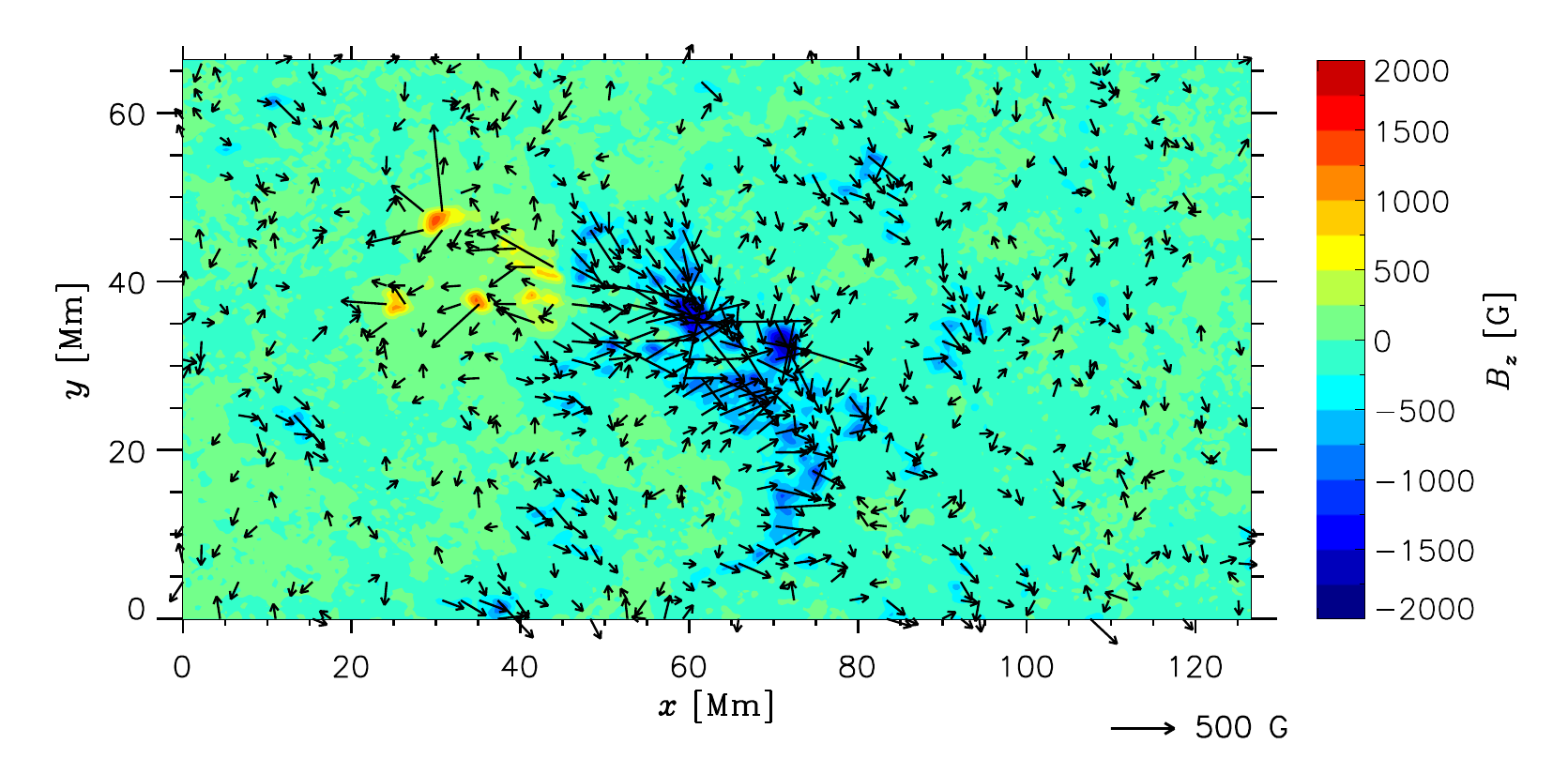}\\[-10pt]
		    {\tiny \hspace{1.8cm}  2015 Aug 11, 18:47  \hspace{4.4cm}  2015 Aug 11, 22:48}\\[10pt]
        \caption{The origin of AR 12400: images in continuum, maps of tracks overlaid with continuum images and full-vector maps of magnetic field for times indicated below each group of three panels.}\label{Fig12400}
	\end{figure}

	At Jan 06, 07:21 and 09:13, faint signs of activity in the form of a few pores can be detected. They are mainly notable in one supergranule, where the positive (trailing) magnetic polarity is initially present, while a pore of the negative (leading) polarity nucleates by the second time, below and to the right of the largest pore, approximately at the point (130, 60) Mm. By times 11:06 and 12:58, the positive polarity is enhanced more strongly than the negative one, the latter having formed very small pores. Thus, the two magnetic polarities develop asymmetrically from the very beginning. No significant disruptions of the supergranulation pattern can be seen.

    At 14:51 and 16:43, as previously, the positive polarity is stronger and more compact than the negative one, forming a small cluster of pores. The negative polarity is represented by a short chain of pores. This evolution occurs against the background of the supergranulation pattern that does not experience significant perturbations but exhibits only usual processes of formation and destruction of individual supergranules. The process continues in a qualitatively similar way at 18:40 and 20:40, and a chain of pores of positive polarity becomes notable by Jan 06, 22:32---Jan 07, 00:25. During this interval, the largest pores gradually develop penumbras, becoming classical sunspots. These features become especially noticeable by 02:17 and 04:10, when a bipolar sunspot group forms. Only at this development stage, the magnetic polarities in the group become comparable in their fluxes and visual appearance. At 06:04 and 07:57, the sunspot group and bipolar magnetic region are well developed. And even now, the supergranulation pattern preserves its usual appearance. More precisely speaking, something resembling a divergent flow around the young sunspot group (in essence, the flow in a very large supergranule) can be identified at 20:40, Jan 06. However, this is \textit{not the case at preceding times}, even as recently as two hours before: then, smaller supergranules are still present in this area, although they are quite expectedly perturbed by the magnetic fields. Therefore, no spreading flow does here precede the pore formation.
 	
    The essential features of the AR 12257 development can be summarised as follows:
    \begin{itemize}
        \item The origin of the first pores is associated with the emergence of magnetic elements of mostly the same, trailing polarity.
        \item The development of the AR and sunspot group is a process asymmetric with respect to magnetic polarities, especially at its early stage.
        \item The two polarities initially nucleate within the same supergranule.
        \item At the early stage of the process, the regular supergranulation pattern is preserved for a fairly long time.
        \item The last mentioned fact implies the absence of any spreading flow on the scale of the entire AR before the formation of pores.
    \end{itemize}

   	\subsection{Active region 12400}\label{12400}

    An early development stage of AR 12400 is shown in Figure~\ref{Fig12400}. Initially, there exists an extensive area of magnetic fields having the leading (negative) polarity. By time 2015 Aug 11, 02:51, two pores, both also of the leading polarity, have formed. At that time, the trailing (positive) polarity is represented by two very small magnetic elements near the point (40, 40) Mm, whose presence is almost unnoticeable in white light. The growth of the large pore is not accompanied by any qualitative changes in its surroundings (see time 06:47), and the only new feature formed by time 14:49 is a second distinct pore---again, of the leading polarity. By 18:47, first trailing-polarity pores become visible. At 22:48, the leading-polarity pores can ultimately be classified as sunspots with very weakly developed penumbras, but this still cannot be said about the trailing-polarity features. Throughout the whole time interval considered, no radical changes in the supergranulation pattern occur, and the growth of new magnetic elements fits into the existing cellular network.

    To summarise the most important details of the evolution of AR 12400, the following should be noted:
    \begin{itemize}
        \item The development pattern of the two magnetic polarities is enormously asymmetric.
        \item The development of sunspots starts against the background of the widely distributed magnetic field of the leading polarity.
        \item The emergence of the sunspots does not disrupt the existing supergranulation pattern.
        \item In particular, no spreading flow is observed.
        \item The trailing-polarity sunspots develop with a very long delay relative to those of the leading polarity.
    \end{itemize}

    \begin{figure}
		\centerline{
			\includegraphics[width=0.5\textwidth,bb=0 80 680 742,clip]{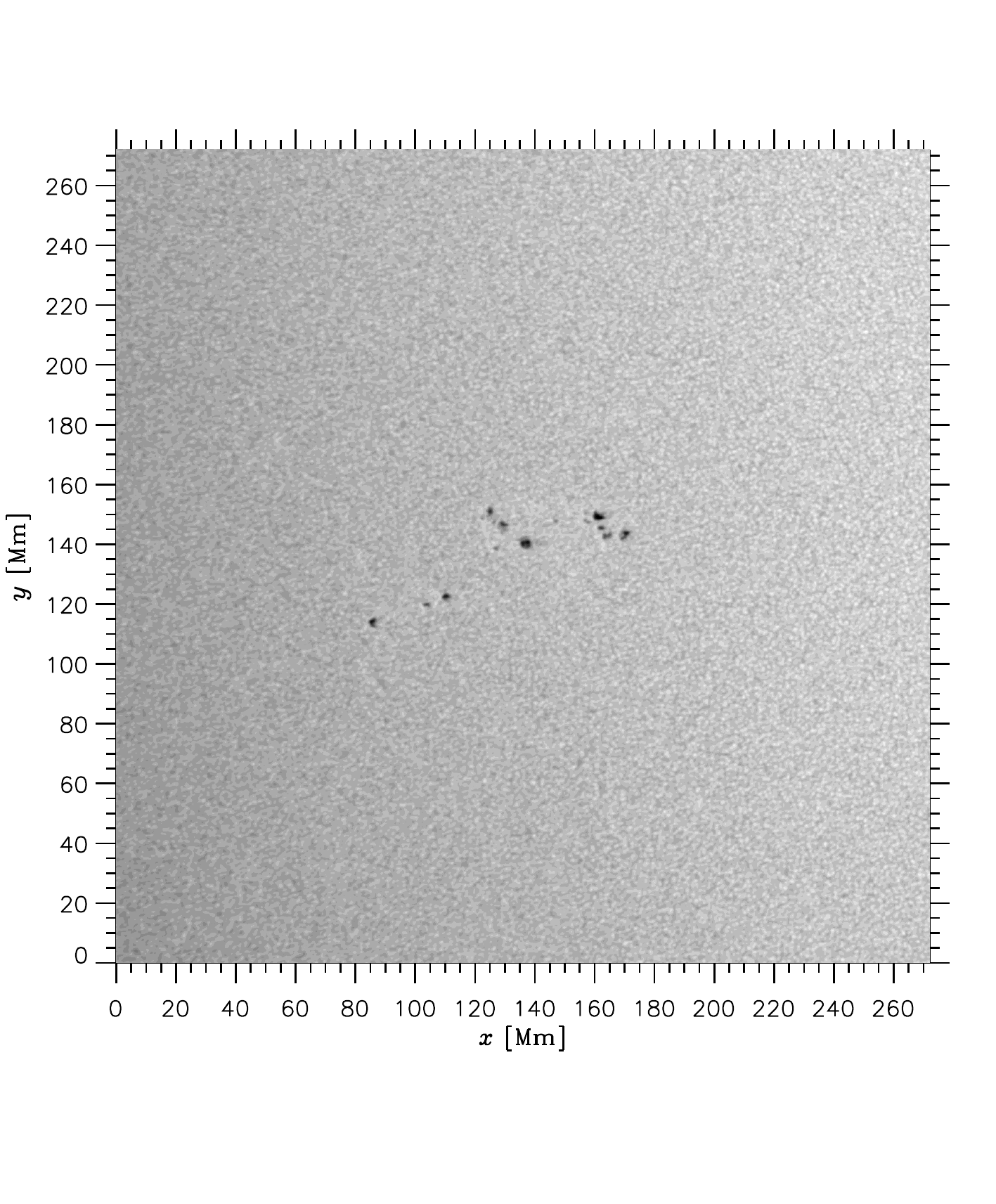}
			\includegraphics[width=0.5\textwidth,bb=0 80 680 742,clip]{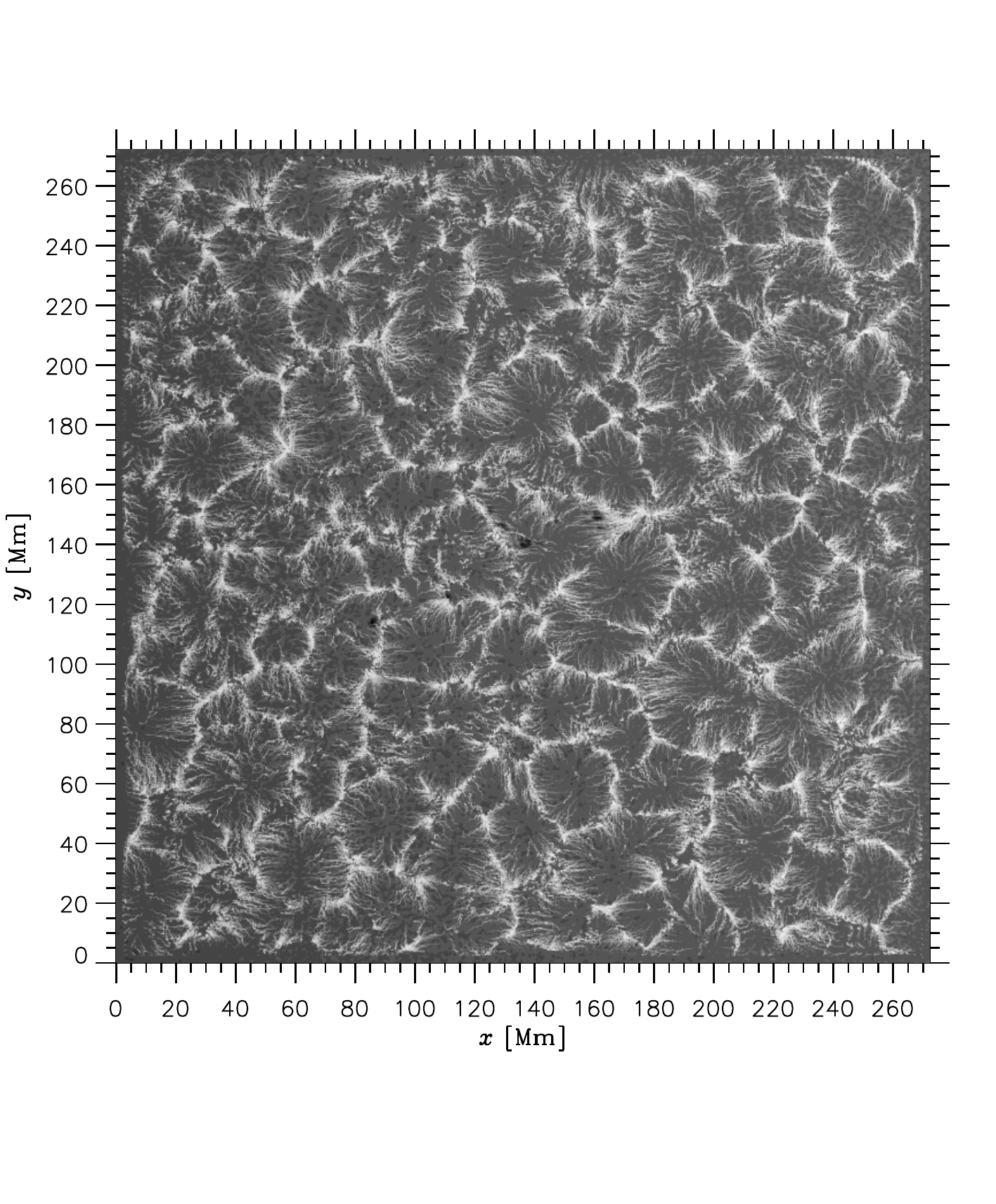}}
            \vspace{-12pt}\centerline{\tiny 2011 Feb 11, 04:44}
		\centerline{
			\includegraphics[width=0.5\textwidth,bb=0 80 680 742,clip]{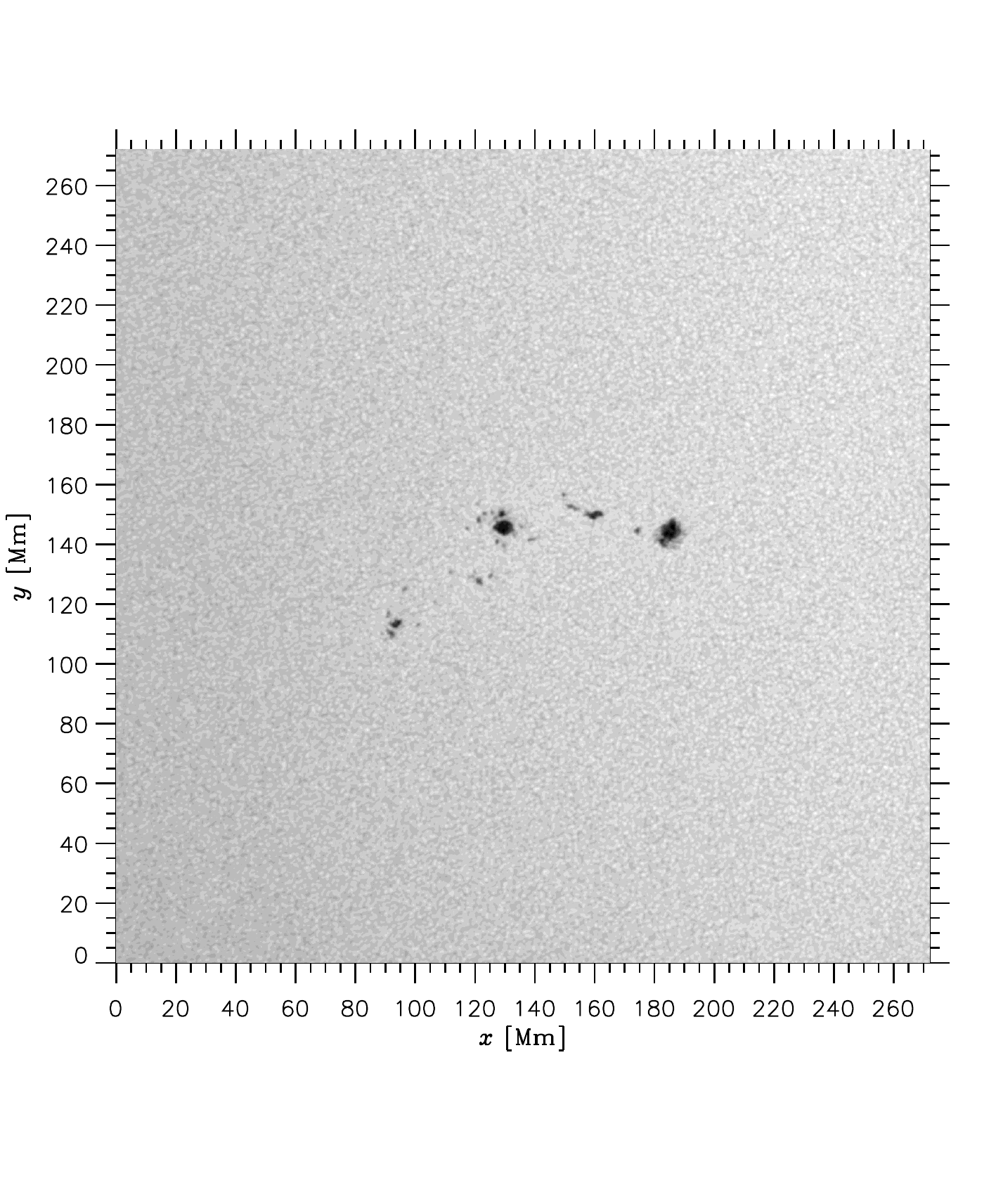}
			\includegraphics[width=0.5\textwidth,bb=0 80 680 742,clip]{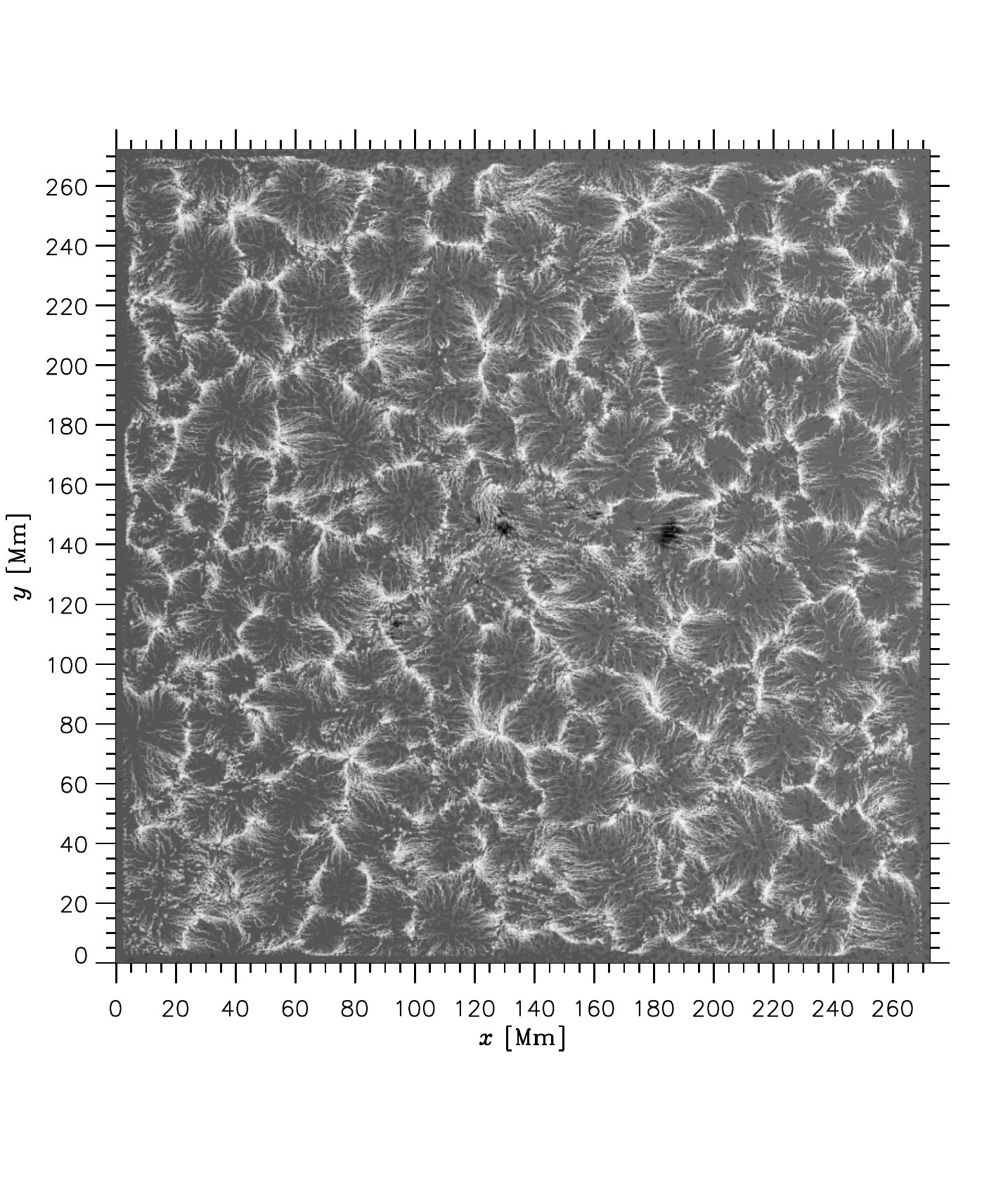}}
            \vspace{-12pt}\centerline{\tiny 2011 Feb 11, 23:29}
		\centerline{
			\includegraphics[width=0.5\textwidth,bb=0 80 680 742,clip]{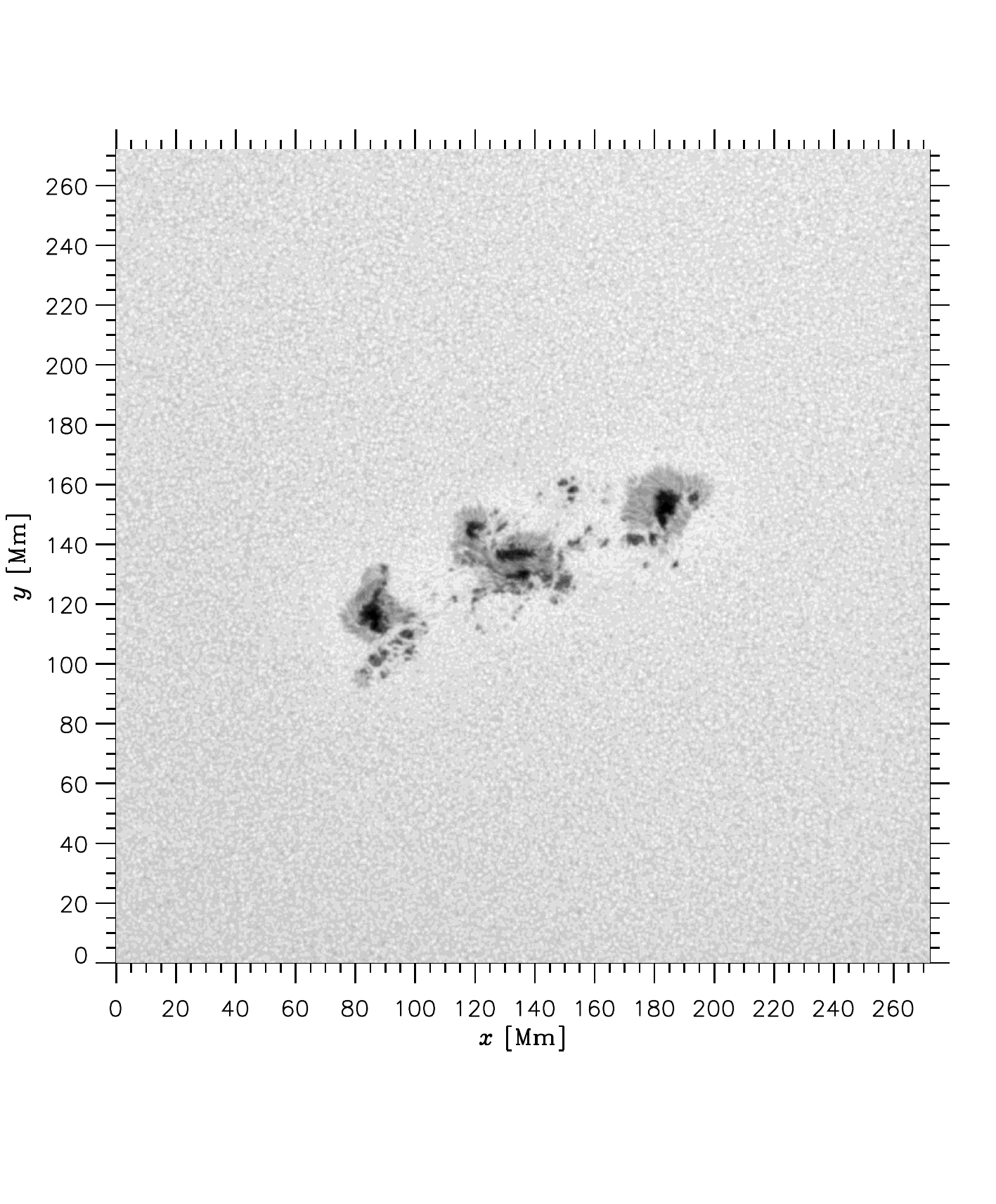}
			\includegraphics[width=0.5\textwidth,bb=0 80 680 742,clip]{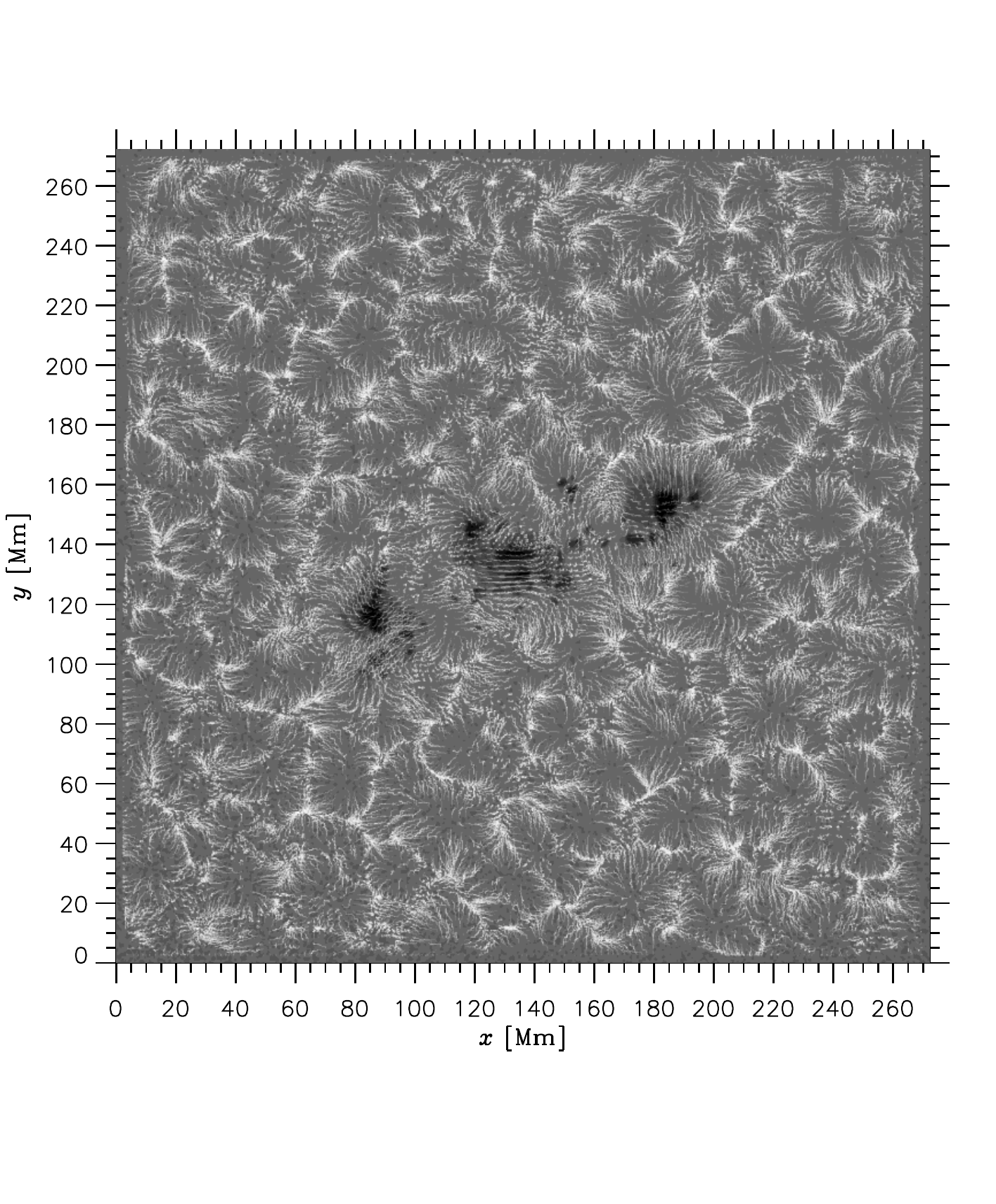}}
            \vspace{-12pt}\centerline{\tiny 2011 Feb 14, 00:00}
    \caption{Three development stages of AR 11158: images in continuum (left column) and maps of tracks overlaid with continuum images (right column) at times indicated below each row.}\label{Fig11158}
	\end{figure}
    
    \subsection{Active region 11158}\label{11158}

    In contrast to the two preceding cases, where the very nucleation of ARs was considered, let us now take a quick look at the history of an AR over a longer time interval during which it undergoes more complete development (Figure~\ref{Fig11158}). The duration of the tracking interval is here 3~h~45~m.

    First, about a dozen of pores have emerged against the background of a visually unperturbed supergranulation pattern (2011~Feb~11, 04:44); therefore, no divergent flow has preceded the pore formation. During the development of the first sunspots having penumbras (23:29), it can be seen that the supergranulation pattern, undergoing some evolution under the influence of magnetic fields, nevertheless retains its cellular structure. Finally, on Feb 14, at time 00:00, a well-developed sunspot group is present with a divergent flow \textit{around each of the large spots}---it can be classified as a moat flow. 

    Thus, the following development features of AR 11158 can be noted:
    \begin{itemize}
        \item The supergranulation pattern does not disrupt due to the emergence of strong magnetic fields, which evolve in mutual agreement with the convective velocity field.
        \item In particular, there are no signs of large-scale spreading, which would occur before the formation of pores if a strong-flux-tube loop emerged.
    \end{itemize}

    \subsection{Active region 11875}\label{11875}

    Until now, we have not touched upon the twist of the hypothetical flux tube because we considered only the early development stages of ARs, when sunspots are small and the magnetic-field twist can hardly be revealed. In this subsection, we will look at an example of a sunspot group in which the twist is quite distinct.
    
\begin{figure}
	\centerline{
		\includegraphics[width=0.5\textwidth,bb= 50 50 530 500,clip]{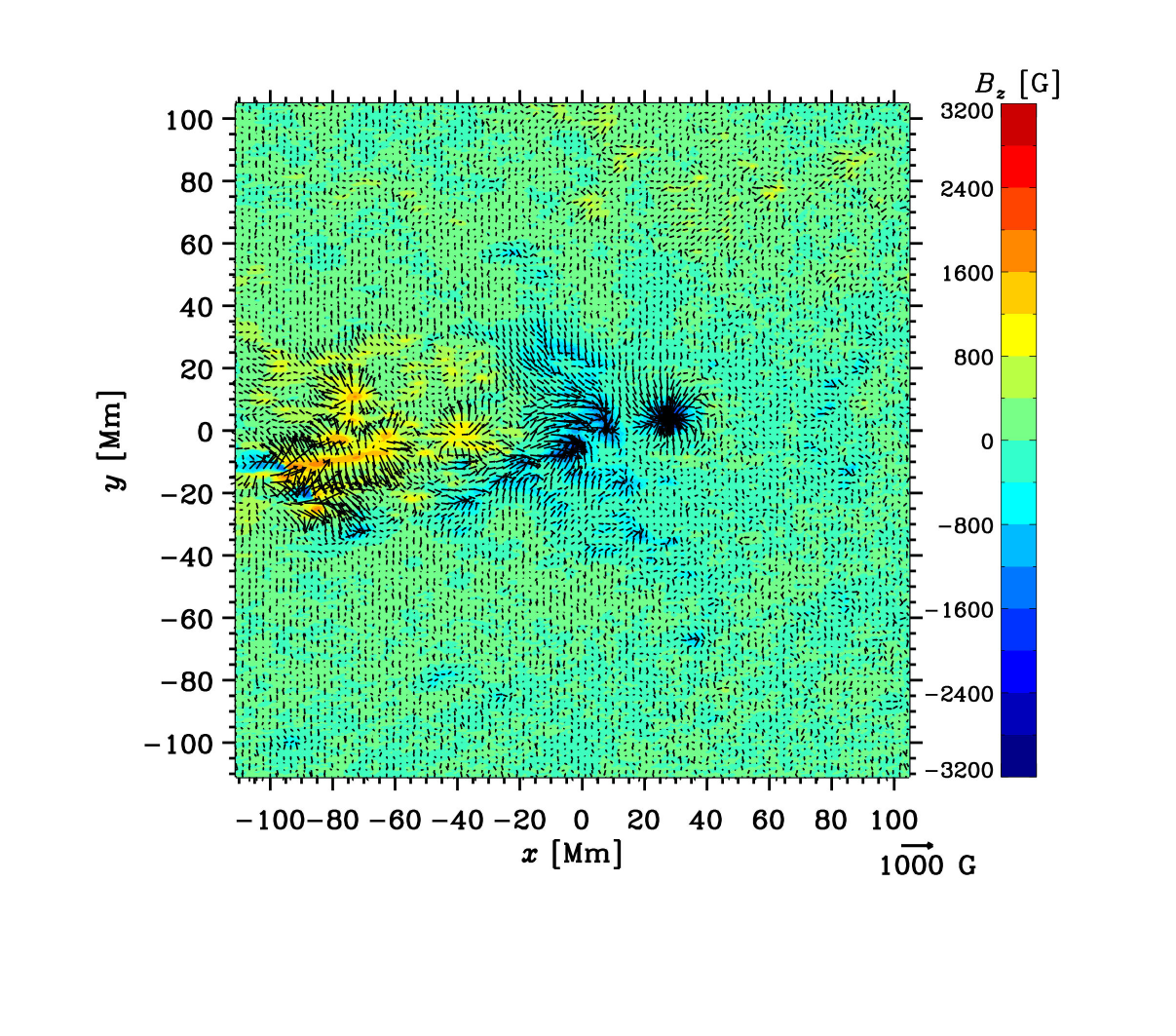} 
		\includegraphics[width=0.5\textwidth,bb= 50 50 530 500,clip]{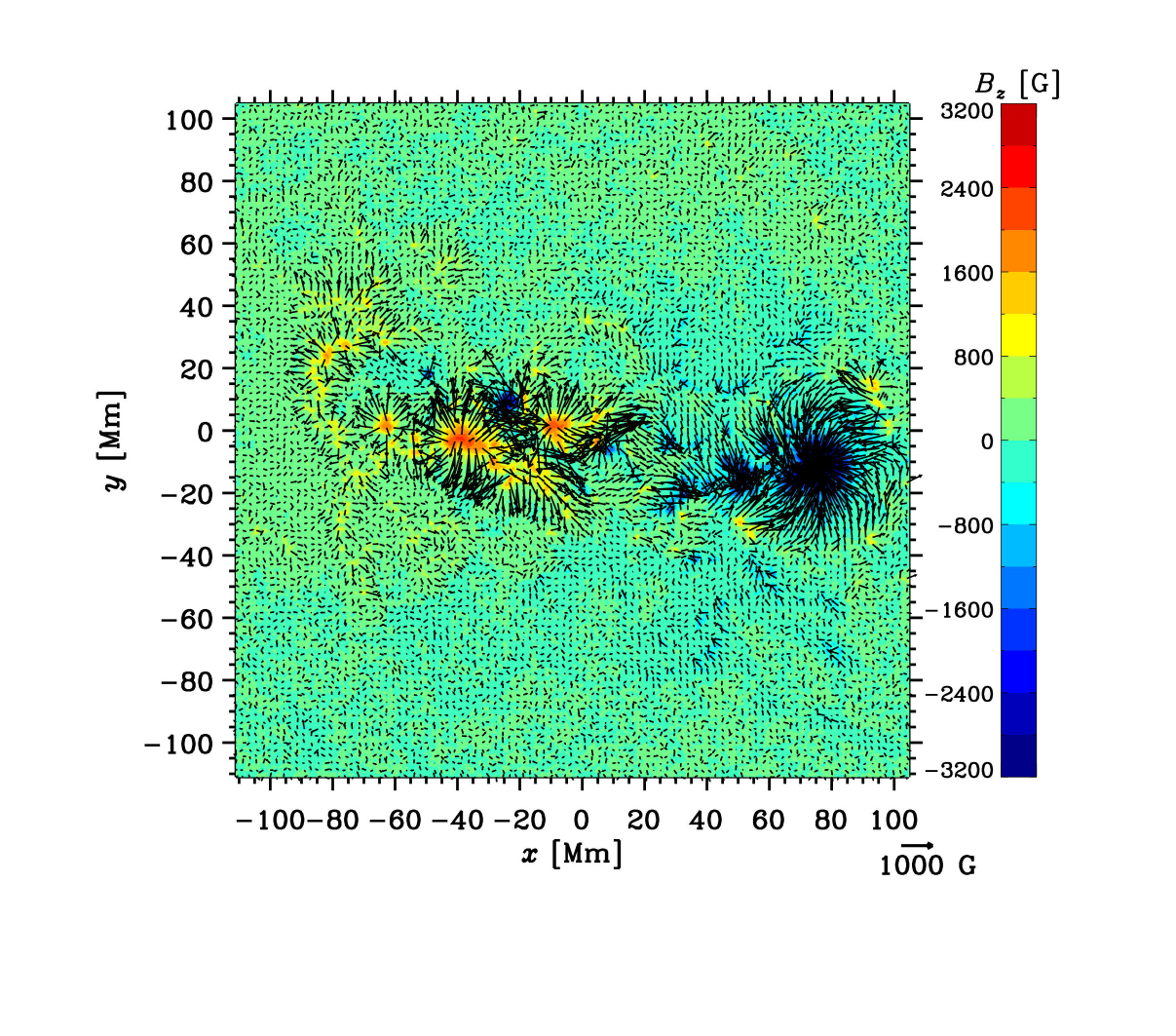}}\vspace{-12pt} 
		{\tiny \hspace{2.2cm}  2013 Oct 18, 07:00  \hspace{4.4cm}  2013 Oct 24, 23:00}
        \caption{Full-vector maps of the magnetic field of AR 11875 at two times.}\label{Fig11875}
    \end{figure}

    In Figure~\ref{Fig11875}, full-vector maps of the magnetic field in AR 11875 are presented for two times separated by almost a week. The AR is already well developed at the first time, and at the second, it contains a large leading sunspot with a very strong magnetic field, a smaller but also mature trailing sunspot, and a number of even smaller ones. What catches one's eye is a pronounced helical structure of the leading-sunspot magnetic field at the second time, which  is not distinguishable at the first time. Therefore, the magnetic-field of the leading sunspot (and only this one!) became helical when this sunspot was already approaching its late development stage. This directly indicates that the formation of the AR was not associated with the rise of a twisted flux tube.

    \begin{figure}
		 \centerline{\includegraphics[width=0.5\textwidth,bb= 0 0 504 250,clip]{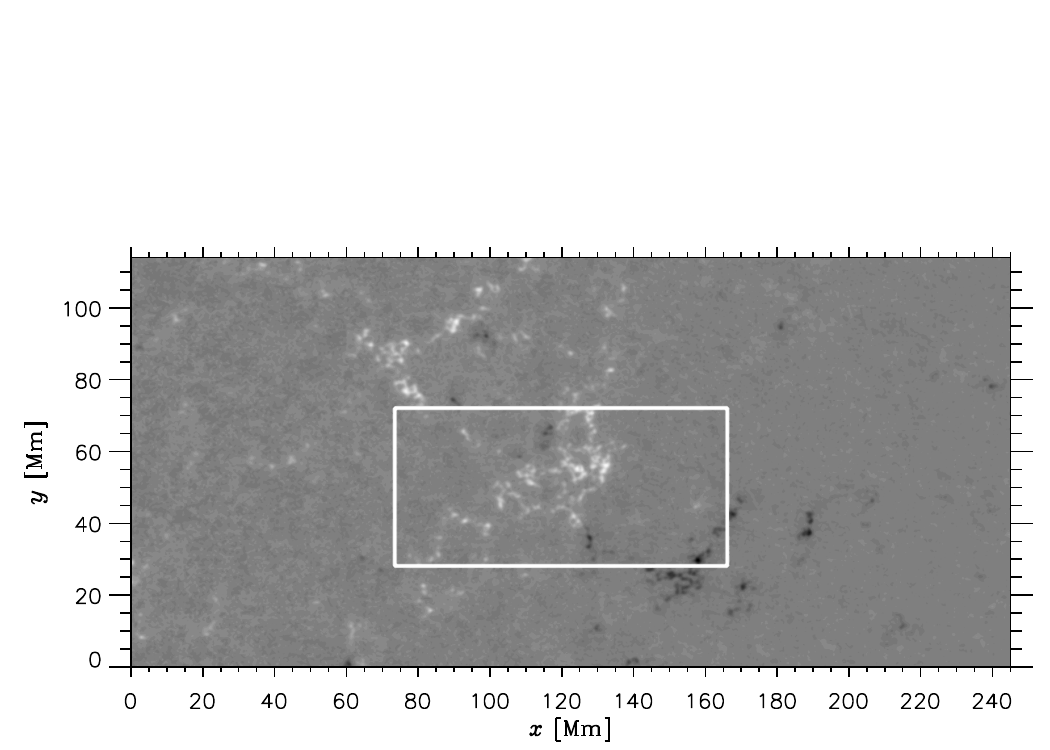}
                  \includegraphics[width=0.5\textwidth,bb= 0 0 504 250,clip]{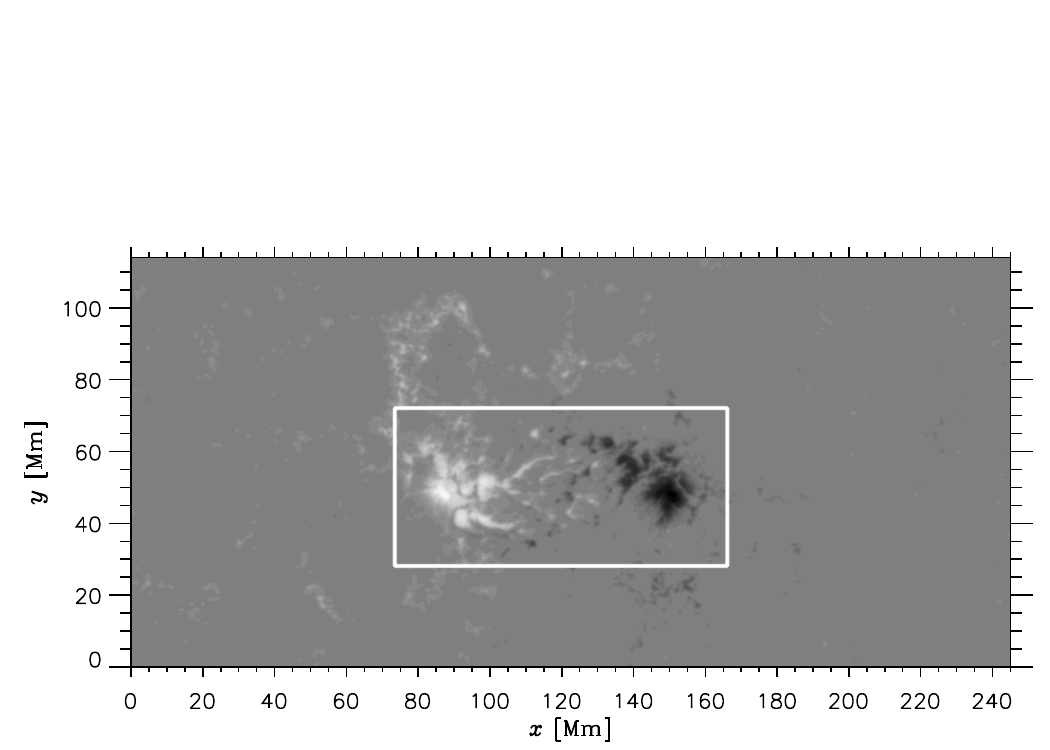}}
      \centerline{\includegraphics[width=0.5\textwidth,bb= 0 0 566 360]{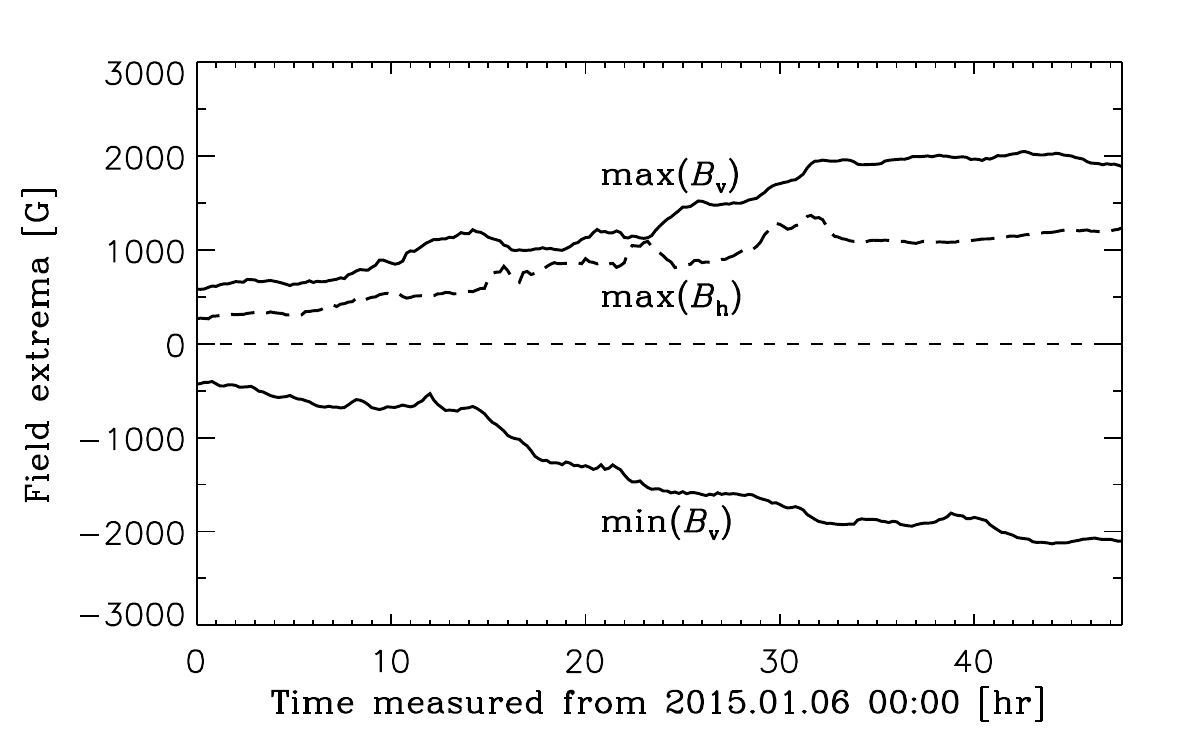}
                 \includegraphics[width=0.5\textwidth,bb= 0 0 566 360]{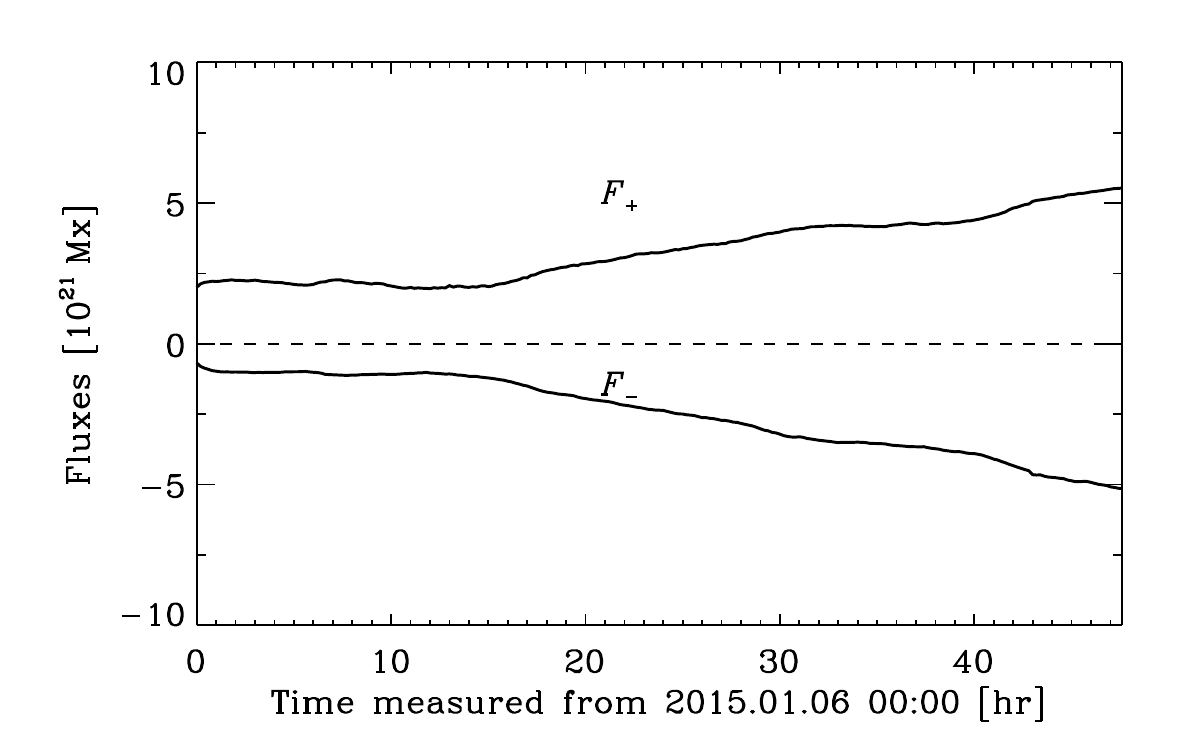}}
		 \caption{Evolution of the magnetic field of AR 12257. Top: maps of the vertical magnetic-field component for 2015 Jan 06, 00:00 (left) and 2015 Jan 08, 00:00 (right); the white frame delineates the area for which the extrema are taken and the fluxes are calculated. Bottom: time variation of the the extreme values of the vertical and horizontal magnetic-field components (left) and the positive and negative fluxes (right) through the area marked in the top panels.}\label{asymm}
	\end{figure}

	    \subsection{Asymmetry of magnetic fluxes}\label{asym}

    Among other things, \citet{Bumba1965} and \citet{Bumba1967} noted that in many cases, during the first days of AR development, the fluxes of the two magnetic polarities were not in balance. Later, an analysis of AR 12548 \citep{Getling_Buchnev_2019} revealed a strong dissymmetry between the positive and negative magnetic fields in their time variation. It was found that during the 20 hours preceding the onset of AR development, the positive magnetic flux exceeded the negative one by an order of magnitude, and these quantities remained very different in their absolute values over 20 hours after the start. A similar behaviour was demonstrated by the extrema of the two polarities. Subsequently, \citet{Grigoryev_etal_2020} reported magnetic-flux imbalances at the time of AR origin, based on the consideration of 48 emerging ARs. 

    Here, we exemplify one more evolutionary scenario of asymmetric development of the magnetic fluxes of the two signs. It refers to AR 12257, which we considered in Subsection~\ref{12257} (Figure~\ref{asymm}). Although the imbalance of magnetic-field fluxes is somewhat smaller in this case than in AR 12548, it is nevertheless quite notable. It can easily be seen that, for a long time, there is no symmetry between the positive and negative fluxes, and only by the end of the considered time interval do they become comparable in their absolute values.

    \section{Discussion and conclusion}\label{conclusion}

    In our opinion, the above-described results are arguments in favour of view B on the AR origin. Before summarising them, let us make some comments on a number of studies that may, at first glance, seem contradicting this view.
    
    \citet{Toriumi_etal_2014} use the Doppler velocity values to determine the horizontal divergent flows (HDFs). To this end, they employ a technique described previously \citep{Toriumi_etal_2012} and based on the construction of LoS-velocity histograms. Averaged quantities are used, and the considered flows are assumed to be mainly horizontal. The measured LoS velocities are interpreted as the LoS-projected horizontal velocities. However, the vertical (radial) velocity can make a significant contribution to the resultant values, all the more so because knowing the depths for the measured Doppler velocities is a fairly subtle point. (Likely, additional uncertainty can arise from independently shifting the histograms apexes for the AR and quiet regions to zero.) Thus, the authors' determinations are too indirect and therefore disputable. The LCT-based visualisation employed in my study offers a more direct and more detailed qualitative description.
    
    \citet{Attie_etal_2018} describe, going into fine details, the moat flow around the mature sunspot in the very peculiar AR 12673 and give particular attention to the perturbations of the moat flow before strong fresh magnetic flux emerges. The mature sunspot in this AR was already two solar rotations old, and a very complex pattern of magnetic field developed around this spot \citep[as described in particular by][]{Getling_2019}. The flux emergence occurs during the formation of a new supergranule-like cell, and the authors base their considerations on ``the hypothesis of emerging flux transported by the supergranulation''. This fits quite clearly into view B and does not contradict our results.
    
    Based on the consideration of about 60 emerging active regions, \citet{Birch_etal_2019} have found that their emergence is, on average, preceded by weak (20--40 m\,$\text{s}^{-1}$) \textit{converging} (rather than diverging) horizontal flows. It is worth noting that the velocities of supergranulation flows, as they can be estimated from the maps presented in our paper, are of order
    $$\frac{\text{cork's path}}{\text{tracking interval}} \sim \frac{10\ \text{Mm}}{6700\ \text{s}} \sim 1500\ \text{m}\,\text{s}^{-1},$$
    which exceeds the converging-flow speeds by two orders of magnitude. For this reason, the latter can hardly be caught against the background of the supergranulation flows in my flow maps. The authors also note that the flow pattern associated with emergence can be understood as a consequence of a correlation between the locations of flux emergence and the supergranulation pattern. This is in essence also a statement in line with view B.
    
    The evolutionary scenarios of the considered ARs reveal a number of remarkable features that sharply disagree with the rising-tube mechanism. Let us summarise these disagreements.

    \begin{itemize}
        \item As our consideration of the origin of ARs 12257, 12400 and 11158 (Figures~\ref{Fig12257}, \ref{Fig12400} and \ref{Fig11158}, respectively) demonstrates, the tracking of photospheric brightness elements (in essence, granules) definitely confirms the conclusion by \citet{Bumba1965} about the character of magnetic-flux emergence on the photospheric surface. Specifically, the supergranulation pattern is not destroyed by the emerging magnetic field of the developing AR, which would be inevitable if a loop of the strong-magnetic-field flux tube rose through the subphotospheric layers. Next, the first nucleating pores may be of the same magnetic polarity. On the other hand, even pores differing in magnetic polarity may originate within the same supergranule.
        
        \item The dissymmetry between the leading and trailing parts is typical of many ARs and may have diversified manifestations. The development of a sunspot group asymmetric with respect to magnetic polarities may be striking in some cases: we saw that AR~12400 originated giving rise to trailing-polarity pores against the background of widely distributed leading polarity. In many cases, the AR evolution is characterised by a strong asymmetry of magnetic fluxes of the two signs \citep[this makes us again remember findings by][]{Bumba1965}. Moreover, we saw sunspot configurations in the developing ARs far from symmetry between the leading and trailing parts from the very beginning of their development. All such asymmetries are obviously impossible in the case of the emergence of an $\Omega$-shaped flux-tube loop. 

        \item We have seen no signs of solar-plasma spreading on scales no less than the size of the whole AR. Such spreading would inevitably result from the rise of the flux tube.

        \item The full-vector magnetic-field maps of the considered ARs at early development stages do not exhibit signs of a helical magnetic-field structure, which would be indicative of flux-tube twisting. It turns out that a helical pattern may arise even much later than the AR originates. Moreover, in a bipolar sunspot group, a pronounced helical structure may be present only in one sunspot.
    \end{itemize}

    It is also worth remembering some other features of AR dynamics that have already been noted previously \citep{Getling_etal_2015,Getling_etal_2016} and that contradict the RTM. 
    \begin{itemize}
        \item Along with flows diverging from the expected location of the flux-tube apex, a strong horizontal field---a predominant magnetic feature on the scale of the entire active region---should appear as a precursor to the sunspot-group origin.
        \item The presence of ``parasitic'' polarities within the
          area filled with a predominant magnetic polarity is
          not accounted for by the RTM.
        \item The coexistence of differently directed vertical
          velocities inside the regions of a given magnetic
          polarity also appears to be inconsistent with this model.
  \end{itemize}
 
    Thus, it can be concluded from our analyses that the described scenarios are in sharp contradiction with the RTM. The development of the magnetic fields of the above-described ARs appears to be controlled by the flows of solar plasma and can be attributed to class-B scenarios. We present here a brief list of some possible alternatives to the RTM, which involve \textit{in situ} amplification and structuring of the magnetic field. 
  \begin{itemize}
      \item First of all, a class of mechanisms usually denoted as local dynamos should be mentioned. The idea of such processes traces back to \citet{Gurevich_Lebedinsky_1946}, who did not specify, however, any particular type of motion. To this class, the simple kinematic model suggested by \citet{Tverskoi_1966} also belongs. Realistic simulations of the convection-driven formation process for bipolar magnetic configuration without an initially present coherent flux tube of strong magnetic field were carried out by \citet{Stein_Nordlund_2012}.
      \item The process of self-aggregation of magnetic elements into clusters similar to that described by \citet{Chumak_2013} seems quite plausible. Such magnetic elements can arise as a result of the upward transport of small loops of relatively weak magnetic field by convective flows.
      \item A mechanism of magnetic-flux concentration due to a specific hydromagnetic instability was considered by \citet{Kitchatinov_Mazur_2000}. This instability crucially depends on the presence of fluid motion and on the  quenching of eddy diffusivity by the enhanced magnetic field, with the plasma cooling down. It can produce magnetic-flux concentrations similar to those observed in sunspots.
      \item One more possible local formation mechanism for bipolar configurations of strong magnetic field is related to the so-called negative effective-magnetic-pressure instability \citep[NEMPI, see][]{Warnecke_etal_2013,Warnecke_etal_2016}. It arises because the mean magnetic field suppresses turbulent hydrodynamic pressure more significantly than it enhances turbulent magnetic pressure. As a result, the total turbulent pressure (the sum of the hydrodynamic and magnetic components) is reduced. 
  \end{itemize}

   Strictly speaking, our reasoning cannot completely rule out the action of a RTM-like mechanism in some cases. However, the material presented both by many authors previously and in this paper demonstrates that the RTM is neither a unique nor a dominant mechanism of AR and sunspot formation.

   The results described here can hopefully stimulate purposeful studies of possible formation mechanisms for active regions and sunspot groups alternative to the RTM and based on view-B approaches. Such efforts may prove to be promising  in terms of understanding one of the key links in the chain of diverse phenomena collectively known as solar activity.
		
	\begin{acks}
        The helioseismological data used here were derived from HMI observational data available courtesy of the NASA/SDO and HMI science teams. The author conducted his study under the state assignment of Lomonosov Moscow State University. 
	\end{acks}
	%
	

	\bibliographystyle{spr-mp-sola}
	\bibliography{Getling}  

\end{document}